\documentclass{aa}

\usepackage[varg]{txfonts}

\usepackage{lineno}
\usepackage{natbib}
\usepackage{amsmath}
\usepackage{graphicx}
\usepackage{xspace}
\usepackage{xcolor}

\usepackage{hyperref}
\hypersetup{
    colorlinks=true,
    linkcolor={red!50!black},
    citecolor={blue!70!black},
    urlcolor={blue!80!black}
}

\graphicspath{{./figs/}{./subs/}}

\newcommand{\swift}{\textsl{Swift}\xspace}
\newcommand{\maxi}{\textsl{MAXI}\xspace}
\newcommand{\nicer}{\textsl{NICER}\xspace}
\newcommand{\xmm}{\textsl{XMM-Newton}\xspace}
\newcommand{\hxmt}{\textsl{Insight-HXMT}\xspace}
\newcommand{\ginga}{\textsl{Ginga}\xspace}
\newcommand{\rxte}{\textsl{RXTE}\xspace}
\newcommand{\cyg}{\mbox{Cyg~X-1}\xspace}

\usepackage[normalem]{ulem}

\usepackage{etoolbox}
\newtoggle{thesis}

\begin{document}

\title{Long term variability of Cygnus X-1. VIII. A spectral-timing look at low energies with \nicer}

\offprints{O.~K\"onig, present \email{ole.koenig@cfa.harvard.edu}}

\author{Ole~K\"onig\inst{1} \and Guglielmo~Mastroserio\inst{2,3} \and
  Thomas~Dauser\inst{1} \and Mariano~M\'endez\inst{4} \and Jingyi~Wang\inst{5} \and
  Javier~A.~Garc\'ia\inst{7,3} \and James~F.~Steiner\inst{6} \and Katja~Pottschmidt\inst{7,8} \and
  Ralf~Ballhausen\inst{7,9} \and Riley~M.~Connors\inst{10,3} \and Federico~Garc\'ia\inst{11} \and
  Victoria~Grinberg\inst{12} \and David~Horn\inst{1} \and
  Adam~Ingram\inst{13} \and Erin~Kara\inst{5} \and
  Timothy~R.~Kallman\inst{7} \and Matteo~Lucchini\inst{15} \and
  Edward~Nathan\inst{3} \and
  Michael~A.~Nowak\inst{14} \and
  Philipp~Thalhammer\inst{1} \and Michiel~van~der~Klis\inst{15} \and
  J\"orn~Wilms\inst{1}}

\institute{Dr. Karl Remeis-Observatory and Erlangen Centre for Astroparticle Physics, Friedrich-Alexander-Universit\"at Erlangen-N\"urnberg, Sternwartstr.~7, 96049 Bamberg, Germany \and
  Dipartimento di Fisica, Universit\`a degli Studi di Milano, Via~Celoria~16, 20133 Milano, Italy \and
  Cahill Center for Astronomy and Astrophysics, California Institute of Technology, Pasadena, CA 91125, USA \and
  Kapteyn Astronomical Institute, University of Groningen, P.O. Box 800, 9700 AV Groningen, The Netherlands \and
  MIT Kavli Institute for Astrophysics and Space Research, Cambridge, MA, 02139, USA \and
  Center for Astrophysics \textbar\ Harvard \& Smithsonian, 60 Garden Street, Cambridge, MA 02138, USA \and
  NASA Goddard Space Flight Center, Astrophysics Science Division, 8800 Greenbelt Road, Greenbelt, MD 20771, USA \and
  CRESST and Center for Space Sciences and Technology, University of Maryland Baltimore County, 1000 Hilltop Circle, Baltimore, MD 21250, USA \and
  CRESST and Department of Astronomy, University of Maryland, College Park, MD 20742, USA \and
  Department of Physics, Villanova University, 800 E Lancaster Avenue, Villanova, PA 19085, USA \and
  Instituto Argentino de Radioastronom\'{\i}a (CCT La Plata, CONICET; CICPBA; UNLP), C.C.5, (1894) Villa Elisa, Buenos Aires, Argentina \and
  European Space Research and Technology Centre (ESTEC), European Space Agency (ESA), Keplerlaan 1, NL-2201 AZ Noordwijk, the Netherlands \and
  School of Mathematics, Statistics and Physics, Newcastle University, Herschel Building, Newcastle upon Tyne NE1 7RU, UK \and
  Department of Physics, Washington University in St. Louis, Campus Box 1105, One Brookings Drive, St. Louis, MO 63130-4899, USA \and
  Anton Pannekoek Institute for Astronomy, University of Amsterdam, Science Park 904, NL-1098XH Amsterdam, the Netherlands
  }

\abstract{The Neutron Star Interior Composition Explorer (\nicer) monitoring campaign of \cyg allows us to study its spectral-timing behavior at energies ${<}1$\,keV across all states. 
The hard state power spectrum can be decomposed into two main broad Lorentzians with a transition at around 1\,Hz. The lower-frequency Lorentzian is the dominant component at low energies. The higher-frequency Lorentzian begins to contribute significantly to the variability above 1.5\,keV and dominates at high energies.
We show that the low- and high-frequency Lorentzians likely represent individual physical processes. The lower-frequency Lorentzian can be associated with a (possibly Comptonized) disk component, while the higher-frequency Lorentzian is clearly associated with the Comptonizing plasma.
At the transition of these components, we discover a low-energy timing phenomenon characterized by an abrupt lag change of hard (${\gtrsim}2$\,keV) with respect to soft (${\lesssim}1.5$\,keV) photons, accompanied by a drop in coherence, and a reduction in amplitude of the second broad Lorentzian. 
The frequency of the phenomenon increases with the frequencies of the Lorentzians as the source softens and cannot be seen when the power spectrum is single-humped. 
A comparison to transient low-mass X-ray binaries shows that this feature does not only appear in \cyg, but that it is a general property of accreting black hole binaries.
In \cyg, we find that the variability at low and high energies is overall highly coherent in the hard and intermediate states. The high coherence shows that there is a process at work which links the variability, suggesting a physical connection between the accretion disk and Comptonizing plasma. 
This process fundamentally changes in the soft state, where strong red noise at high energies is incoherent to the variability at low energies.}


\date{Received 24 January 2024 / Accepted 12 May 2024}

\maketitle 

\section{Introduction}
\label{sec:introduction}

Black hole X-ray binary systems, where a stellar-mass black hole
accretes from a donor star, show rich phenomenology in the spectral
and timing domains. These systems can exhibit drastic changes in their
emission behavior on timescales of hours to weeks. Their behavior can
be broadly classified into a (high) soft, (low) hard, and intermediate
state \citep[][and references therein]{Belloni10a}. In the soft state,
the emission is dominated by thermal radiation from an accretion disk,
while in the hard state most of the radiation is due to inverse
Compton scattering of soft photons in a ${\sim}100$\,keV hot plasma,
which is often (but potentially misleadingly) called the ``corona'' \citep{LiangPrice1977a,SunyaevTruemper1979a,Wilms06a}. The physical reasons and the
changes of the accretion flow that occur during the state changes are
still debated, as is the nature of the Comptonizing plasma
\citep[e.g.,][]{Markoff05a,Kara19a,Zdziarski21a}.

\cyg \citep{Bowyer1965Sci} is a high-mass X-ray binary (HMXB) with a
$(21.2\pm 2.2)\,\mathrm{M}_\odot$ black hole \citep{MillerJones21a}
that accretes matter from the stellar wind of the supergiant
HDE~226868 \citep{Walborn73a}. \cyg is a persistent source that is
located on the lower branch of the q-shaped
hardness-intensity diagram \citep[HID; see][]{Belloni10a,Nowak12a},
where it transitions from a 
hard state to a softer state
that still shows a power law component at energies ${>}10$\,keV \citep[e.g.,][]{Wilms06a}.
Each of the states shows unique variability properties at short timescales. 

The power spectral density (PSD) in the hard state of \cyg is generally well-modeled using a superposition of up to four Lorentzian components (\citealt{Nowak2000a} and see \citealt{Miyamoto91a} for an earlier application of Lorentzians to GX~339$-$4) with a total root mean square (RMS) variability amplitude of 30--40\% (2--13\,keV; \citealt{Pottschmidt03a}). In the intermediate state, the RMS drops and the PSD becomes single-humped \citep{Pottschmidt03a}. The soft-state PSD is dominated by red noise \citep{Cui97a} at ${\sim}20\%$ RMS (2--15\,keV; \citealt{Grinberg14a}), which is higher than what is usually seen in low-mass X-ray binary (LMXB) black holes in the soft state \citep[e.g.,][]{Belloni2005a}. In this state, the RMS variability of \cyg strongly increases with energy to more than 30\% RMS at ${\gtrsim}10$\,keV \citep{Pottschmidt2006a,Grinberg14a,Zhou22a}.

In addition to power spectra, the phase of the cross power spectral density is used to analyze the arrival times of photons in different energy bands, so-called ``time lag'' spectra. Measuring such spectra is possible if the light curves in the two energy bands are sufficiently coherent \citep[][and references therein]{VaughanNowak97a,Nowak99a}. Throughout this paper we denote high-energy photons lagging low-energy photons as ``hard lags'', and use the convention that they are positive. Soft lags, where the soft energy photons lag the high energy ones, are therefore negative.

Hard lags between photons in the 1.5--3\,keV and 12--42\,keV bands were first identified in \cyg by \citet{Priedhorsky79a}. At energies ${\gtrsim}2$\,keV, the lag-frequency spectrum roughly follows a $\nu^{-0.7}$ shape and can show shelf-like structures coinciding with the overlap region of the Lorentzians in the PSD \citep{miyamoto:1989,Nowak99a}.
Furthermore, in the LMXB GX~339$-$4, \citet{Nowak1999c} showed that these structures can coincide with drops in coherence by up to 10\% in the frequency range where the Lorentzians overlap. \citet{Nowak1999c} argue that this behavior indicates that the components are incoherent with respect to each other.

Time lags are of use when investigating the geometrical changes of the Comptonizing medium or other time-delaying effects such as diffusion. 
In accreting stellar mass black holes, the amplitude of hard lags increases as the source moves (fully or ``failed'') towards the soft state. This behavior is accompanied by an overall drop in coherence and an increase in the frequencies of the Lorentzians that describe the power spectrum (see \citealt{Cui97a,Pottschmidt03a,Grinberg14a} for \cyg and \citealt{Wang22a} for several LMXB black holes). 

The increase of the hard lags in the transition between the hard and the softer state is connected with the observation of radio flares. This simultaneous change of the radio and X-ray properties has been interpreted as an indication of a change in the size of the Comptonizing plasma \citep{Cui97a,Pottschmidt03a,Wilms07a}.
Similarly, amplitude and frequency changes of soft lags are also seen in a number of LMXB black holes, and have again been interpreted as changes in the size of the Comptonizing plasma \citep{Kara19a,Wang22a}, or a
change in the inner accretion disk radius \citep{DeMarco21a}.
The interpretation of lags in black holes as size changes of the Comptonizing plasma or other  changes in the accretion flow geometry such as reprocessing in the accretion disk (``reverberation''), however, is not unique. For \cyg, \citet{Lai22a} showed that up to 50\,ms soft lags at low frequencies 
can also be attributed to scattering or recombination processes in the stellar wind. 
In the state transition of the LMXB Swift~J1727.8$-$1613, polarimetric properties associated with the Comptonizing plasma are not seen to change \citep{Ingram2023b_ARXIV}.
Note, however, that state transitions in LMXB black holes may be due to different physics than the state transitions in \cyg.

\cyg has been subject of a decades-long effort to characterize its spectral and
timing properties, which was mainly based on data by the Rossi X-Ray Timing Explorer (\rxte) mission. The results of this effort have been published in the ``Long term variability of Cygnus~X-1'' paper series
\citep{Pottschmidt03a,Gleissner04b,Gleissner04a,Wilms06a,Grinberg13a,Grinberg14a,Grinberg15a}. This \rxte monitoring campaign has revealed a wealth of variability phenomena due to both the accretion flow and the interaction of the X-rays with the surrounding stellar wind, such as the imprints of strong transient absorption events in the lightcurve called ``dipping'' \citep[][and references therein]{Grinberg15a,Hirsch19a,Lai22a}.
In this eighth paper of the series, we revisit the timing behavior of \cyg using \nicer \citep{Gendreau16a}, which offers sufficient energy and timing resolution
to permit spectral and timing studies with high signal-to-noise
observations
\citep[e.g.,][]{StieleKong18,Kara19a,Zhang20a,Wang20a}. In particular,
the design of \nicer's X-ray Timing Instrument allows us to investigate the variability
on timescales of ms without pile-up or deadtime distortions.
We analyze the \nicer archive of \cyg up to April 2022 (cycle~4). 
The main emphasis is to characterize the variability of the accretion disk at low X-ray energies ${\lesssim}2$\,keV and its connection to the Comptonizing medium, as well as putting the timing phenomenology into context of the previous \rxte work.

The paper is organized as follows. In Sect.~\ref{sec:cygx1_data_reduction}, we discuss the data reduction. In Sect.~\ref{sec:evolution_and_hardness}, we classify the spectral states and hardness evolution, emphasizing the extension of previous monitoring campaigns to low X-ray energies. In Sect.~\ref{sec:spectra_timing_analysis}, we analyze three representative hard, intermediate, and soft state observations, and report the discovery of previously unknown time lag and coherence phenomena, both in the hard and soft state. We discuss and interpret our results in Sect.~\ref{sec:discussion} and summarize the paper in Sect.~\ref{sec:summary}.

\section{Data Reduction}
\label{sec:cygx1_data_reduction}

Our \nicer dataset of \cyg comprises 52 observations performed between
2017 June 13 and 2022 April 15 (see observation log in
Table~\ref{tab:observation_log}). The data were reduced with HEASOFT
6.29c and NICERDAS 2021-08-31\_V008c. We exclude data taken inside the South
Atlantic Anomaly (\texttt{saafilt=YES}) and all data taken with detectors~14 and~34, which have exhibited anomalous behavior.
We filter out data measured below an angle of $15^\circ$ to the Earth's X-ray bright limb to avoid contamination from fluorescence in the atmosphere, as well as data taken below $30^\circ$ with respect to Earth's horizon in case the surface is illuminated by the Sun to prevent optical loading by reflected optical light.
We do not filter out all data taken during orbit day. Instead, we limit our analysis to observations which have a per-detector undershoot rate below $200\,\mathrm{counts}\,\mathrm{s}^{-1}$, as recommended by the
\nicer analysis guide\footnote{\url{https://heasarc.gsfc.nasa.gov/docs/nicer/data_analysis/workshops/NICER-Workshop-Filtering-Markwardt-2021.pdf}}.
Furthermore, to ensure that no low-energy noise intrudes the data products, we constrain our analysis to the energy range above 0.5\,keV.
Overshoots are constrained to the default range of less than $30\,\mathrm{counts}\,\mathrm{s}^{-1}$ per detector to exclude periods of high particle background.
We then extract lightcurves in multiple energy bands between 0.5--10\,keV with a time resolution of 1\,ms. 
These lightcurves have full-instrument count rates ranging from 500--$30\,000\,\mathrm{counts}\,\mathrm{s}^{-1}$, while typical background rates of \nicer are estimated to be on the order of $1\,\mathrm{count}\,\mathrm{s}^{-1}$ \citep{Remillard2022a}.
Given the brightness of \cyg, no background correction is necessary for the spectral and timing products given the overshoot restriction.
The data analysis is performed with ISIS 1.6.2-47 \citep{ISISHouck02}.

For the hardness diagrams in Sect.~\ref{subsec:state_classification} and \ref{subsec:color-color}, we rebin the lightcurves to a time resolution of 100\,s. 
For the Fourier analysis in Sect.~\ref{sec:spectra_timing_analysis}, we compute the timing products by splitting the lightcurves into segments of 16.384\,s and, in case of gaps, only include data that fully
cover a segment to avoid aliasing. This approach rejects
${\sim}5.6$\,ks of data, such that we remain with a total of
${\sim}211$\,ks of exposure. 
We do not further subdivide the 52 observations based on breaks in the data which arise due to the orbit of the ISS (``good time intervals'').
All PSDs in this study are normalized to fractional RMS units \citep{BelloniHasinger90b,Miyamoto91a} and have the Poisson-noise component, $2/\bar{x}$, subtracted, where $\bar{x}$ is the mean count rate.
Generally, the impact of deadtime on the Fourier products in \nicer is low, especially at low frequencies \citep[e.g.,][]{Stevens18a}. 
Since we concentrate on frequencies below 10\,Hz in this study, we do not perform a deadtime correction.
The coherence and its
uncertainty are determined using Eqs.~2 and~8 of \citet{VaughanNowak97a}\footnote{We note that there is an incorrect factor $\gamma^2$ in the numerator of the intrinsic coherence calculation in 
  \citet[][Eq.~8]{Nowak99a}, see \citet[][Eq.~79]{BendatPiersol86} and
  the text after Eq.~2 of \citet{VaughanNowak97a}.}. The RMS and its uncertainty are computed using Eqs.~5 and 11 of \citet{Vaughan03a}\footnote{For the uncertainty
  of the RMS we note that a number of slightly different definitions are also
  used, e.g., \citet[][Eq.~14]{Uttley14a} and
  \citet[][Eq.~22]{Ingram19a}.}.
The uncertainty of the
(segment-averaged) lags for each Fourier frequency is calculated from
the coherence \citep[][their Eq.~16]{Nowak99a}. Finally, we
logarithmically rebin all timing products in frequency.

\section{Evolution of the Flux and Hardness over all States}
\label{sec:evolution_and_hardness}

\subsection{Position of \cyg in the q-diagram}
\label{subsec:cygx1_in_q-diagram}

\begin{figure}
  \resizebox{\hsize}{!}{\includegraphics{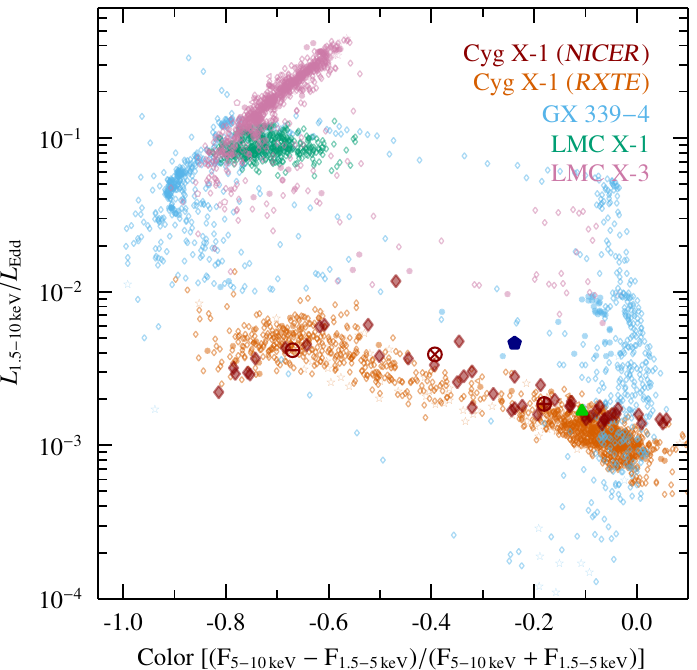}}
  \caption{Color-luminosity diagram of \cyg and other black hole X-ray
    binaries from the \rxte archive. The transient source GX~339$-$4
    shows a \texttt{q}-shaped pattern, while \cyg only populates the
    lower branch of the \texttt{q}. The soft and hard bands are
    1.5--5\,keV and 5--10\,keV, respectively, and the color is derived
    from (absorbed) fluxes of phenomenological model fits. The \nicer
    data (red diamonds) cover the entire color range seen in \cyg with \rxte. 
    Red circles with a plus, times, and minus sign within it denote \nicer observation 2636010101, 0100320110, and 1100320122 of \cyg, respectively.
    The blue pentagon and green triangle denote \nicer observations 1200120268 of MAXI~J1820+070 and 2200530129 of MAXI~J1348$-$630, respectively.}
  \label{fig:q_diagram}
\end{figure}

We start our analysis by putting the \nicer data into the context of previous monitoring campaigns of \cyg and other black hole binaries observed with \rxte, using a color-luminosity diagram (Fig.~\ref{fig:q_diagram}). 
To be able to compare the \nicer and \rxte data, we derive the fluxes and hardnesses of each observation from simple empirical spectral fits rather than employing the traditional approach of showing a hardness intensity diagram in terms of mission-dependent count rates and hardness ratios.

To model the continuum, we use a power law plus disk black body. The absorption of this continuum in the stellar wind and the interstellar medium is described with \texttt{TBabs} \citep{Wilms00a}, and the iron line at around 6.7\,keV is modeled with a broad Gaussian emission line (modeling the detailed shape of the relativistic line is not necessary). Two further narrow Gaussians are added at 6.4\,keV and 7.1\,keV, respectively, to model Fe~K$\alpha$/$\beta$ emission from neutral iron.
As we are interested only in the behavior of the overall broad continuum components, we include a 5\% systematic uncertainty to account for more subtle features not at our focus which are omitted by our spectral model, and which is also sufficiently large to encompass any calibration uncertainties of \nicer \footnote{\url{https://heasarc.gsfc.nasa.gov/docs/nicer/data_analysis/nicer_analysis_tips.html}}. Furthermore, we fit the data only above 2.5\,keV in order to be able to better compare it to the \rxte results.
We then display the luminosity in units of the Eddington luminosity (assuming the black hole mass and distance estimate of \citealt{MillerJones21a}) as a function of the flux ratio for the 1.5--5\,keV and the 5--10\,keV bands. This way the diagram is independent of the detector used and also permits us to directly compare different sources \citep[see][and references therein]{Wilms06a,Barillier2023a}.

Figure~\ref{fig:q_diagram} shows that the \nicer
data cover the hard and soft state of \cyg, connected by observations
in the parameter space of the lower transition from the soft back to the hard state in transient black hole
binaries. Contrary to the other sources, \cyg always remains on the lower horizontal track in this diagram. The \nicer-measured luminosity is
approximately 20\% higher than that of \rxte, which is likely due to systematics in the flux calibration of \nicer and \rxte/PCA data \citep{Jahoda2006a,Shaposhnikov2012a,Garcia2014a}. See
\citet{Barillier2023a} for a detailed discussion of the fitting
approach of the \rxte data and a comparison of the different sources used in the preparation of the color-luminosity diagram.

\subsection{State Classification and Long-Term Monitoring}
\label{subsec:state_classification}

\begin{figure*}                   
  \sidecaption                  
  \includegraphics[width=12cm]{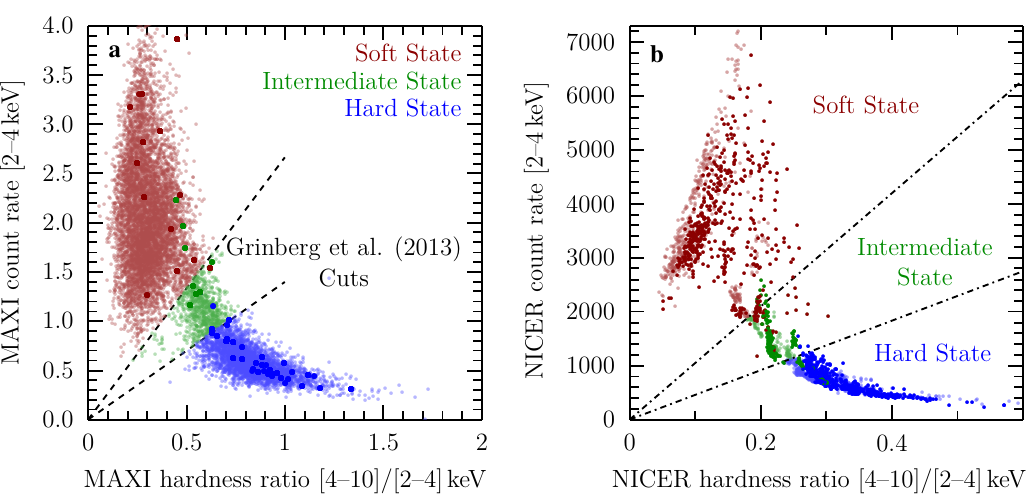}
  \caption{Hardness-intensity diagrams of \cyg. \textbf{a)} HID using \maxi count rates from 6\,h binned lightcurves. Light points are classified according to \citet[][dashed line]{Grinberg13a}. Solid points denote \maxi time bins which have a \nicer bin falling into the 6\,h time frame, colored according to our \nicer cuts. \textbf{b)} HID using \nicer count rates from 100\,s binned lightcurves. Light points are colored according to our \nicer criterion. Solid points denote \nicer time bins which fall into the 6\,h \maxi time frame and are classified according to the \maxi cuts. The \nicer cuts (dash-dotted line) are chosen such that the \maxi criterion is reproduced with minimal contamination.}
  \label{fig:hid}
\end{figure*}

In order to classify the \nicer observations, we place them into the context of existing state classifications of \cyg. Such
spectral states were introduced to roughly characterize the \textit{continuous} changes in brightness as well as the spectral and timing properties of black hole X-ray binaries \citep{Tananbaum1972a,MendezVanderKlis1997a,Belloni10a}. Specifically, we use the \maxi criteria given by \citet{Grinberg13a} to classify \cyg into a hard, intermediate, and soft state (Fig.~\ref{fig:hid}a). 
For each observation, we assign a state based on those bins in 100\,s resolved \nicer lightcurves which are strictly simultaneous to the 6\,h binned \maxi lightcurves\footnote{\url{http://maxi.riken.jp/pubdata/v7lrkn/J1958+352/}}. In order to verify the precision of the mapping, we color strictly simultaneous time bins in the HIDs according to the state definition of the other mission. We define the cuts in the \nicer HID (Fig.~\ref{fig:hid}b) such that the \maxi criterion of \citet{Grinberg13a} is best reproduced. Specifically, we determine that \cyg is in the soft state for a 2--4\,keV \nicer count rate, $R$, that satisfies $R\geq 10500\cdot h$ and is in the hard state for $R< 4580\cdot h$, where $h$ is the count rate ratio of the 4--10\,keV to the 2--4\,keV band.

The derived \nicer criterion is then applied to all \nicer data to classify the states in the HID. The overall behavior in the HID is similar to previous monitoring campaigns from \rxte \citep{Grinberg13a} and \maxi \citep{Sugimoto16a}. \cyg exhibits a smooth transition from the hard to the intermediate state and shows a kink when the source is in the soft state.

We emphasize that the state definitions used so far follow the traditional names for the states of \cyg, which are inconsistent with state designations used for other black hole binaries.
In addition, state classifications in \cyg and other sources should only be taken to be approximate as they attempt to discretize a continuous source evolution. 
We therefore regard the photon index, $\Gamma$, inferred from our power law fits of Sect.~\ref{subsec:cygx1_in_q-diagram}, as the most salient property characterizing the observations, rather than the state designation.

\begin{figure}
  \resizebox{\hsize}{!}{\includegraphics{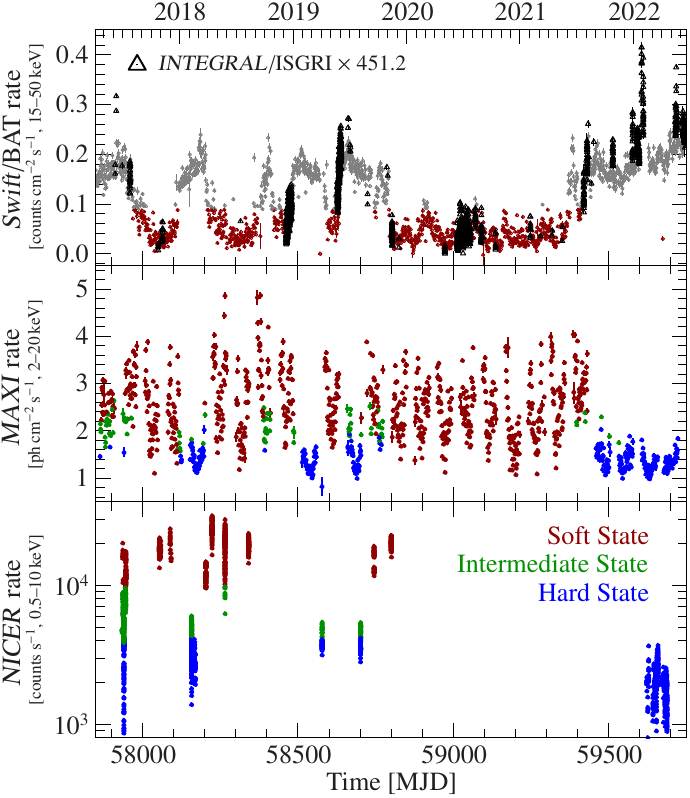}}
  \caption{\swift/BAT, \textsl{INTEGRAL}/ISGRI (top panel), \maxi (center), and \nicer (bottom) light curves of \cyg. In the top panel, gray
      data points denote either the hard or intermediate state as
      \swift/BAT's energy range does not allow a further
      differentiation. Black triangles show \textsl{INTEGRAL}/ISGRI
      count rates in the 30--50\,keV band, showing each science window as one data point. These data are scaled by the ratio of the
      average BAT and ISGRI count rates over the whole campaign in
      order to align them with the \swift/BAT data.}
    \label{fig:longterm_lc}
\end{figure}

Finally, in Fig.~\ref{fig:longterm_lc}, we put the \nicer data into perspective of the long-term evolution of \cyg. We show the \nicer and \maxi lightcurves and also add \swift/BAT data, using the state classification methodology of \citet{Grinberg13a}, as well as \textsl{INTEGRAL}/ISGRI count rates (Thalhammer et al., to be submitted).
The long-term lightcurves show that \cyg underwent several state transitions during the \nicer observations. 

\subsection{Color-Color Behavior throughout the Spectral States}
\label{subsec:color-color}

\begin{figure}
  \resizebox{\hsize}{!}{\includegraphics{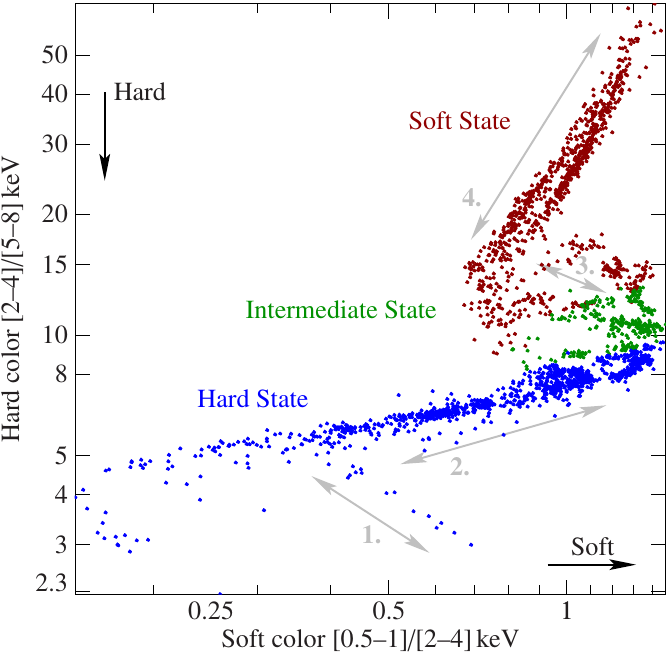}}
  \caption{Color-color diagram of \cyg. The diagram 
  shows distinct regions where the source undergoes several turns. Numbered gray arrows indicate preferred tracks which \cyg follows when transitioning between the hard and the soft state. Some observations in the hard state show ``nose''-shaped tracks where the color-color bends down towards a harder spectral shape (``1.''), which is called dipping.   When \cyg softens, the source shows large variations dominantly in soft color (``2.''). As the source transitions through the intermediate into the soft state, it follows a distinct zig-zag shape driven by the soft color (``3.''). When \cyg is in the pure soft state, it still ranges significantly in hard color (``4.'').}
  \label{fig:ccd}
\end{figure}

The color-color diagram of the \nicer observations is shown in Fig.~\ref{fig:ccd}.
The hard state follows a broad lower track which varies most strongly in soft color. The diagram shows at least two outlier tracks, which deviate from the straight line (gray arrow labeled ``1.''). 
These ``nose''-shaped variations are well known and can be attributed to dipping events where absorption of soft X-rays in the clumps of the stellar wind leads to a hardening of the spectrum \citep{Nowak11a,Hirsch19a,Grinberg20a,Lai22a}. 

When \cyg undergoes a full transition to the soft state, it first softens and moves rightward in the diagram, mainly changing in soft color (``2.''). The track then bends around and the source hardens, again changing predominantly in the soft color (``3.'').
In the soft state, the track bends around again and softens, both in the soft and hard color (``4.''). This overall behavior creates a zig-zag shape in the color-color diagram.

In summary, the \nicer color-color diagram shows a rather complex pattern with a distinctive zig-zag shape in the transition that has not been seen from monitoring campaigns of \cyg limited to the hard X-ray regime \citep{DoneGierlinski2003a,Zdziarski2016a}\footnote{While many color-color diagrams of \cyg are widely available, also at softer energies, they often only cover individual spectral states and not the full transition cycle \citep[e.g.,][]{Nowak11a,Grinberg20a}.}. This shape disappears for the \nicer data when we use the harder energy bands in the color-color diagram typical of previous campaigns. The more complex soft behavior is therefore a property that only emerges at soft X-ray energies and is probably due to absorption and ionization effects in the foreground material and/or complex behavior of the accretion disk emission.

\begin{figure}
  \resizebox{\hsize}{!}{\includegraphics{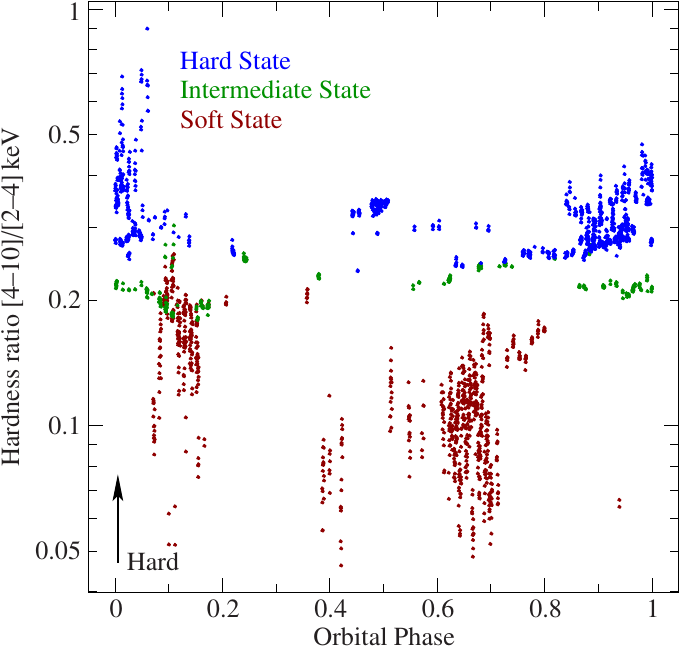}}
  \caption{Orbital phase coverage in the \nicer dataset of \cyg.}
  \label{fig:orbital_phase_diagram}
\end{figure}

A further complication in the analysis of observations of the
intrinsic spectral and timing properties of the accretion flow in
Cyg~X-1 is that HDE~226868 has a strong stellar wind which has an
influence both on the observable spectral and the timing properties
of \cyg \citep[][and references therein]{Hirsch19a,Lai22a}. It is
especially strong at superior conjunction
($\phi_\mathrm{orb}=0$), where the line of sight passes
through regions of the wind that cause ``dipping'' by
inhomogeneities in the wind \citep[e.g.,][and references
therein]{balucinska-church:2000,Grinberg15a}. While these
observations are ideal to study dipping effects due to absorption in
the stellar wind, they complicate the analysis of the intrinsic
variability. While the \nicer archive contains observations at
almost all orbital phases (Fig.~\ref{fig:orbital_phase_diagram}), many
hard state observations cluster at superior/inferior conjunction
($\phi_\mathrm{orb}=0/0.5$). We identify observations strongly affected by dipping by looking for the presence of the characteristic tracks (``nose-track''; \citealt{Grinberg20a}) in the color-color diagrams\footnote{Examples are observation 0100320106
  (Fig.~\ref{fig:app:0100320106}) and 4690010111
  (Fig.~\ref{fig:app:4690010111}).} and mark them in the observation log in Table~\ref{tab:observation_log}. For the
further analysis in Sect.~\ref{sec:spectra_timing_analysis}, we exclude these observations in order to reduce contamination of the variability
behavior by foreground effects.

\section{Spectral-Timing Analysis: Energy-resolved Timing Properties for each Spectral State}
\label{sec:spectra_timing_analysis}

In order to illustrate the general behavior and emphasize the extension of the variability to low X-ray energies compared to the earlier \rxte work, we concentrate on three selected observations that are representative of the overall source behavior.
To illustrate the behavior when the source spectral shape is hard, we choose observation 2636010101 at $\Gamma\sim 1.8$. 
This observation has an orbital phase of $\phi_\mathrm{orb}=0.94$--0.04 but does not show prominent signatures of dipping in the lightcurve or color-color diagram. 
Observation 0100320110 ($\phi_\mathrm{orb}=0.05$--0.21) is formally classified as soft state according to the \nicer to \maxi mapping discussed in Sect.~\ref{subsec:state_classification} but exhibits many properties of a transition between typical hard and soft state behavior, such as a photon index of $\Gamma\sim 2.3$. With $\Gamma\sim 3.1$, observation 1100320122 ($\phi_\mathrm{orb}=0.61$--0.69) is characteristic of the classical soft state of \cyg. 
These three observations are also marked in the q-diagram (Fig.~\ref{fig:q_diagram}). 

In this section, we will first investigate the power spectra of these observations and connect them to energy spectral components (Sect.~\ref{sec:power_spectral_density}). We then study the coherence to probe the linear correlation between the low- and high-energy lightcurves (Sect.~\ref{sec:coherence}), before we analyze the time lags (Sect.~\ref{subsec:time_lag_change}) as they can be reliably interpreted only for sufficiently high coherence \citep[see, e.g., ][]{Nowak99a}.

For reference, we also show the energy-resolved PSDs, lag-frequency spectra, and coherence of all \nicer observations studied here
in Appendix~\ref{appendix:summary_plots}. These figures also include \nicer products for the typical \rxte bands.

\subsection{Power Spectral Density}
\label{sec:power_spectral_density}

\begin{figure}
  \resizebox{\hsize}{!}{\includegraphics{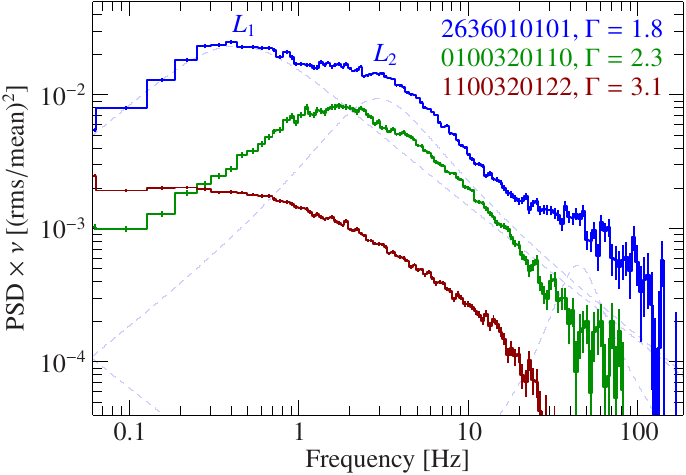}}
  \caption{Power spectral densities in the 0.5--10\,keV band of three representative observations of \cyg associated with the hard (blue) and soft spectral state (red). Observation 0100320110 at $\Gamma\sim 2.3$ is in the transition region between these states (green). The hard state
    can be characterized by strong broad band noise that can be
    resolved into multiple Lorentzian components (blue dashed
    lines). The two most prominent Lorentzians, $L_1$ and $L_2$, are labeled. In the transition, \cyg exhibits roughly a single broad
    component. The soft state shows almost featureless red noise.}
  \label{fig:psds}
\end{figure}
Figure~\ref{fig:psds} shows
representative 0.5--10\,keV
PSDs in the spectral states of \cyg.
Consistent with earlier work \citep[e.g.,][and references therein]{miyamoto:1989,Nowak99a}, 
multiple broad components can be clearly identified in the hard state. As the source transits to the soft state, these components merge to a single-humped structure \citep[e.g.,][]{Cui97b,Pottschmidt03a,Grinberg14a}. 
The PSD in the soft state is red noise-like \citep[e.g.,][]{Cui97b,Cui97a} with a 0.5--10\,keV RMS of around 10\% (0.06--500\,Hz), which is much higher than what is seen in the soft state of LMXB black holes.
As shown, e.g., by \citet[][and references therein]{Grinberg14a},
the structure of the PSD is a function of the spectral continuum shape (and therefore state). 

Following \citet{Nowak2000a} and \citet{Pottschmidt03a}, we model the hard state PSD in the 0.06--20\,Hz range with two Lorentzian functions peaking at
roughly 0.2\,Hz and 1.8\,Hz, plus a zero-centered
Lorentzian. As in these earlier works, we call the Lorentzians $L_1$, $L_2$, and $L_0$, respectively. Observation 2636010101 potentially shows an additional high-frequency component at around 50\,Hz, which has been seen before \citep{Pottschmidt03a} and which is also included in the model fits shown in 
Fig.~\ref{fig:psds}.
Another Lorentzian, which was clearly visible 
at around 8\,Hz in \rxte data taken before 1998 and then vanished \citep{Pottschmidt03a} is not significantly detected in the \nicer data. 
\citet{AxelssonDone2018a} found that the high-frequency variability components (${\gtrsim}10$\,Hz) in one \rxte observation of \cyg in the hard state have a hard energy spectrum with most contribution from above 10\,keV. In this paper, we concentrate on the frequency range below 10\,Hz and do not address the significance of possible Lorentzians at high frequencies.

\begin{figure*}
    \resizebox{\hsize}{!}{\includegraphics{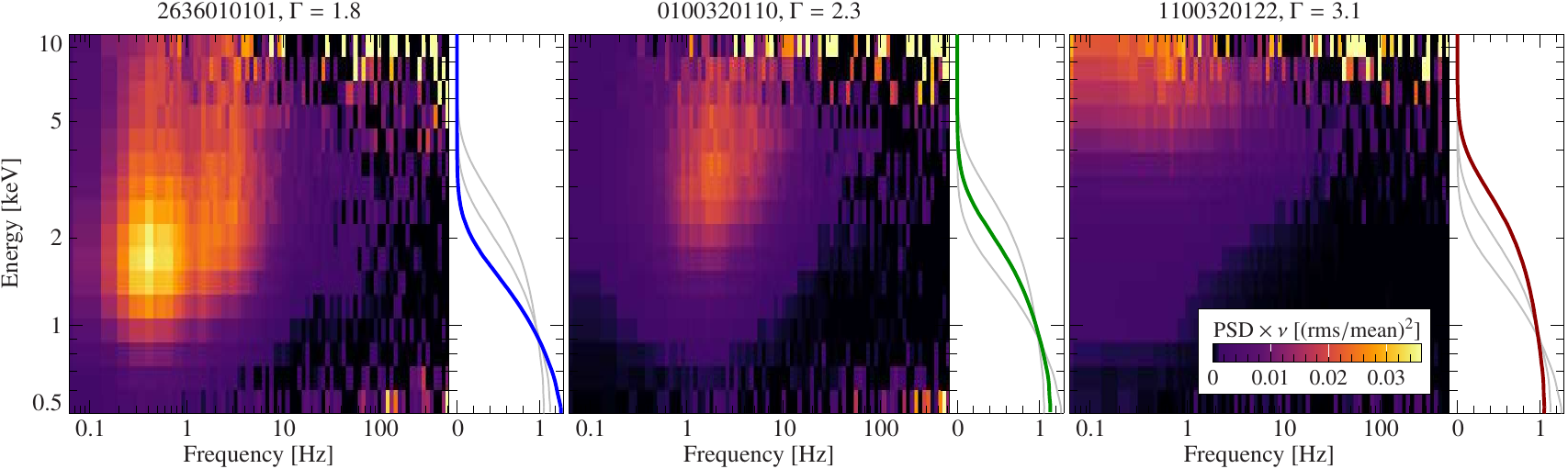}}
  \caption{Energy-resolved PSDs for three representative observations of \cyg at different hardness. The blue, green, and red thick lines in the side panels show the relative contribution from the accretion disk (that is, the quantity $F(\texttt{diskbb})/F(\texttt{simpl}\otimes\texttt{diskbb})$) for the hard ($\Gamma\sim 1.8$), intermediate ($\Gamma\sim 2.3$), and soft state ($\Gamma\sim 3.1$), respectively. The gray lines show the curves of the other states for comparison. 
  \textit{Left:}
  In the hard state, the power spectrum peaks at ${\sim}2$\,keV. It is double-humped at energies ${\gtrsim}1.5$\,keV while the second hump reduces in amplitude at lower energies. \textit{Center:} At $\Gamma\sim 2.3$, the PSD has a single-humped structure that peaks at ${\sim}4$\,keV. The variability peak shifts to higher frequencies compared to the hard state.
  \textit{Right:}
  In the soft state, the PSD is red noise-like and is strongest at the highest energies covered by \nicer.}
  \label{fig:energy_power_map}
\end{figure*}

It is well known that the power spectra of all states are energy dependent \citep[e.g.,][]{Nowak99a,Grinberg14a,Zhou22a}. In order to study this energy dependence, we calculate an energy-resolved power map for each observation (Fig.~\ref{fig:energy_power_map}). These maps show the PSDs computed for narrow energy bands in a color coded way, such that we can study the relative contribution of each Lorentzian component. 

In the hard state observation, the main components seen in the broad energy band PSDs (Fig.~\ref{fig:psds})
can be well identified in the power map (Fig.~\ref{fig:energy_power_map}, \textit{left}) as yellow/orange peaks. 
While their peak frequencies do not change with energy, their strength strongly depends on the energy. At the lowest energies below ${\sim}1.5$\,keV, $L_1$ dominates the variability, albeit with low amplitude. At energies ${\gtrsim}1.5$\,keV, 
$L_2$ starts to become apparent.  The strength of $L_1$ peaks at around 1.7\,keV and then decreases at harder energies. On the other hand, at energies above ${\sim}1.5$\,keV, the strength of $L_2$ remains fairly constant (for completeness, we show the RMS spectrum of $L_1$ and $L_2$ in Appendix~\ref{subsec:rms_spectrum}). 
Integrating the power spectrum over all frequencies,
we derive a fractional RMS of $(18.5\pm 0.4)\%$ in 0.5--1\,keV and $(30.0\pm 1.7)$\% in 5--8\,keV.

In the data taken during the transition between the soft and hard state ($\Gamma\sim 2.3$), the PSD is single-humped at a slightly lower RMS compared to the hard state (Fig.~\ref{fig:energy_power_map}, \textit{center}). The hump can be described by a broad Lorentzian peaking at around 1--2\,Hz in frequency and 3--4\,keV in energy space. The RMS at softer energies (${\lesssim}2$\,keV) is significantly lower compared to harder energies, but the overall shape of the PSD is similar. Again, the position and width of the Lorentzian components do not depend on the energy. 

The pure soft state observation ($\Gamma\sim 3.1$, Fig.~\ref{fig:energy_power_map}, \textit{right}) shows the typical red noise behavior reported previously for the spectrally softest observations of \cyg. 
In this observation the variability increases drastically with energy, as also seen in, e.g., \citet{Grinberg14a} or \citet{Zhou22a}, from $(6.57\pm 0.21)\%$ fractional RMS in 0.5--1\,keV to $(30.6\pm 1.2)\%$ in 5--8\,keV.

In order to understand the physical origin of the variability, we first consider
physical processes underlying the dominant
emission in the energy band where the variability is strongest.
Therefore, we perform spectral fits to the time-averaged X-ray
spectrum in the 0.5--10\,keV range, to determine the contribution of
the accretion disk emission and of Comptonization to each energy bin. For all states,
the spectrum of \cyg can be approximately described as a combination
of thermal emission from the accretion disk and a non-thermal
component. We describe this spectral shape with a model consisting of
emission from a multi-temperature accretion disk (\texttt{diskbb}),
which is partly Compton-upscattered following \citep[\texttt{simpl}
model; ][]{Steiner2009a}. As before, we include a 5\% systematic
uncertainty to account for possible residuals due to absorption in the
ionized stellar wind, calibration uncertainties of \nicer, and the simplicity of the model.

To see whether the variability is due to the disk emission or to effects related to the Comptonization of that radiation, for each energy we compute the ratio between the (absorbed\footnote{The difference between the ratio curves computed from absorbed and unabsorbed fluxes is minor in our energy range of interest.}) fluxes emitted by the accretion disk component $F(\texttt{diskbb})$, and the Comptonized flux $F(\texttt{simpl}\otimes\texttt{diskbb})$. Note that at low energies this fraction can become larger than unity because Comptonization redistributes photons from lower to higher energies. 

We display the energy-dependent disk fraction for the three example observations in the vertical side panels of Fig.~\ref{fig:energy_power_map}. 
In general, and in line with more detailed spectral analyses \citep{Nowak11a,Tomsick14a}, the disk dominates the emission at low energies, while above 1--3\,keV the Comptonized emission begins to dominate. This transition region 
where the Comptonization starts to dominate  
shifts from ${\sim}1.5$\,keV in the hard state to ${\sim}2.5$\,keV in the soft state. 
Likewise, the variability shifts to higher energies. 
At $\Gamma\sim 2.3$, the disk fraction is below 10\% at the peak of the variability around 3--4\,keV and in the soft state ($\Gamma\sim 3.1$), the contribution of the disk is essentially zero in the energy range of the largest variability. 
This behavior clearly suggests that the variability is due to the behavior of the Comptonizing medium in the soft state. 

\subsection{Coherence}
\label{sec:coherence}

\begin{figure*}
  \centering
  \includegraphics[width=.93\textwidth]{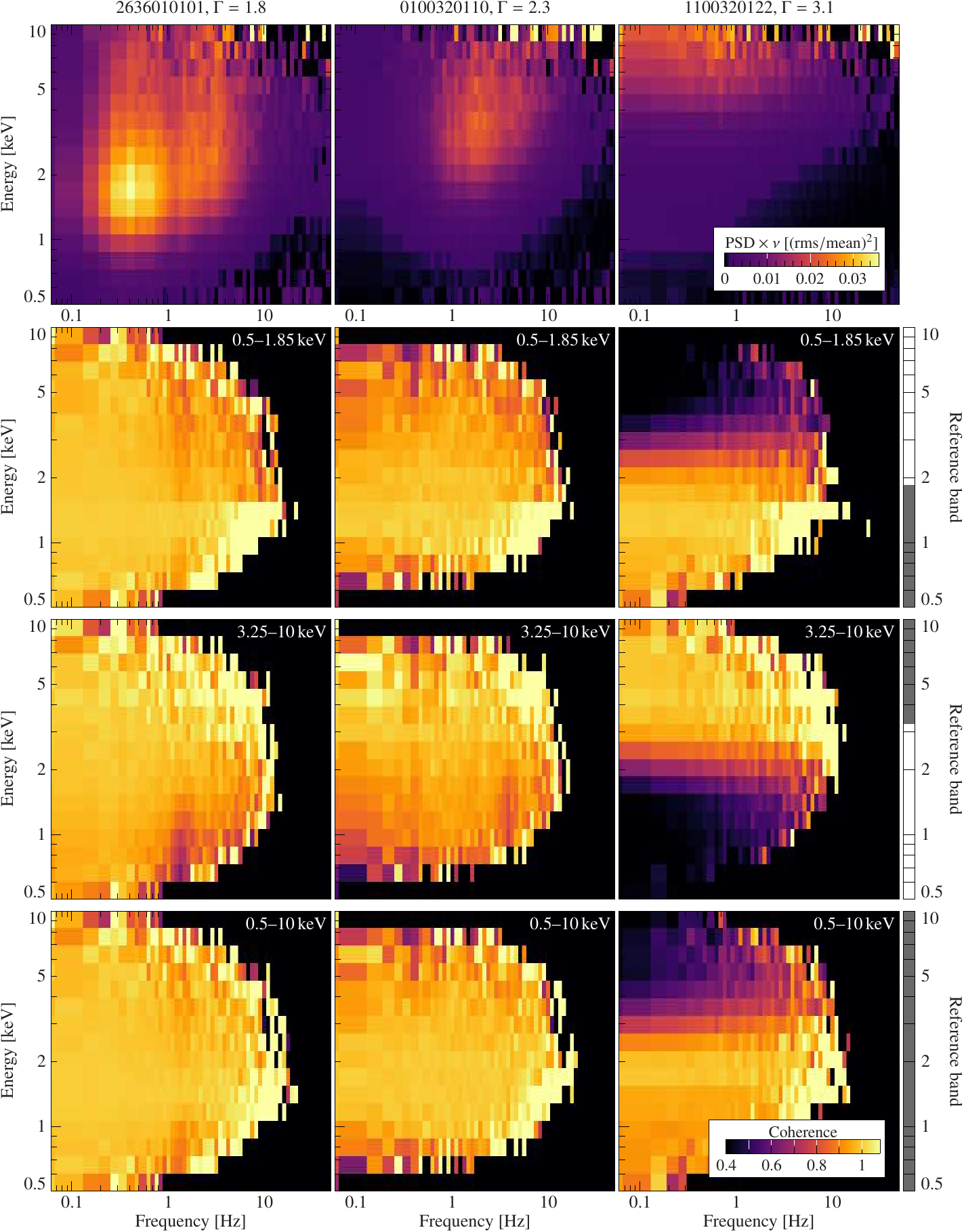}
  \caption{Energy-resolved coherence maps of \cyg with respect to three
    different reference bands (written in the top right corner of each
    panel and indicated via the shaded gray region on the right). The power map
    for each observation is shown in the top row for reference. \textit{Left
      column:} In the hard state, the coherence is generally close to
    unity in the probed frequency range, except for a drop between 1--2\,Hz. \textit{Center
      column:} When the PSD is single-humped at $\Gamma\sim 2.3$, the coherence is still close to
    unity. This behavior is largely independent to the chosen
    reference band. \textit{Right column:} In the soft state, the high and low
    energy variability is incoherent with respect to each other. This
    can be isolated by choosing a low or high energy reference
    band. We emphasize that the choice of the reference band has a
    drastic influence on the derived coherence values for soft state.}
  \label{fig:coherence_map}
\end{figure*}

Figure~\ref{fig:coherence_map} shows the energy-resolved intrinsic coherence,
calculated between small energy bins and a reference band. In
order to test the influence of the choice of the reference band on the behavior of the coherence function, we calculate maps
with three different reference bands: A broad 0.5--10\,keV reference
band, following the typical procedure of the lag-energy spectrum
calculation \citep{Uttley14a}, a 3.3--10\,keV reference band, tracing the 
variability of the Comptonized emission, and 0.5--1.9\,keV, which traces variability dominated by emission from the accretion disk.
Figure~\ref{fig:coherence_map} also shows the PSD map in the top row such that the amount of variability at the relevant energy can be directly related to the coherence. 

Generally, in all \nicer data of \cyg, the coherence can
only be constrained for frequencies up to ${\sim}10$\,Hz for energies around 2\,keV, beyond which the coherence begins to fluctuate wildly. At
lower and higher energies, where the power is lower, the coherence becomes unconstrained earlier
(in all states). For clarity, coherence values with fractional uncertainties
larger than 50\% are set to zero. 

In the hard state observation, the coherence can be well constrained
to near unity at almost all frequencies in the 0.06--10\,Hz band. This
behavior is largely independent of the chosen reference band. The
coherence is reduced only in the 1--2\,Hz band, which can be seen as a
purple stripe at 0.5--1\,keV in the map with the 3.3--10\,keV
reference band. This frequency band is the location where the two
variability components in the PSD have an overlap. We will
systematically characterize the implications of this coherence drop in Sect.~\ref{subsec:time_lag_change}.

For the data taken during an intermediate spectral shape,
$\Gamma\sim 2.3$, similar to the hard state, the coherence is again
close to unity at frequencies where the variability in the PSD map peaks,
and is also largely independent of the reference band (Fig.~\ref{fig:coherence_map}, \textit{center-column}).

This behavior changes in the soft state observation
($\Gamma\sim 3.1$). Here, the global variability is strongly dominated
by the hard X-rays above 5\,keV
(Sect.~\ref{sec:power_spectral_density}). However, looking at the
coherence with respect to the low-energy reference band
(Fig.~\ref{fig:coherence_map} \textit{upper-right}), the coherence of this
variability with this lower energy band is very low. Note that the
coherence with respect to the hard reference band
(Fig.~\ref{fig:coherence_map} \textit{middle-right}) shows identical
behavior, but in the converse. Namely, the measured coherence
indicates that the low-energy variability is \textit{incoherent}
with respect to the high-energy reference band, i.e., the emission in both bands varies independently.

Figure~\ref{fig:coherence_map} also shows that in the soft state the value of the coherence increases with frequency. In the coherence maps, this effect also depends on energy and the reference band but it occurs whenever bands between the soft (${\lesssim}2$\,keV) and hard X-rays (${\gtrsim}3$\,keV) are involved\footnote{For simplicity, we also refer to the corresponding coherence-frequency spectrum in  Fig.~\ref{fig:app:1100320122} in the appendix, where the increase is shown in the traditional one-dimensional way. We also note that the coherence can increase up to unity at high frequencies in other observations in the soft state (e.g., Fig.~\ref{fig:app:1100320119}).}. For instance, the coherence at 5\,keV (0.5--1.9\,keV reference band, Fig.~\ref{fig:coherence_map}, \textit{upper-right}) increases from below 40\% at 0.1\,Hz to above 70\% at 5\,Hz. This shape traces the red noise variability seen in the PSD: The coherence with respect to soft X-rays is lowest at the energy and frequency bins where the variability is strongest. From this overall behavior we conclude that the red noise is not linearly correlated to the low-energy variability.

As we have seen above, especially in the soft state, the choice of the
reference band drastically influences the measured coherence values
(Fig.~\ref{fig:coherence_map}, \textit{right column}). While this is not
surprising given that different physical processes are at play in
different energy bands (this is the reason why studies like \citealt{Nowak99a} or \citealt{Kara2013a} use a reference band with a small energy range), in order to increase the signal-to-noise ratio
many studies of the coherence and other energy dependent quantities
choose a broad reference band that often spans the entire range of the
data \citep[see, e.g.,][]{Uttley11a,Kara19a,DeMarco21a}. 
In the absence of an accepted model for X-ray binary variability that includes non-unity coherence between variability components, a broad energy band may lead to a misinterpretation of derived quantities such as the coherence or time lags.
Choosing narrow energy bands, one has better control and understanding of which energy bands are coherent or incoherent with respect to each other, which leads to a potential physical interpretation. 
A good example for the importance of choosing interpretable energy bands is shown in Fig.~\ref{fig:coherence_map}. Here, the coherence map for the broad reference band 0.5--10\,keV is very similar to the one for the low energy reference
band (0.5--1.85\,keV). 
This is due to instrumental effects and not due to physics:
X-ray instruments tend to have a large effective area at low energies. In addition, Cyg~X-1 and other black hole X-ray binaries have energy spectra which decline precipitously with energy. Combining both effects means that taking a broad reference band gives more weight to the measurement towards the peak of the instrumental effective area -- in reality, the effective reference band is not broad, as the event-energy distribution for a broad bandpass selection is actually relatively narrowly peaked due to these effects. 
%As a consequence,  the
%coherence and any quantities derived from variability with respect to
%a broad energy band (such as time lags) may be very difficult to
%interpret. 
We therefore specifically choose reference bands such that
either the accretion disk emission or Comptonization dominates (see previous section
and Fig.~\ref{fig:energy_power_map}) in order to be able to test the
connection of the variability between these physical processes.

\subsection{Time Lags} 
\label{subsec:time_lag_change}

In the previous section, we found that the low- and high-energy variability is highly coherent in the hard and intermediate states, while the coherence drops significantly in the soft state. 
In this section, we analyze the time lags of the three example observations. In particular, we concentrate on the changes of the time lag phenomenology when the energy bands are extended to soft X-rays below 2\,keV.  We show the lag behavior of all observations in Appendix~\ref{appendix:summary_plots}, including lags between harder, \rxte-like, energy bands.

In the soft state observation at $\Gamma\sim 3.1$, time lags between bands chosen in the \nicer energy range, including low energies below 2\,keV, have low amplitude 
${\lesssim}10$\,ms. This is consistent with previous studies of \cyg above 2\,keV \citep{Grinberg13a,Grinberg14a} and shows that the behavior does not change significantly for lower energies.
As shown in Sect.~\ref{sec:coherence}, the coherence of this observation is low and an interpretation of the time lags is, therefore, difficult.
We do not discuss the lags in the soft state further.

\begin{figure}
  \resizebox{\hsize}{!}{\includegraphics{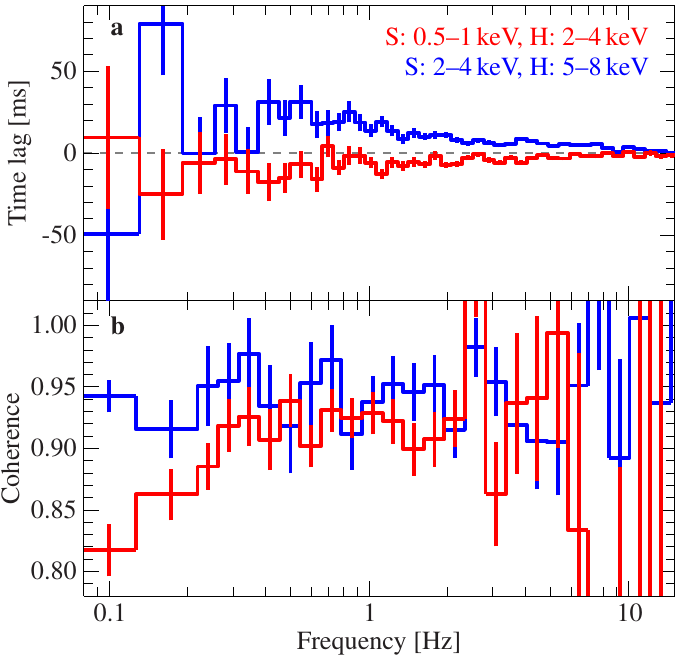}}
  \caption{Time lag and coherence of observation 0100320110 of \cyg at $\Gamma\sim 2.3$. Time lags at hard X-ray energies (blue) show power law-like hard lags at a high coherence. In soft X-rays (red), the time lags are predominantly soft with a relatively low amplitude at a coherence ${>}0.9$ in the 0.3--5\,Hz range. At lower frequencies, the coherence is reduced.}
  \label{fig:0100320110_TIMING_indiv}
\end{figure}

Figure~\ref{fig:0100320110_TIMING_indiv} shows the time lags of the transition observation with $\Gamma\sim 2.3$. We consider two pairs of energy bands, a high-energy pair of bands which resembles \rxte bands, and a lower-energy pair to explore behavior out of reach to \rxte. We show the corresponding coherence. Again, the time lags strongly depend on the choice of the energy bands. 
The lag-frequency spectrum between the 5--8\,keV and 2--4\,keV bands shows the familiar, highly coherent power law-like hard lags \citep{Cui97a}. 
At low energies (2--4\,keV and 0.5--1\,keV), there is a soft lag at most frequencies, albeit with low amplitude. In the 0.3--5\,Hz range, the coherence is ${>}0.9$, while it drops to ${\sim}80\%$ at low frequencies where the power is reduced (see again the PSD map in Fig.~\ref{fig:energy_power_map}). The time lag behavior is qualitatively similar to the phenomenology seen in the hard to soft transition of LMXB black holes (e.g., \citealt{Wang22a}, their Fig.~3f, and also \citealt{Uttley11a}, \citealt{Kara19a}, and \citealt{DeMarco21a}), keeping in mind the different state terminology for these sources and the fact that \cyg is on the lower branch of the q-diagram. 
A full interpretation of the soft lag involves complex modeling of the spectral-timing products \citep[e.g.,][]{Ingram19a,Mastroserio21a,Kawamura22a}, which is beyond the scope of this paper, and we concentrate on its empirical behavior here.

\begin{figure}
  \resizebox{\hsize}{!}{\includegraphics{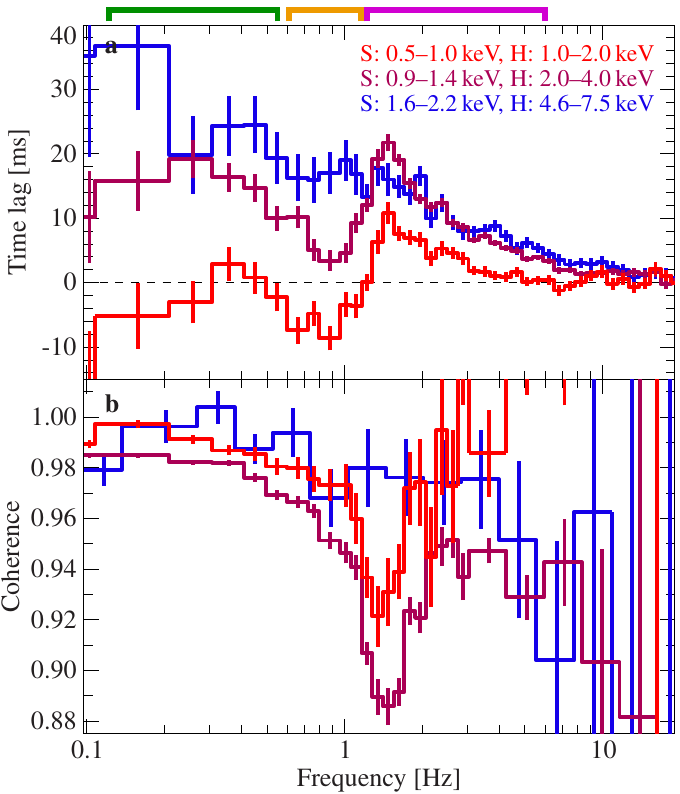}}
  \caption{Lag-frequency and coherence spectra of observation 2636010101 of \cyg in the hard state for various energy bands. \textbf{a)} Time lags between energy bands ${\gtrsim}1.5$\,keV show the well-known power law-like behavior (blue line), while an abrupt change in time lag with soft X-ray data ${\lesssim}1.5$\,keV is seen (red and purple lines). \textbf{b)} The coherence between high energy bands, when power law-like hard lags are seen, is close to unity. When the energy range is extended to below 1.5\,keV, the coherence drops significantly between 1--2\,Hz. Colored brackets on top denote the frequency ranges chosen for the lag-energy spectrum in Fig.~\ref{fig:2636010101_LES}.}
  \label{fig:sliding_window}
\end{figure}

We show the time lags between different energy bands for the hard state observation with $\Gamma\sim 1.8$ in Fig.~\ref{fig:sliding_window}. 
For bands above 1.5\,keV, highly coherent hard lags with the typical power law-like frequency dependence are seen, as has been extensively documented before \citep[e.g.,][]{Nowak99a}.
Moving to softer energies, the time lag drops towards or even below zero around 1\,Hz, before it slightly increases again at lower frequencies. Interestingly, this feature appears simultaneously with a drop in coherence (see Fig.~\ref{fig:sliding_window}b) and increases in strength for softer energy bands. The frequency of the coherence drop coincides with the peak of the hard time lag around 1.6\,Hz. 
Since the lag phenomenon is strongly correlated with the coherence, we avoid a naming that relates to the time lag only and refer to the whole structure as a \textit{timing feature}. 
We focus the remainder of this section on the analysis of this timing feature in the hard state observation as this low-energy phenomenon has not previously been found in \cyg.

\begin{figure}
  \resizebox{\hsize}{!}{\includegraphics{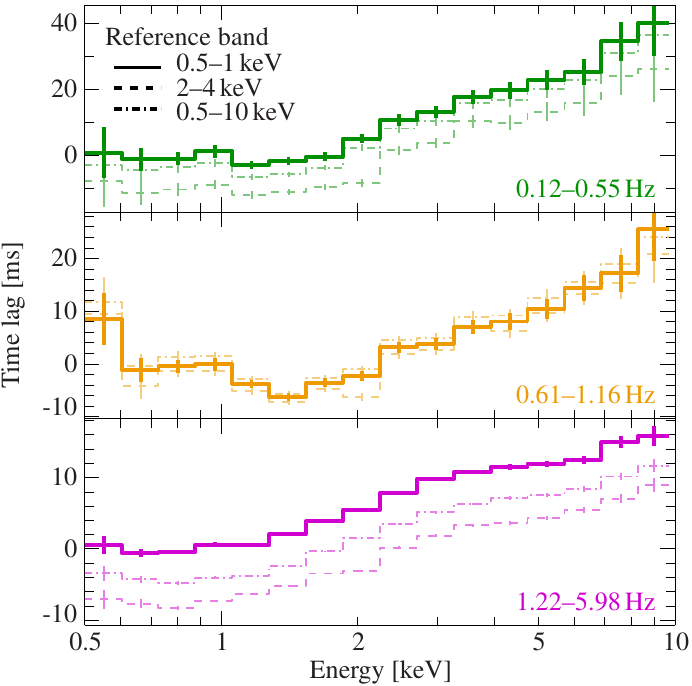}}
  \caption{Lag-energy spectra of observation 2636010101 of \cyg in the hard state. Frequency ranges are chosen to trace the timing feature (colored brackets in Fig.~\ref{fig:sliding_window}). The low-frequency range, 0.1--0.6\,Hz, shows hard lags. The frequency range 0.6--1.2\,Hz during the ``dip'' in the lag-frequency spectrum shows an upturn at low energies (soft lags). This upturn vanishes at higher frequencies, 1.2--6\,Hz, which covers the peak of the timing feature. Solid, dashed, and dash-dotted lines denote spectra for a broad, and two narrow reference bands.}
  \label{fig:2636010101_LES}
\end{figure}

To better understand the energy dependency of this timing feature, we show the lag-energy spectrum in Fig.~\ref{fig:2636010101_LES}. We choose frequency ranges according to the structure in the lag-frequency spectrum (Fig.~\ref{fig:sliding_window}a): the first frequency range (0.1--0.6\,Hz) corresponds to the first peak, the second (0.6--1.2\,Hz) covers the drop in time lag, and the third frequency range, (1.2--6\,Hz), corresponds to the peak of the hard lag and the largest drop in coherence. Furthermore, the time lags are calculated for different reference bands. From the results found in Sect.~\ref{sec:power_spectral_density}, we choose bands dominated by the accretion disk emission (0.5--1\,keV), or by Comptonization (2--4\,keV), as well as a broad reference band (0.5--10\,keV). 

The lag-energy spectrum in Fig.~\ref{fig:2636010101_LES} shows hard lags at high energies in all three frequency bands. While the general behavior at these high energies is similar, the curvature of the hard lag changes for different frequencies. As in \citet{Kotov01a} or \citet{Mastroserio19a}, we do not detect any significant iron line lag feature either. For each frequency band, we see that the overall shape of the lag-energy spectrum is very similar for the different reference bands and mainly the zero-crossing of the time lag changes.

Additionally, we detect soft lags at energies ${\leq}1.5$\,keV, but only in the second frequency range (0.6--1.2\,Hz). In the literature such low-energy soft lags are often referred to as ``reverberation lags'' and they are interpreted as light-crossing time delays between hard photons from the Comptonizing medium and soft photons from the accretion disk (note, however, that there might be other physical processes causing soft lags, see, e.g., \citealt{Veledina2018a} and \citealt{Kawamura2023a}). A prediction for reverberation lags is that they continue being present at higher frequencies \citep[e.g.,][]{Ingram19a}. In observations of the LMXBs GX~339$-$4 \citep{Uttley11a} and MAXI~J1820+070 \citep{Wang21a}, for instance, low-energy soft lags are consistently seen above ${\sim}2$\,Hz.
Such a behavior is not found in the \nicer observation of \cyg with $\Gamma\sim 1.8$. Therefore, the low-energy soft lag in the 0.6--1.2\,Hz range is likely not due to reverberation.

\begin{figure*}
  \sidecaption
  \includegraphics[width=12cm]{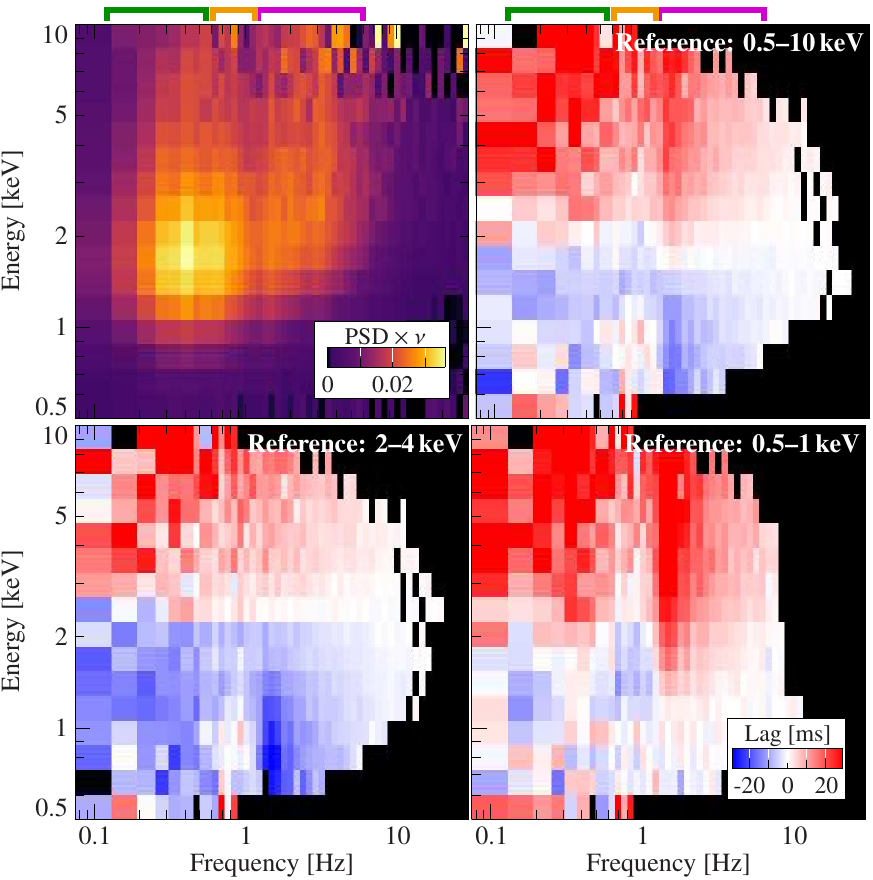}
  \caption{Energy-resolved time lag maps of observation 2636010101 of \cyg in the
    hard state. The maps are calculated
    between small energy bins and three reference bands, written in
    the top right corner of each plot. For reference, we also show the
    power map in the top left panel. Pixels where the coherence has an uncertainty larger than 50\% of its value are set to black. Colored brackets on top denote the frequency ranges chosen for the lag-energy spectrum in Fig.~\ref{fig:2636010101_LES}.}
  \label{fig:lag_map}
\end{figure*}

In Fig.~\ref{fig:lag_map}, we show a two-dimensional map of the time lag depending on energy and frequency. Again, the maps are shown for the different reference bands and include the PSD map to allow for a better comparison. The time lag peak of the timing feature can be identified as a red or blue stripe-pattern at around 1.6\,Hz, depending on if we use the low- or high-energy reference band. As can be seen in comparison with the PSD, the peak of the timing feature is at a frequency where the two variability components $L_1$ and $L_2$ overlap, and does not depend on energy. As mentioned previously, the peak of the feature shows a hard lag
and is seen as a positive time lag when using a low-energy reference band (red stripe in Fig.~\ref{fig:lag_map}, \textit{bottom-right}). As expected, the sign of this lag changes to negative if a high-energy reference band is used (blue stripe in Fig.~\ref{fig:lag_map}, \textit{bottom-left}), but still remains a hard lag (the soft photons come before the hard reference band). 

We note that lag maps can only be reliably interpreted when both, the time lags and the coherence are well constrained. Using the same approach as for the coherence map in Fig.~\ref{fig:coherence_map}, we  set pixels to black where the relative uncertainty of the coherence  is ${>}50$\%. The coherence is ${\gtrsim}70$\% in the frequency range of the timing feature (1--2\,Hz) and near unity outside of this range.

For completeness, we also derive a time lag map for a broad reference band (0.5--10\,keV; Fig.~\ref{fig:lag_map}, \textit{top-right}). As is also the case for the coherence map based on a broad reference band (Fig.~\ref{fig:coherence_map}, \textit{bottom}), this map is very similar to the time lag map for the low energy reference band, showing a positive time lag for the feature at high energies. In addition a negative lag at lower energies appears. As discussed in Sect.~\ref{sec:coherence}, this is an effect of the detector's effective area, which peaks near 1.5\,keV. In case of the time lags,
this peak of the effective area
determines the energy of zero-crossing of the lag. We emphasize again that for the physical interpretation of the absolute value of the time lag, it is important to understand the variability process that dominates in the reference band. 
Due to the influence of the effective area, interpreting absolute time lag values using a broad reference band is particularly challenging when time lags are compared across different instruments, as the relative contribution of variability to the broad reference band changes. We will therefore focus our interpretation on the soft (0.5--1\,keV) reference and hard (2--4\,keV) subject bands, corresponding specifically to where the emission from the accretion disk and from the Comptonizing plasma, respectively, dominate.
Specifically, the Comptonization-dominated hard band was constrained to 2--4\,keV to also avoid signal from relativistic reflection from iron, which can contribute in the 5--8\,keV range.

In order to rule out an instrumental origin of the timing feature, we performed a number of detailed tests adopting different selection criteria, such as selecting only one continuous data segment (to exclude potential time shifts between good time intervals) obtained during International Space Station night with a low undershoot range. Even with the most conservative data extraction cuts, the timing feature is visible in the data. We therefore exclude an instrumental origin. 

\section{Discussion}
\label{sec:discussion}

Previous X-ray timing missions which performed monitoring campaigns of \cyg include \textsl{Ginga} \citep{Miyamoto1988a}, \textsl{EXOSAT} \citep[e.g.,][]{BelloniHasinger90}, \rxte \citep[e.g.,][]{Cui97a,Nowak99a,Revnivtsev2000a,Pottschmidt03a,Axelsson2005a,Axelsson2006a,Grinberg14a}, \textsl{AstroSAT} \citep{Misra17a,Maqbool19a}, \xmm \citep{Lai22a}, and \hxmt \citep{Zhou22a}. Most of these missions have limited sensitivity to \cyg's disk emission, being limited to the hard X-ray regime above 1.0--2.7\,keV. The extension to softer X-rays provided, e.g., by \nicer and \xmm is, however, essential to understand the accretion disk variability, which contributes below ${\sim}2$\,keV (see Sect.~\ref{sec:power_spectral_density}), and its connection to the Comptonizing plasma. However, due to the brightness of \cyg, \xmm observations can be affected by telemetry drop outs and suffer from pile-up.  As a result, \xmm spectral-timing studies with soft X-ray coverage is challenging and limited to \cyg's fainter hard state\footnote{For the brighter states of \cyg, the only option is to use the modified timing mode of the EPIC-pn cameras \citep{Duro11a}, which, however, ignores data ${<}2.8$\,keV.}. Work with \xmm has mainly concentrated on the timing effects introduced by the stellar wind \citep{Lai22a}.

The data from the \nicer monitoring give a unique opportunity to analyze the general spectral-timing behavior of \cyg down to 0.5\,keV for all spectral states for the first time. In the following, we discuss our spectral-timing results presented in the previous section and put them into context with previous monitoring campaigns. We start in Sect.~\ref{subsec:interpretation_of_variability} with a general overview of the variability behavior, followed by an overview of the the coherence behavior in Sect.~\ref{subsec:coherence} and a discussion of the time lags in Sect.~\ref{subsec:lags}.

\subsection{Variability Components in the Power Spectrum}
\label{subsec:interpretation_of_variability}

In the hard state, the high-energy PSDs of \cyg exhibit two or three prominent Lorentzians, which are well studied \citep[see, e.g., ][and references therein]{Nowak2000a,Grinberg14a}. 
In order to be able to interpret the physical origin of these components, we need to understand the processes that are dominating the flux at each energy. In particular, by extending the energy band to below 2\,keV, where the accretion disk emission dominates, we are able to study the contribution of the disk to the variability.

In Sect~\ref{sec:power_spectral_density}, we showed that the peak frequency of the $L_1$ and $L_2$ Lorentzians is independent of the energy and both components have a unique energy signature. This result supports the interpretation that the Lorentzian components represent individual variability components each of which has a unique physical origin. The total variability seen is the superposition of these independent components.
In the following, we show that the data are consistent with
the interpretation that 
$L_1$ originates in the accretion disk and is then modulated by Comptonization, while $L_2$ can be directly associated with the Comptonizing medium. 

For the strongest variability component in the \nicer energy range, $L_1$, we find that it peaks at around 1.7\,keV, while also contributing significantly to the variability above 3\,keV.
From our model fits (Fig.~\ref{fig:energy_power_map}), we have seen that Comptonization dominates the emission above ${\sim}1.5$\,keV. Together with the fact that $L_1$ is also strong above 3\,keV, where the disk contribution to the total flux is negligible, we conclude that this component has to be connected to the Comptonizing plasma. 
Below 1.7\,keV, $L_1$ is the main source of variability, although it drops in strength. 
The peak strength of $L_1$ lies
in the energy band
where accretion disk emission and Comptonization both contribute roughly equally. 
This behavior could suggest
that at least part of the variability in the accretion disk contributes to $L_1$. Possible physical mechanisms were suggested in earlier work by
\citet{WilkinsonUttley09a}, who attributed the variability of $L_1$ to an
unstable accretion flow in the disk, while theoretical work by \citet{MummeryBalbus2022a} shows that the Wien tail of
the disk can vary significantly in luminosity. 
A process such as propagating fluctuations, which connects the intrinsic 
disk variability with the emission of the Comptonizing plasma, may be a physical model to explain this behavior (see \citealt{Lyubarskii1997a}, \citealt{Kotov01a}, \citealt{Ingram13a}, and \citealt{Rapisarda2017a} for an application to \cyg in the soft and hard state).
An alternative interpretation is that $L_1$ may originate from a region of ``soft Comptonization'', that is, an optically-thin plasma with a temperature of a few keV \citep[e.g.,][]{AxelssonDone2018a}. Such a process matches the energy dependence of $L_1$ well.

In contrast, we have shown in Sect.~\ref{sec:power_spectral_density} that the strength of $L_2$ decreases at energies below 2\,keV and that it only contributes significantly above ${\sim}1.5$\,keV. This suggests that $L_2$ is due to processes in the Comptonizing plasma.
This interpretation is also consistent with \rxte data, which show that $L_2$ is the dominant component above 10\,keV \citep{Grinberg14a}, and \hxmt data shown by \citet[][their Fig.~11]{Zhou22a}, where $L_2$ is detected up to ${\sim}90$\,keV.

In the soft state, the variability properties fundamentally change. Red noise dominates the PSD (see Fig.~\ref{fig:psds},
and, e.g., \citealt{Cui97b,Gilfanov2000a,Churazov2001a,Axelsson2006a}) and the variability strongly increases with energy, peaking above 7--10\,keV (see Sect.~\ref{sec:power_spectral_density} and \citealt{Grinberg14a,Zhou22a}). At these energies, the soft state spectrum of \cyg shows a power law component, which is associated with non-thermal Comptonization \citep[e.g.,][]{Wilms06a}. As the contribution of the accretion disk is negligible at these energies, the variability in the soft state must be linked to the Comptonizing plasma. This is consistent with earlier work by \citet{Churazov2001a} who linearly decomposed \rxte lightcurves into stable and variable
components associated with the disk and Comptonizing plasma, respectively.  

While the variability drops by almost a factor of 5 to around 6\% RMS at low energies (0.5--1\,keV), it is still significantly detected. At these energies, the fraction of the disk black body contributing to the spectrum is very high (see Fig.~\ref{fig:energy_power_map}).
In principle, it could be possible that, while the disk dominates the flux, the Comptonizing plasma dominates the variability even below 1\,keV. 
However, since only a low fraction of the soft emission is due to Comptonization, we propose that this low-energy variability in the soft state can also directly originate from the accretion disk. That the soft state has a lower RMS (${\sim}6$\%) at low energies compared to the hard state (${\sim}18$\%) would be consistent with the picture that a soft state disk is more stable than a hard state disk, as proposed by, e.g.,
\citet{nowak:95a} and \citet{Churazov2001a}. 

\subsection{Coherence between Low and High Energies}\label{subsec:coherence}

The coherence of the variability of \cyg between energy bands above ${\sim}2.5$\,keV has been investigated in detail with  \rxte. The coherence is typically between 0.95 and 1.0 in the hard state \citep{Nowak99a,Grinberg14a}. We find for all hard and intermediate state \nicer observations in our sample that the 0.06\,Hz to ${\sim}10$\,Hz coherence is close to unity between bands chosen in the 2--10\,keV range, consistent with those previous results.

With the \nicer monitoring we can extend the coherence measurements to the soft X-rays. This has not been studied before.
As shown in Fig.~\ref{fig:energy_power_map}, the low-energy (${\lesssim}1.5$\,keV) and high-energy bands (${\gtrsim}2$\,keV) can be identified with the accretion disk and the Comptonizing plasma, respectively. 
For the $\Gamma\sim 1.8$ and $\Gamma\sim 2.3$ observations, we find that the coherence between these bands is close to unity (see Sect.~\ref{sec:coherence}), with the exception of a drop at 1--2\,Hz, which we further discuss in the next section. The high coherence between the bands where disk emission and Comptonization dominate the flux strongly suggests a physical connection of the variability of these two components
%\footnote{The coherence only measures the degree of linear correlation and one has to be careful attributing causality based on it \citep{VaughanNowak97a,BendatPiersol2010}. If the intrinsic coherence is below unity, and in the absence of dilution, it is possible that the two time series are either uncorrelated (and have no physical connection) or that there is a non-linear transfer function which couples the signals. In the latter case, the coherence is low although there is an underlying correlation.
%On the other hand, a coherence of unity does show that there is at least a linear correlation and we can make statements about causality.}. 
(see \citealt{VaughanNowak97a} for an explanation of the coherence).
A model such as propagating fluctuations would explain such a connection. In the soft state ($\Gamma\sim 3.1$), the coherence map shows clearly that the high-energy variability, which we identified with Comptonization, is not coherent with respect to the variability at low energies any more. As there is significant RMS at a 6\% level at 0.5--1\,keV, the loss of coherence between low and high energies is statistically robust and not an effect of the low RMS. Therefore, regardless of the nature of the physical processes that produce the coherent variability in the hard state and the transition, this process changes in the soft state. 
%Following the interpretation that the low- and high-energy variability is due to the disk and the Comptonizing medium, respectively, this result demonstrates that the variability from these regions is not linearly connected in the soft state. 
%If there is intrinsic disk variability propagating into the Comptonizing plasma, the variability information appears to be diluted.

\subsection{Evidence for an Abrupt Time Lag Change and Drop of Coherence in the Hard State}\label{subsec:lags}

We measured time lags with respect to the soft energy band below
2\,keV for three example observations at $\Gamma \sim 1.8$, 2.3, and 3.1 
(Sect.~\ref{subsec:time_lag_change}). While the time lags of the $\Gamma\sim 2.3$ and 3.1 observations show overall consistent behavior
with previous intermediate and soft state observations, respectively, in the hard state observation at $\Gamma\sim 1.8$, the time lags contain a strong feature together with a drop in coherence at around 1--2\,Hz where the $L_1$ and $L_2$ components overlap. 
That this feature has not
been seen in \cyg before is mainly due to the fact that, as a low-energy
phenomenon, it was difficult to access with \rxte and similar missions with hard X-ray
timing capabilities.
%\footnote{The only data set of \cyg with low-energy coverage ${<}1$\,keV to compare the timing feature to is the \xmm monitoring campaign presented in \citet{Lai22a}. Their coarsely re-binned lag-frequency spectrum in Fig.~6, using 0.3--1\,keV as soft and 2--10\,keV as hard band, does not show a peak-like timing feature. The PSDs of these \xmm observations have a less pronounced low-frequency hump, and the relativistic reflection fitting (their Table~C1) indicates that \cyg was observed in the harder tail of the hard state. Observations in the \nicer archive (e.g., Fig.~\ref{fig:lag_frequency_spectra_LHS}c) with a comparable PSD, lag spectrum, and photon index suggest that the timing feature appears as a shelf in this region of the hard state. Such a shelf may be present in their Fig.~6, potentially with a coherence drop. We note, however, that such a signature can easily be missed if the Fourier products are re-binned too heavily, if the data quality is lower, or if one averages over long data segments (100\,ks in \citealt{Lai22a}) in case the source properties change.}.  
In our example observation (obs. ID 2636010101), the feature
disappears when only taking into account photons above 1.5\,keV
(Fig.~\ref{fig:sliding_window}).
Similarly narrow drops in coherence at the overlap of structures in the power spectrum have, however, been seen in GX~339$-$4 \citep{Nowak1999c} at around 1\,Hz and in GRS~1915+105 \citep{Ji2003a} at 0.03\,Hz using \rxte data. While no time lags are shown in \citet{Ji2003a}, the time lags in \citet{Nowak1999c} show no peaked feature but a shelf-like structure at 1\,Hz (such shelves have also been seen in \cyg; \citealt[e.g.,][]{Nowak99a,Uttley14a}). We consider it likely that the underlying physical mechanism of these previously detected lag shelves and coherence drops, and the timing feature discussed in this paper, are similar. 
%Furthermore, the harder energy dependence of the phenomenon in GX~339$-$4 and GRS~1915+105 (\rxte data was limited to ${\gtrsim}2.5$\,keV, while we cannot detect the feature above ${\sim}1.5$\,keV) may be attributed to different parameters in these sources. 

To further understand the origin of the feature in \cyg, in the
remainder of this section we will consistently use the 0.5--1\,keV
band as the reference band and the 2--4\,keV band as the subject
band. We note that these bands are mainly
dominated by the accretion disk emission, and Comptonized photons,
respectively (see also Sect.~\ref{subsec:time_lag_change}).

\begin{figure*}
  \resizebox{\hsize}{!}{\includegraphics{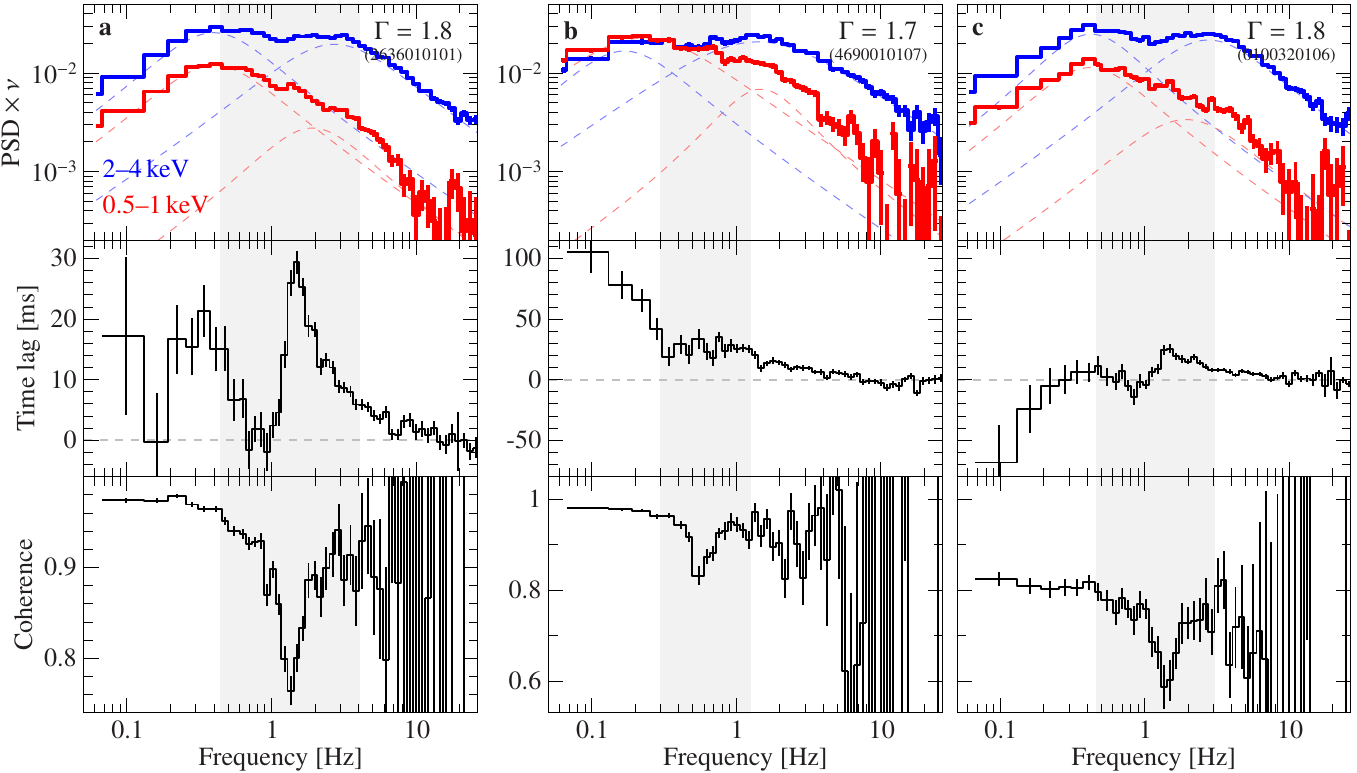}}
  \caption{Timing behavior in the hard state of \cyg. The time lag and coherence are calculated between the 2--4\,keV and 0.5--1\,keV bands.
  To guide the eye, the gray shaded region roughly indicates the frequency range of the feature, as indicated by the reduced coherence.
  \textbf{a)} Observation 2636010101 ($\Gamma\sim 1.8$) shows the clearest signature of the timing feature in the \nicer data. A localized ${\sim}30$\,ms hard lag at ${\sim}1.6$\,Hz is present at the overlap region of the $L_1$ and $L_2$ components. The amplitude of $L_2$ is reduced at soft energies. The coherence drops at the peak frequency of the timing feature's time lag, and appears asymmetric with respect to the time lag.
  \textbf{b)} Observation 4690010107 is slightly harder at $\Gamma\sim 1.7$ and shows coherent hard lags at low frequencies. At 0.5\,Hz, a coherence drop indicates the presence of the timing feature. The time lag shows a shelf-like structure at this frequency with an amplitude around 30\,ms (note that the y-axis is different).
  \textbf{c)} Observation 0100320106 is affected by dipping and shows long soft lags at low frequencies at an overall reduced coherence. The timing feature is blended into the lag spectrum with similar amplitude as in the other observations.}
  \label{fig:lag_frequency_spectra_LHS}
\end{figure*}

We have 
systematically searched all \nicer data of \cyg taken up to cycle~4 (until April 2022) to study the occurrence of the timing feature in detail. In general, the feature is not present in observations with soft
spectra with $\Gamma \gtrsim 2.2$. In the following, we will show that observations with spectra harder
than $\Gamma \sim 2.0$, on the other hand, show significantly different
time lag behavior than the softer ones, especially at frequencies ${\lesssim}1$\,Hz.

In Fig.~\ref{fig:lag_frequency_spectra_LHS}, we compare the power spectra, time lag, and coherence for three selected hard state observations and show that despite major differences in the lag-frequency spectra between those observations the feature occurs in all of them. All of the observations have in common that hard photons lag behind soft photons by approximately 30\,ms at the frequency of the coherence drop. This drop in coherence is generally largest at the peak of the feature, albeit its shape is found to be different in these observations due to the low-frequency time lag behavior. 
In the hard state observation ($\Gamma\sim 1.8$) that we have shown in Sect.~\ref{sec:spectra_timing_analysis} and which represents the observation with the strongest timing feature in the \nicer data, the peak of the hard time lag and the strongest coherence drop is located at 1.6\,Hz (Fig.~\ref{fig:lag_frequency_spectra_LHS}a). The coherence drop appears asymmetric with respect to the time lag in the sense that the lag shows a sharp drop below the peak with a tail towards higher frequencies, while the coherence drop extends to lower frequencies.

The data shown in Fig.~\ref{fig:lag_frequency_spectra_LHS}b (obs. ID
4690010107) were taken when \cyg was slightly harder ($\Gamma\sim 1.7$). Here, pronounced hard lags with a
shelf-like structure are seen, resembling the shape found with \ginga
\citep[][their Fig.~1c]{Miyamoto92a} and \rxte (e.g.,
\citealt{Nowak99a} or \citealt{Grinberg14a}, their Fig.~11). These
structures are significantly less peaked compared to
Fig.~\ref{fig:lag_frequency_spectra_LHS}a. In the \nicer data of
Fig.~\ref{fig:lag_frequency_spectra_LHS}b, a coherence drop can be
seen at the frequency of the shelf, which is not apparent in the \rxte
data of \cyg at higher energies \citep[][their Fig.~10]{Grinberg14a}.
A similar coherence drop was found in GX~339$-$4 by \citet[][their
Fig.~6c--d]{Nowak1999c}, who used ${\sim}2.5$--3.9\,keV and
10.8--21.9\,keV bands. Due to the frequency match of the overlap
region of $L_1$ and $L_2$, the lag shelf, and the coherence drop for \cyg, the
underlying process responsible for this shelf-like structure could be
the same as for the timing feature. We also note that the drop in
coherence is at significantly lower frequency, roughly 0.5\,Hz,
compared to the previous observation and that, while the time lag
shelf is rather broad, the absolute time lag is again around 30\,ms.

Finally, Fig.~\ref{fig:lag_frequency_spectra_LHS}c illustrates the
variability behavior of one of the observations that shows strong
``nose''-shaped color-color variations (obs. ID 0100320106,
$\Gamma\sim 1.8$). As discussed in Sect.~\ref{subsec:color-color},
these variations originate from absorption of X-rays in clumps of the
stellar wind. During such absorption events, additional modulation of
the X-ray variability by the absorbing foreground material imprints
long soft lags at low frequencies and reduce the coherence
(see Appendix~\ref{app:sec:stellar_wind} and \citealt{Lai22a}).
In this example, the timing feature blends with the time lag induced by the
stellar wind, while its overall shape is very similar to
Fig.~\ref{fig:lag_frequency_spectra_LHS}a, again with a maximum hard
lag of $\sim$30\,ms.

\begin{figure}
    \resizebox{\hsize}{!}{\includegraphics{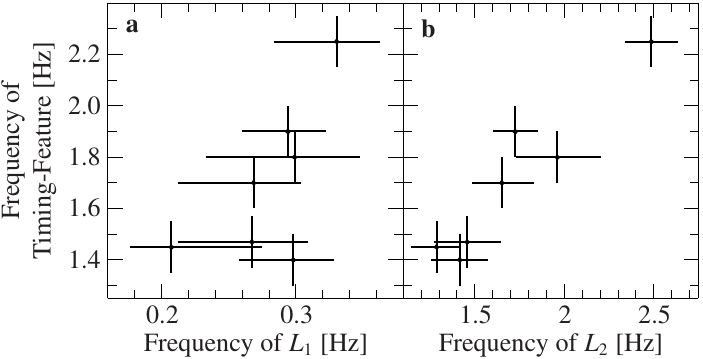}}
  \caption{Frequency of the timing feature versus the peak frequencies of $L_1$ and $L_2$ for seven observations of \cyg with a clear signature of the timing feature (marked in Table~\ref{tab:observation_log}). The frequency of the feature is parameterized by the peak in the hard time lag, assuming a systematic uncertainty of 0.1\,Hz. Uncertainties on the Lorentzian frequencies and $\Gamma$ are at the 90\% confidence level. The frequency of the feature correlates positively with the position of the Lorentzians, which are known to increase in frequency as the source softens.}
  \label{fig:lag_correlation_lorentzians}
\end{figure}

\begin{figure}
  \resizebox{\hsize}{!}{\includegraphics{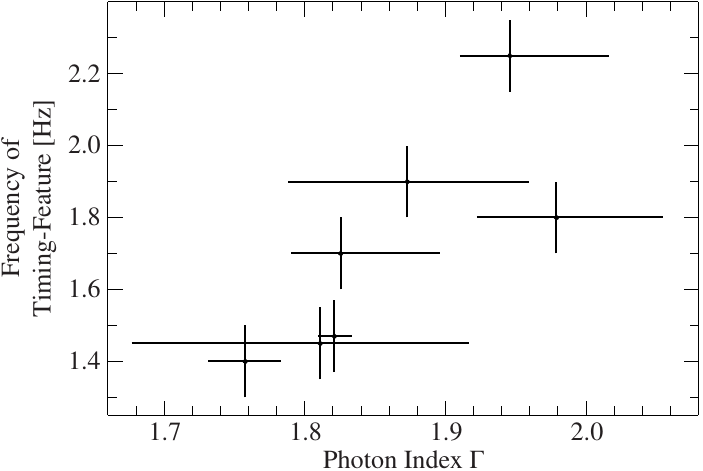}}
  \caption{The frequency of the timing feature increases as \cyg softens.}
  \label{fig:lag_correlation_gamma}
\end{figure}

Having established how the feature impacts the \cyg hard state data, we next quantify its frequency behavior for the different observations. 
By visually inspecting the \nicer products for a peaked lag structure at the overlap region of the Lorentzians in combination with a coherence drop at the same frequency (see Appendix~\ref{appendix:summary_plots} for reference), we identify seven observations in the hard state ($\Gamma \lesssim 2.0$) where we confidently detect the feature; we explicitly exclude observations with a shelf-like lag structure.
These observations are marked in  
Table~\ref{tab:observation_log}.
We then read the feature's frequency off from the peak in the lag-frequency spectrum and check that the coherence drop is at the same frequency. 
As this is a manual process, we adopt a systematic uncertainty of 0.1\,Hz. We find that the frequency of the feature is correlated with the peak frequencies of $L_1$ and $L_2$ (Fig.~\ref{fig:lag_correlation_lorentzians}), with
Pearson correlation coefficients of 0.95 and 0.99, respectively. Similar to the peak frequencies of the Lorentzians \citep{Pottschmidt03a,Grinberg14a}, the frequency of the feature increases as the spectral continuum softens (Fig.~\ref{fig:lag_correlation_gamma}).
These results indicate a connection between the feature and
the two Lorentzians, which is also consistent with the fact that we do
not find observations  with a single-humped or
red noise-like PSD where the feature is present.

\begin{figure*}
  \centering
  \resizebox{\hsize}{!}{\includegraphics{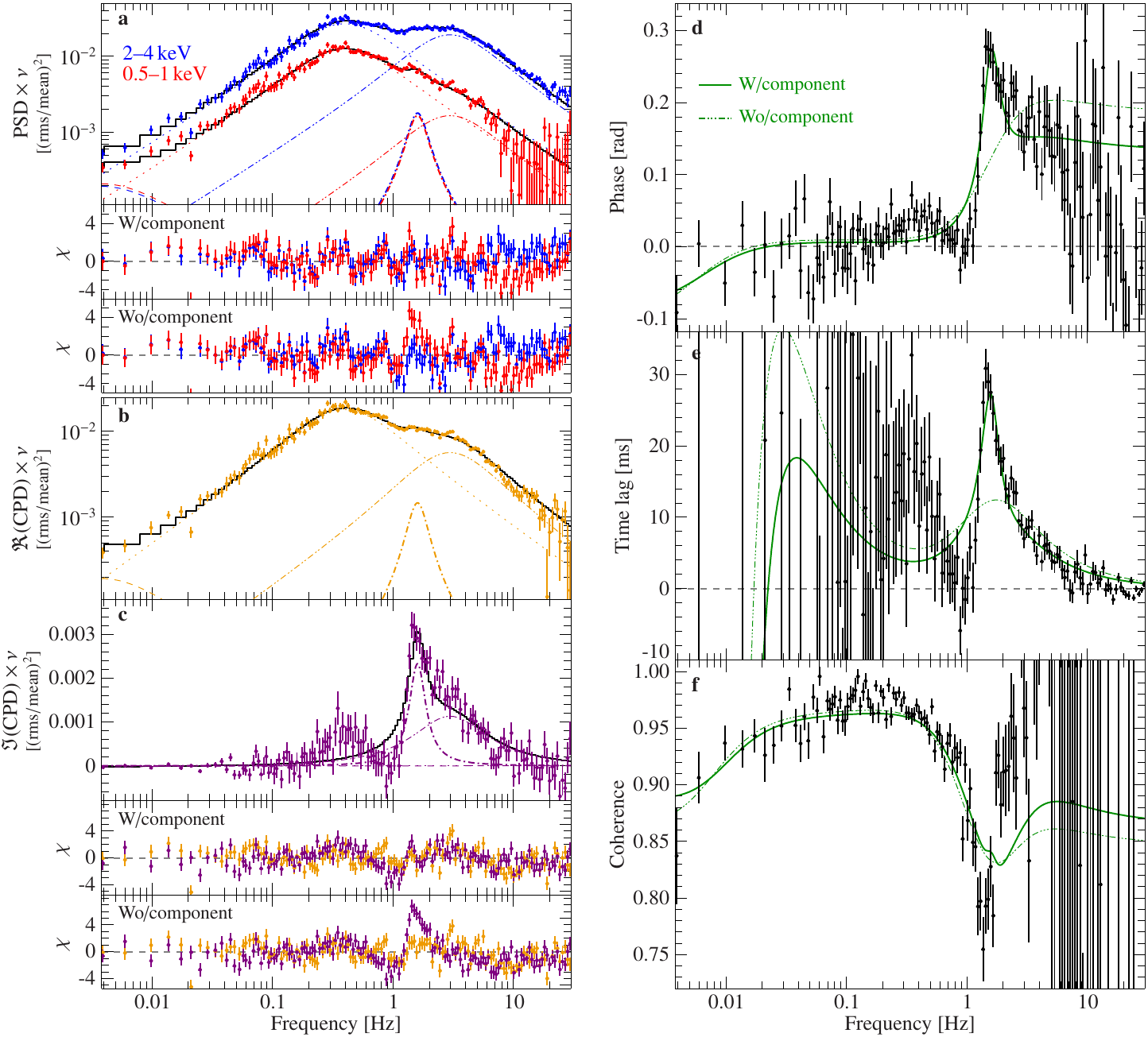}}
  \caption{Simultaneous fit of the power and cross spectra of \cyg for \nicer observation 2636010101. The data can be modeled with one zero-centered Lorentzian (dashed line), two broad Lorentzians (dotted and dash-dot-dotted lines), and one narrow Lorentzian at 1.57\,Hz (dash-dotted lines). The total fit model is shown as a black line. The fit model with and without the narrow component evaluated on the lags and coherence is shown as a solid and dash-dotted green line, respectively. \textbf{a)} Power spectral density and fit residuals with and without the narrow component at 1.6\,Hz. \textbf{b)--c)} Real and imaginary part of the cross spectrum (soft band: 0.5--1\,keV, hard band: 2--4\,keV) and the fit residuals with and without the narrow component. The additional Lorentzian is very significant in the imaginary part. \textbf{d)--e)} Model evaluated (not fitted) on the phase and time lag spectra. The model without the narrow component is shown as green dash-dotted line. 
  \textbf{f)} The coherence spectrum changes only slightly when including the narrow component. The dip at the overlap of $L_1$ and $L_2$ emerges because the Lorentzians are assumed to be incoherent with respect to each other.}
  \label{fig:2636010101_multifit_4lor_plag}
\end{figure*}

In order to quantify this relationship between the timing feature and the Lorentzians further, we use the approach of \citet[][motivated by earlier work, e.g., \citealt{Nowak1999c,Nowak2000a}]{Mendez2024a}, who describe the variability of black hole candidates through the sum of incoherent multiple Lorentzian functions that have a constant phase lag 
$\Delta \phi(\nu)$
between 
the energy bands. 
In such a simple empirical model, we expect high coherence within the frequency band dominated by a single Lorentzian, and a decreased coherence in frequency bands where there are contributions by more than one variability component. 
We 
fit the data from observation 2636010101 using two broad and a zero-centered Lorentzians (Fig.~\ref{fig:2636010101_multifit_4lor_plag}). The model successfully reproduces the reduction in coherence in the overlap region of the $L_1$ and $L_2$ components (1--2\,Hz). However, strong fit residuals remain in the imaginary part of the cross spectrum and the 0.5--1\,keV PSD at around 1--2\,Hz and the fit is unable to describe the peaked lag feature (dashed line in Fig.~\ref{fig:2636010101_multifit_4lor_plag}d--e). 
Adding an additional Lorentzian component at $1.57\pm0.04$\,Hz with a width of $0.703\pm 0.018$\,Hz improves the $\chi^2$ from 1420 to 1066 while adding six free parameters (see Table~\ref{tab:2636010101_multifit_4lor_plag} for all parameters with uncertainties). 
The reduced $\chi^2$ of our final model with four Lorentzians is 1.95 for 546 degrees of freedom. 
This model fits the imaginary part of the cross-spectrum better and, when evaluated on the lags, therefore also reproduces the abrupt lag change at 1--2\,Hz. 

Measuring all time lags with respect to $L_1$, i.e., setting $\Delta\phi(L_1)=0$, we find that $L_0$ has a soft lag of $-7^\circ$ (as this component is zero-centered, the phase lag cannot be transformed to a time lag).
$L_2$ has a slight hard lag of $+11^\circ$, which corresponds to 18\,ms for the centroid frequency of $L_2$. In the additional narrow Lorentzian at 1.57\,Hz, the 2--4\,keV photons arrive with a long hard lag of roughly $+58^\circ$ with respect to the 0.5--1\,keV photons, corresponding to 103\,ms. While this value may appear large, we re-iterate that the timing feature is most likely not due to reverberation (see Sect.~\ref{subsec:time_lag_change}) and that our phenomenological model assumes a constant phase relationship between $L_1$ and $L_2$. More realistic physical models are likely to impose more complex phase relationship between individual variability components. Such models may also explain the remaining residuals in the cross spectrum at around 0.4\,Hz and 3\,Hz that we do not attempt to fit here. 
Alternatively, the timing feature might also be a real, distinct physical feature, similar to the variability components discussed by \citet{Mendez2024a}, which were also only seen in the cross-spectrum but not in the PSD.
While we illustrate here that using one additional narrow Lorentzian reproduces the basic behavior, \citet{Mendez2024a} use several Lorentzians to obtain a precise description of the lag and coherence behavior in the overlap region for a similar observation of MAXI~J1820+070.
% It is also interesting to note that \citet{Rutledge99a} reported a QPO-like feature in the \cyg hard state power spectrum at around 1\,Hz using \ginga data (their Sect.~3.2.1 and Fig.~5), which was also strongest at low energies, 2.3--4.6\,keV, and had a quality factor of roughly 1, similar to what we find for the timing feature.

\section{Summary and Conclusions}
\label{sec:summary}

In this paper we analyzed
211\,ks of \nicer data from \cyg across all spectral states to study the spectral-timing behavior below 1\,keV.
This energy range is important to understand the contribution of the accretion disk to the variability, however, it has not been comprehensively addressed in previous monitoring campaigns. 
We find that significantly more complex phenomenology emerges at
soft X-rays compared to the harder X-ray bands. When investigating
the overall state evolution, for instance, we find that the
color-color diagram of \cyg splits up into a zig-zag track when 
soft energies are included (Sect.~\ref{subsec:color-color}).

The properties of the power spectrum are highly energy dependent
(Sect.~\ref{sec:power_spectral_density}). Consistent with earlier
results, the hard state PSD has two main variability components, which
are commonly described with Lorentzian functions. The first
Lorentzian, $L_1$, dominates the low energy variability where
accretion disk emission contributes the most, while the second
Lorentzian, $L_2$, is the dominant component at high energies where
Comptonization dominates the X-ray spectrum. The second component is
suppressed at low energies.

We emphasize that the existence of these components means that it is
difficult to draw conclusions about X-ray time lags or other timing
quantities when using a broad energy band as the reference band in the time
series analysis.
We directly compare the variability with the disk
contribution, as inferred from the energy spectrum, and show that the
variability at low and high energies has dominant contribution from
the disk and Comptonizing plasma, respectively. 
The energy
dependency of $L_1$ and $L_2$ is therefore consistent with an
interpretation in which the low-frequency Lorentzian originates from
fluctuations in the accretion disk being modulated by Comptonization, while the high-frequency
Lorentzian is solely related to the Comptonizing plasma.

Both components shift to higher frequencies as \cyg softens
\citep{Pottschmidt03a,Grinberg14a}. The variability also shifts to
higher energies in the state transition, clearly showing that the
Comptonized emission becomes the main mechanism associated with the variability. The
soft state of \cyg shows very strong red noise variability at high
energies, with the flux exclusively coming from the Comptonized
emission. At low energies, the source still shows a low but
significant level of variability. We note that the fact that the low-energy
variability likely comes directly from the accretion disk and has lower level of variability compared to the hard state is consistent with the picture that the soft state accretion disk is relatively stable.
Using energy-resolved coherence maps, in Sect.~\ref{sec:coherence} we
investigated the correlation of the variability between low and high
energies. Overall, the hard and intermediate states each show a coherence of
unity or close to unity. This connection changes in the soft state, where the
high-energy red noise is incoherent with low-energy variability. 
%In the propagating fluctuations paradigm, the physical process linking the accretion disk variability and Comptonized emission in the hard and intermediate states is broken up in the soft state -- the propagating fluctuations appear to not reach the emitted red noise variability from the Comptonizing plasma in the \cyg soft state.

We also find a low-energy timing phenomenon in the hard state
(Sect.~\ref{subsec:lags}), which can be characterized by (i) a sudden
jump from a soft to a hard time lag at around 1--2\,Hz, (ii) a
drop in coherence at the peak frequency of this hard time lag, (iii) a
position in between the two broad Lorentzians in the power spectrum, and
(iv) a strong energy dependence as it vanishes when bands
${\gtrsim}1.5$\,keV are considered. The feature increases in frequency as the
source moves into the transition, together with the Lorentzians. If other sources of lags are
present, such as the well-known power law-like hard lags, the feature
is blended into the lag spectrum and emerges as a shelf-like
structure. We do not find the feature in observations with a
single-humped power spectrum softer than $\Gamma\gtrsim 2.3$.

\begin{figure*}
  \resizebox{\hsize}{!}{\includegraphics{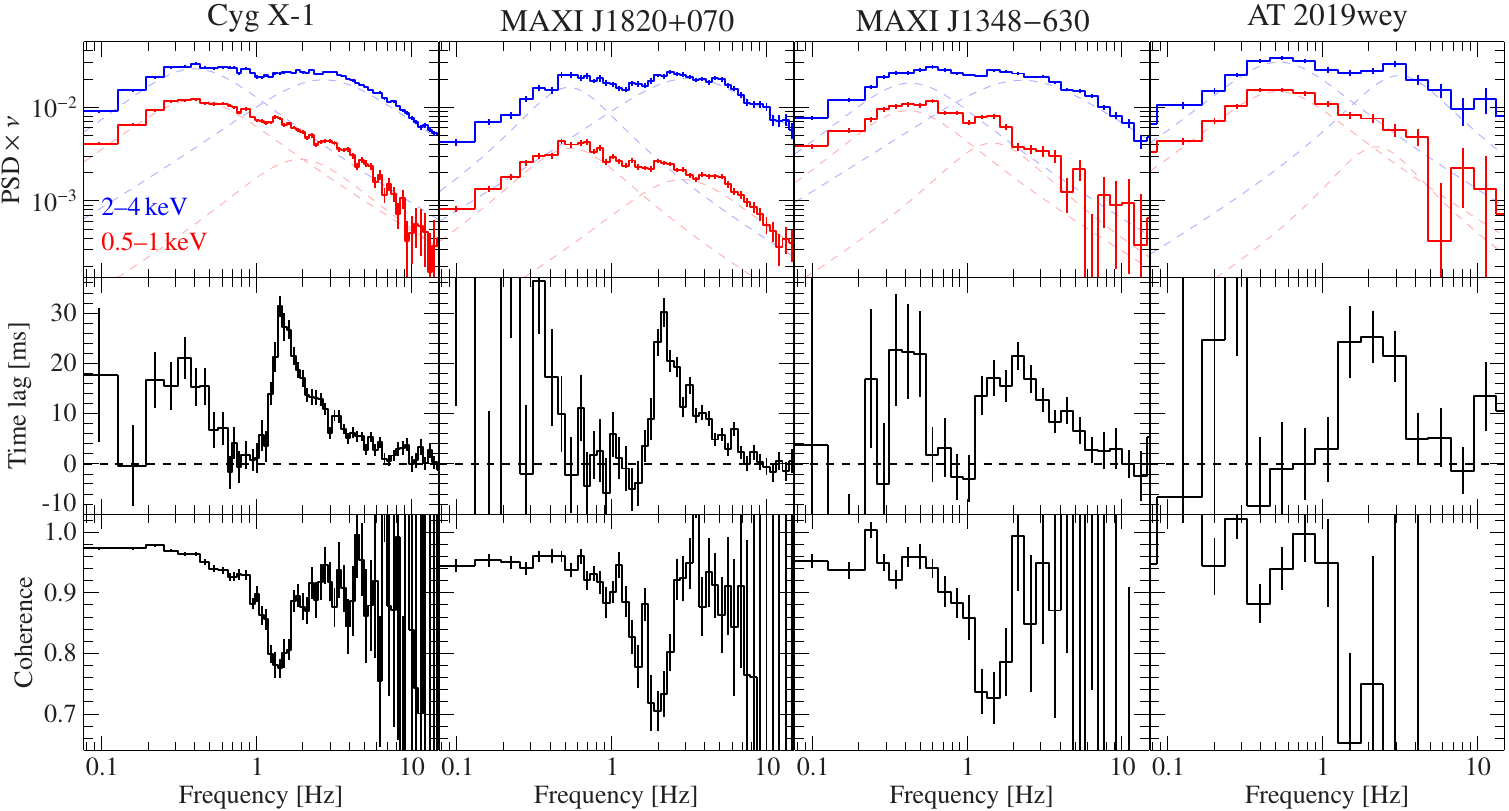}}
  \caption{Comparison of the timing feature in \cyg to similar data for the low-mass X-ray binaries MAXI~J1820+070, MAXI~J1348$-$630, and AT~2019wey. In all four observations, a reduction of $L_2$ at low energies, a time lag jump between soft (0.5--1\,keV) and hard (2--4\,keV) photons, and a coherence drop in the overlap region of the Lorentzians can be seen. The similarity of the data suggests that the timing feature is a ubiquitous property of accreting black hole binaries.}
  \label{fig:timing_feature_source_comparison}
\end{figure*}

We find that similar features are not only present in the
variability of the HMXB \cyg, but also in data from hard state
observations of LMXB black holes.
Figure~\ref{fig:timing_feature_source_comparison} shows a comparison
of the \cyg data to observations of MAXI~J1820+070 (obs. ID
1200120268), MAXI~J1348$-$630 (obs. ID 2200530129), and AT~2019wey
(obs. ID 3201710115), identified using the 
products of \citet{Wang22a}. These data show the exact same
characteristics as the data from \cyg, that is, a reduced amplitude of
$L_2$ at low energies, a time lag in the overlap region of the
Lorentzians, and a drop in coherence. This comparison provides
evidence that the timing feature is a general property intrinsic to
both high- and low-mass X-ray binary black holes. It is
interesting to note that for the sources where reliable mass and
distance estimates are available, such that they can be placed on the
q-diagram, the observations showing the
feature appear to be located on the lower branch of the q-diagram (see blue pentagon and green triangle in Fig.~\ref{fig:q_diagram}).
Further research is required to systematically study the occurrence of
the feature in LMXBs.

Since this complex lag behavior is found in both HMXBs and LMXBs, the
Lorentzian components and the feature are not due to effects from the
interaction of the X-rays with the stellar wind. 
By simultaneously fitting the power and cross spectra (Sect.~\ref{subsec:lags}), we showed that the feature can be modeled as a relatively narrow Lorentzian component 
in between the two well-known Lorentzians constituting the broad band noise. 
While the simplified
modeling in Sect.~\ref{subsec:lags} did not assume
any relationship between the Lorentzians, in reality these components
show correlated behavior, e.g., they shift together
in frequency when the source softens
\citep{Pottschmidt03a,Grinberg14a}. However, they are not perfectly
coupled, as a drop in coherence is observed in the overlap region
\citep[][and Sect.~\ref{sec:coherence}]{Nowak2000a} and because the
strength of each Lorentzian with energy is different between the
components \citep[][and
Sect.~\ref{sec:power_spectral_density}]{Grinberg14a}.

If the Lorentzians are sufficiently coupled, the timing feature might be due to a beat between the two broad Lorentzians, similar to, e.g., models invoked to describe kHz~QPOs in accreting neutron stars \citep[e.g.,][]{AlparShaham1985a,Lamb1985a}, although we note that kHz QPOs show much higher quality factors than $L_1$ or $L_2$, and therefore physical models for them are likely not applicable to black hole candidates.
The beat frequency is roughly given by the difference between the centroid frequencies of the two Lorentzians, that is, $\nu_{L_2}-\nu_{L_1}$. 
For our example hard state observation, the difference frequency of $L_2$ and $L_1$  matches the frequency of the Lorentzian representing the timing feature in Sect.~\ref{subsec:lags}. 
We emphasize that this beat would not necessarily mean the presence of a separate distinct physical process, such as an oscillation at a characteristic frequency, creating the feature. Instead, the beat arises from the interaction of the variability processes constituting $L_1$ and $L_2$.
A theoretical beat model will have to show how these two partially incoherent, heavily damped processes (``low-Q'') can create a narrow (``high-Q''), localized lag at a reduced coherence.

In a similar interpretation, lags due to interference effects of two distinct spectral components have also been discussed by \citet{Veledina2018a}. She shows that such interference can lead to a broad lag structure at around 8\,Hz in an \rxte/PCA observation of \cyg. This observation is located on the soft side of the q-diagram. We also note that the lag structure modeled by \citet{Veledina2018a} corresponds to the enhanced lag seen in \cyg during state transitions at energies above ${\sim}2.5$\,keV \citep{Pottschmidt00a}. It cannot be ruled out, however, that similar interference behavior may also cause the lag structures found here.
It is beyond the scope of this paper
to extend the \citet{Veledina2018a} model for non-unity coherence or develop a beat model that describes the observed coherence loss and lags in order to prove the beat frequency hypothesis.
However, independent of the interpretation of the feature, the
detection in both types of accreting stellar mass black holes X-ray binaries has
intriguing implications. In particular, we stress that the soft lag at low energies associated with the timing feature cannot be interpreted as reverberation
lags, as shown in Sect.~\ref{subsec:time_lag_change}. Instead, this
detection requires a modification of the general picture on how soft
lags can be created, at least for the observations which show this
feature.

\begin{acknowledgements}
We thank the anonymous referee for constructive comments that improved the manuscript.
  We acknowledge funding from the Deutsches Zentrum f\"ur Luft- und
  Raumfahrt contract 50~QR~2103. OK thanks the German Academic
  Exchange Service (DAAD) for the fellowship Forschungsstipendien
  f\"ur Doktorandinnen und Doktoranden, 2021/22 (57556281) that
  enabled this research during a three month visit of Caltech, NASA
  GSFC, CfA/Harvard-Smithsonian, and MIT. GM acknowledges financial support from the European Union’s Horizon Europe research and innovation programme under the Marie Sk\l{}odowska-Curie grant agreement No. 101107057.
MM acknowledges the research
  program Athena with project number 184.034.002, which is (partly)
  financed by the Dutch Research Council (NWO). MM and FG thank the
  Team Meeting at the International Space Science Institute (Bern) for
  fruitful discussions. MAN acknowledges support from NASA Grants 80NSSC21K0960 and 80NSSC23K0994. FG is a CONICET researcher and acknowledges support from PIP 0113 and PIBAA 1275 (CONICET). AI acknowledges support from the Royal Society. 
  MK acknowledges support from the NWO Spinoza Prize.
  The material is based upon work supported by
  NASA under award number 80GSFC21M0002 (CRESST II). This research has
  made use of ISIS functions (ISISscripts) provided by ECAP/Remeis
  observatory and MIT (http://www.sternwarte.uni-erlangen.de/isis/).
\end{acknowledgements}

\bibliographystyle{jwaabib}
\bibliography{mnemonic,aa_abbrv,references}

\begin{thebibliography}{}

\bibitem[\protect\astroncite{{Alpar} \& {Shaham}}{1985}]{AlparShaham1985a}
{Alpar} M.A., {Shaham} J.,  1985, Nat
  \href{http://dx.doi.org/10.1038/316239a0}{316, 239}

\bibitem[\protect\astroncite{{Axelsson} et~al.}{2005}]{Axelsson2005a}
{Axelsson} M., {Borgonovo} L., {Larsson} S.,  2005, A\&A
  \href{http://dx.doi.org/10.1051/0004-6361:20042362}{438, 999}

\bibitem[\protect\astroncite{{Axelsson} et~al.}{2006}]{Axelsson2006a}
{Axelsson} M., {Borgonovo} L., {Larsson} S.,  2006, A\&A
  \href{http://dx.doi.org/10.1051/0004-6361:20054397}{452, 975}

\bibitem[\protect\astroncite{{Axelsson} \& {Done}}{2018}]{AxelssonDone2018a}
{Axelsson} M., {Done} C.,  2018, MNRAS
  \href{http://dx.doi.org/10.1093/mnras/sty1801}{480, 751}

\bibitem[\protect\astroncite{{Ba{\l}uci{\'n}ska-Church}
  et~al.}{2000}]{balucinska-church:2000}
{Ba{\l}uci{\'n}ska-Church} M., Church M.J., Charles P.A., et~al., 2000, MNRAS
  \href{http://dx.doi.org/10.1046/j.1365-8711.2000.03149.x}{311, 861}

\bibitem[\protect\astroncite{{Barillier} et~al.}{2023}]{Barillier2023a}
{Barillier} E., {Grinberg} V., {Horn} D., et~al., 2023, ApJ
  \href{http://dx.doi.org/10.3847/1538-4357/acaeaf}{944, 165}

\bibitem[\protect\astroncite{{Belloni} \&
  {Hasinger}}{1990a}]{BelloniHasinger90b}
{Belloni} T., {Hasinger} G.,  1990a, A\&A 230, 103

\bibitem[\protect\astroncite{{Belloni}}{2010}]{Belloni10a}
{Belloni} T.M.,  2010,
\newblock {States and Transitions in Black Hole Binaries}. In: Belloni T. (ed.)
  The Jet Paradigm, Vol. 794. Lecture Notes in Physics Springer, Berlin, p.~53

\bibitem[\protect\astroncite{{Belloni} \&
  {Hasinger}}{1990b}]{BelloniHasinger90}
{Belloni} T.M., {Hasinger} G.,  1990b, A\&A 227, L33

\bibitem[\protect\astroncite{{Belloni} et~al.}{2005}]{Belloni2005a}
{Belloni} T.M., {Homan} J., {Casella} P., et~al., 2005, A\&A
  \href{http://dx.doi.org/10.1051/0004-6361:20042457}{440, 207}

\bibitem[\protect\astroncite{{Bendat} \& {Piersol}}{1986}]{BendatPiersol86}
{Bendat} J.S., {Piersol} A.G.,  1986, J.\ Sound \& Vibration
  \href{http://dx.doi.org/10.1016/0022-460X(86)90186-0}{106, 391}

\bibitem[\protect\astroncite{{Bowyer} et~al.}{1965}]{Bowyer1965Sci}
{Bowyer} S., {Byram} E.T., {Chubb} T.A., {Friedman} H.,  1965, Sci
  \href{http://dx.doi.org/10.1126/science.147.3656.394}{147, 394}

\bibitem[\protect\astroncite{{Churazov} et~al.}{2001}]{Churazov2001a}
{Churazov} E., {Gilfanov} M., {Revnivtsev} M.,  2001, MNRAS
  \href{http://dx.doi.org/10.1046/j.1365-8711.2001.04056.x}{321, 759}

\bibitem[\protect\astroncite{{Cui} et~al.}{1997a}]{Cui97b}
{Cui} W., {Heindl} W.A., {Rothschild} R.E., et~al., 1997a, ApJL
  \href{http://dx.doi.org/10.1086/310419}{474, L57}

\bibitem[\protect\astroncite{{Cui} et~al.}{1997b}]{Cui97a}
{Cui} W., {Zhang} S.N., {Focke} W., {Swank} J.H.,  1997b, ApJ
  \href{http://dx.doi.org/10.1086/304341}{484, 383}

\bibitem[\protect\astroncite{{De Marco} et~al.}{2021}]{DeMarco21a}
{De Marco} B., {Zdziarski} A.A., {Ponti} G., et~al., 2021, A\&A
  \href{http://dx.doi.org/10.1051/0004-6361/202140567}{654, A14}

\bibitem[\protect\astroncite{{Done} \&
  {Gierli{\'n}ski}}{2003}]{DoneGierlinski2003a}
{Done} C., {Gierli{\'n}ski} M.,  2003, MNRAS
  \href{http://dx.doi.org/10.1046/j.1365-8711.2003.06614.x}{342, 1041}

\bibitem[\protect\astroncite{{Duro} et~al.}{2011}]{Duro11a}
{Duro} R., {Dauser} T., {Wilms} J., et~al., 2011, A\&A
  \href{http://dx.doi.org/10.1051/0004-6361/201117446}{533, L3}

\bibitem[\protect\astroncite{{Feng} et~al.}{2022}]{Feng22a}
{Feng} M.Z., {Kong} L.D., {Wang} P.J., et~al., 2022, ApJ
  \href{http://dx.doi.org/10.3847/1538-4357/ac7875}{934, 47}

\bibitem[\protect\astroncite{{Garc{\'\i}a} et~al.}{2014}]{Garcia2014a}
{Garc{\'\i}a} J.A., {McClintock} J.E., {Steiner} J.F., et~al., 2014, ApJ
  \href{http://dx.doi.org/10.1088/0004-637X/794/1/73}{794, 73}

\bibitem[\protect\astroncite{{Gendreau} et~al.}{2016}]{Gendreau16a}
{Gendreau} K.C., {Arzoumanian} Z., {Adkins} P.W., et~al., 2016,
\newblock In: {den Herder} J.W.A., {Takahashi} T., {Bautz} M. (eds.) Proc.
  SPIE., Vol. 9905. Space Telescopes and Instrumentation 2016: Ultraviolet to
  Gamma Ray, Edinburgh, United Kingdom,
  \href{http://dx.doi.org/10.1117/12.2231304}{p. 99051H}

\bibitem[\protect\astroncite{{Gilfanov} et~al.}{2000}]{Gilfanov2000a}
{Gilfanov} M., {Churazov} E., {Revnivtsev} M.,  2000, MNRAS
  \href{http://dx.doi.org/10.1046/j.1365-8711.2000.03686.x}{316, 923}

\bibitem[\protect\astroncite{{Gleissner} et~al.}{2004a}]{Gleissner04b}
{Gleissner} T., {Wilms} J., {Pooley} G.G., et~al., 2004a, A\&A
  \href{http://dx.doi.org/10.1051/0004-6361:20040280}{425, 1061}

\bibitem[\protect\astroncite{{Gleissner} et~al.}{2004b}]{Gleissner04a}
{Gleissner} T., {Wilms} J., {Pottschmidt} K., et~al., 2004b, A\&A
  \href{http://dx.doi.org/10.1051/0004-6361:20031684}{414, 1091}

\bibitem[\protect\astroncite{{Grinberg} et~al.}{2013}]{Grinberg13a}
{Grinberg} V., {Hell} N., {Pottschmidt} K., et~al., 2013, A\&A
  \href{http://dx.doi.org/10.1051/0004-6361/201321128}{554, A88}

\bibitem[\protect\astroncite{{Grinberg} et~al.}{2015}]{Grinberg15a}
{Grinberg} V., {Leutenegger} M.A., {Hell} N., et~al., 2015, A\&A
  \href{http://dx.doi.org/10.1051/0004-6361/201425418}{576, A117}

\bibitem[\protect\astroncite{{Grinberg} et~al.}{2020}]{Grinberg20a}
{Grinberg} V., {Nowak} M.A., {Hell} N.,  2020, A\&A
  \href{http://dx.doi.org/10.1051/0004-6361/202039183}{643, A109}

\bibitem[\protect\astroncite{{Grinberg} et~al.}{2014}]{Grinberg14a}
{Grinberg} V., {Pottschmidt} K., {B{\"o}ck} M., et~al., 2014, A\&A
  \href{http://dx.doi.org/10.1051/0004-6361/201322969}{565, A1}

\bibitem[\protect\astroncite{{Hirsch} et~al.}{2019}]{Hirsch19a}
{Hirsch} M., {Hell} N., {Grinberg} V., et~al., 2019, A\&A
  \href{http://dx.doi.org/10.1051/0004-6361/201935074}{626, A64}

\bibitem[\protect\astroncite{{Houck}}{2002}]{ISISHouck02}
{Houck} J.C.,  2002,
\newblock In: {Branduardi-Raymont} G. (ed.) High Resolution X-ray Spectroscopy
  with XMM-Newton and Chandra., Mullard Space Science Laboratory, p.~17

\bibitem[\protect\astroncite{{Ingram} et~al.}{2023}]{Ingram2023b_ARXIV}
{Ingram} A., {Bollemeijer} N., {Veledina} A., et~al., 2023, {submitted to ApJ}
  \href{http://dx.doi.org/10.48550/arXiv.2311.05497}{ arXiv:2311.05497}

\bibitem[\protect\astroncite{{Ingram} et~al.}{2019}]{Ingram19a}
{Ingram} A., {Mastroserio} G., {Dauser} T., et~al., 2019, MNRAS
  \href{http://dx.doi.org/10.1093/mnras/stz1720}{488, 324}

\bibitem[\protect\astroncite{{Ingram} \& {van der Klis}}{2013}]{Ingram13a}
{Ingram} A., {van der Klis} M.,  2013, MNRAS
  \href{http://dx.doi.org/10.1093/mnras/stt1107}{434, 1476}

\bibitem[\protect\astroncite{{Jahoda} et~al.}{2006}]{Jahoda2006a}
{Jahoda} K., {Markwardt} C.B., {Radeva} Y., et~al., 2006, ApJS
  \href{http://dx.doi.org/10.1086/500659}{163, 401}

\bibitem[\protect\astroncite{{Ji} et~al.}{2003}]{Ji2003a}
{Ji} J.F., {Zhang} S.N., {Qu} J.L., {Li} T.P.,  2003, ApJL
  \href{http://dx.doi.org/10.1086/368269}{584, L23}

\bibitem[\protect\astroncite{{Kara} et~al.}{2013}]{Kara2013a}
{Kara} E., {Fabian} A.C., {Cackett} E.M., et~al., 2013, MNRAS
  \href{http://dx.doi.org/10.1093/mnras/sts155}{428, 2795}

\bibitem[\protect\astroncite{{Kara} et~al.}{2019}]{Kara19a}
{Kara} E., {Steiner} J.F., {Fabian} A.C., et~al., 2019, Nat
  \href{http://dx.doi.org/10.1038/s41586-018-0803-x}{565, 198}

\bibitem[\protect\astroncite{{Kawamura} et~al.}{2022}]{Kawamura22a}
{Kawamura} T., {Axelsson} M., {Done} C., {Takahashi} T.,  2022, MNRAS
  \href{http://dx.doi.org/10.1093/mnras/stac045}{511, 536}

\bibitem[\protect\astroncite{{Kawamura} et~al.}{2023}]{Kawamura2023a}
{Kawamura} T., {Done} C., {Takahashi} T.,  2023, MNRAS
  \href{http://dx.doi.org/10.1093/mnras/stad2338}{525, 1280}

\bibitem[\protect\astroncite{{Kotov} et~al.}{2001}]{Kotov01a}
{Kotov} O., {Churazov} E., {Gilfanov} M.,  2001, MNRAS
  \href{http://dx.doi.org/10.1046/j.1365-8711.2001.04769.x}{327, 799}

\bibitem[\protect\astroncite{{Lai} et~al.}{2022}]{Lai22a}
{Lai} E.V., {De Marco} B., {Zdziarski} A.A., et~al., 2022, MNRAS
  \href{http://dx.doi.org/10.1093/mnras/stac688}{512, 2671}

\bibitem[\protect\astroncite{{Lamb} et~al.}{1985}]{Lamb1985a}
{Lamb} F.K., {Shibazaki} N., {Alpar} M.A., {Shaham} J.,  1985, Nat
  \href{http://dx.doi.org/10.1038/317681a0}{317, 681}

\bibitem[\protect\astroncite{{Liang} \& {Price}}{1977}]{LiangPrice1977a}
{Liang} E.P.T., {Price} R.H.,  1977, ApJ
  \href{http://dx.doi.org/10.1086/155677}{218, 247}

\bibitem[\protect\astroncite{{Lyubarskii}}{1997}]{Lyubarskii1997a}
{Lyubarskii} Y.E.,  1997, MNRAS
  \href{http://dx.doi.org/10.1093/mnras/292.3.679}{292, 679}

\bibitem[\protect\astroncite{{Maqbool} et~al.}{2019}]{Maqbool19a}
{Maqbool} B., {Mudambi} S.P., {Misra} R., et~al., 2019, MNRAS
  \href{http://dx.doi.org/10.1093/mnras/stz930}{486, 2964}

\bibitem[\protect\astroncite{{Markoff} et~al.}{2005}]{Markoff05a}
{Markoff} S., {Nowak} M.A., {Wilms} J.,  2005, ApJ
  \href{http://dx.doi.org/10.1086/497628}{635, 1203}

\bibitem[\protect\astroncite{{Mastroserio} et~al.}{2019}]{Mastroserio19a}
{Mastroserio} G., {Ingram} A., {van der Klis} M.,  2019, MNRAS
  \href{http://dx.doi.org/10.1093/mnras/stz1727}{488, 348}

\bibitem[\protect\astroncite{{Mastroserio} et~al.}{2021}]{Mastroserio21a}
{Mastroserio} G., {Ingram} A., {Wang} J., et~al., 2021, MNRAS
  \href{http://dx.doi.org/10.1093/mnras/stab2056}{507, 55}

\bibitem[\protect\astroncite{{M{\'e}ndez} et~al.}{2024}]{Mendez2024a}
{M{\'e}ndez} M., {Peirano} V., {Garc{\'\i}a} F., et~al., 2024, MNRAS
  \href{http://dx.doi.org/10.1093/mnras/stad3786}{527, 9405}

\bibitem[\protect\astroncite{{M{\'e}ndez} \& {van der
  Klis}}{1997}]{MendezVanderKlis1997a}
{M{\'e}ndez} M., {van der Klis} M.,  1997, ApJ
  \href{http://dx.doi.org/10.1086/303914}{479, 926}

\bibitem[\protect\astroncite{{Miller-Jones} et~al.}{2021}]{MillerJones21a}
{Miller-Jones} J.C.A., {Bahramian} A., {Orosz} J.A., et~al., 2021, Sci
  \href{http://dx.doi.org/10.1126/science.abb3363}{371, 1046}

\bibitem[\protect\astroncite{{Misra} et~al.}{2017}]{Misra17a}
{Misra} R., {Yadav} J.S., {Verdhan Chauhan} J., et~al., 2017, ApJ
  \href{http://dx.doi.org/10.3847/1538-4357/835/2/195}{835, 195}

\bibitem[\protect\astroncite{{Miyamoto} et~al.}{1991}]{Miyamoto91a}
{Miyamoto} S., {Kimura} K., {Kitamoto} S., et~al., 1991, ApJ
  \href{http://dx.doi.org/10.1086/170837}{383, 784}

\bibitem[\protect\astroncite{Miyamoto \& Kitamoto}{1989}]{miyamoto:1989}
Miyamoto S., Kitamoto S.,  1989, Nat
  \href{http://dx.doi.org/10.1038/342773a0}{342, 773}

\bibitem[\protect\astroncite{{Miyamoto} et~al.}{1992}]{Miyamoto92a}
{Miyamoto} S., {Kitamoto} S., {Iga} S., et~al., 1992, ApJL
  \href{http://dx.doi.org/10.1086/186389}{391, L21}

\bibitem[\protect\astroncite{{Miyamoto} et~al.}{1988}]{Miyamoto1988a}
{Miyamoto} S., {Kitamoto} S., {Mitsuda} K., {Dotani} T.,  1988, Nat
  \href{http://dx.doi.org/10.1038/336450a0}{336, 450}

\bibitem[\protect\astroncite{{Mummery} \& {Balbus}}{2022}]{MummeryBalbus2022a}
{Mummery} A., {Balbus} S.,  2022, MNRAS
  \href{http://dx.doi.org/10.1093/mnras/stac2844}{517, 3423}

\bibitem[\protect\astroncite{{Nowak}}{1995}]{nowak:95a}
{Nowak} M.A.,  1995, PASP 107, 1207

\bibitem[\protect\astroncite{{Nowak}}{2000}]{Nowak2000a}
{Nowak} M.A.,  2000, MNRAS
  \href{http://dx.doi.org/10.1046/j.1365-8711.2000.03668.x}{318, 361}

\bibitem[\protect\astroncite{{Nowak} et~al.}{2011}]{Nowak11a}
{Nowak} M.A., {Hanke} M., {Trowbridge} S.N., et~al., 2011, ApJ
  \href{http://dx.doi.org/10.1088/0004-637X/728/1/13}{728, 13}

\bibitem[\protect\astroncite{{Nowak} et~al.}{1999a}]{Nowak99a}
{Nowak} M.A., {Vaughan} B.A., {Wilms} J., et~al., 1999a, ApJ
  \href{http://dx.doi.org/10.1086/306610}{510, 874}

\bibitem[\protect\astroncite{{Nowak} et~al.}{1999b}]{Nowak1999c}
{Nowak} M.A., {Wilms} J., {Dove} J.B.,  1999b, ApJ
  \href{http://dx.doi.org/10.1086/307189}{517, 355}

\bibitem[\protect\astroncite{{Nowak} et~al.}{2012}]{Nowak12a}
{Nowak} M.A., {Wilms} J., {Hanke} M., et~al., 2012, Mem. S.A. It. 83, 202

\bibitem[\protect\astroncite{{Pottschmidt} et~al.}{2000}]{Pottschmidt00a}
{Pottschmidt} K., {Wilms} J., {Nowak} M.A., et~al., 2000, A\&A 357, L17

\bibitem[\protect\astroncite{{Pottschmidt} et~al.}{2006}]{Pottschmidt2006a}
{Pottschmidt} K., {Wilms} J., {Nowak} M.A., et~al., 2006, Adv. Space Res.
  \href{http://dx.doi.org/10.1016/j.asr.2005.04.032}{38, 1350}

\bibitem[\protect\astroncite{{Pottschmidt} et~al.}{2003}]{Pottschmidt03a}
{Pottschmidt} K., {Wilms} J., {Nowak} M.A., et~al., 2003, A\&A
  \href{http://dx.doi.org/10.1051/0004-6361:20030906}{407, 1039}

\bibitem[\protect\astroncite{{Priedhorsky} et~al.}{1979}]{Priedhorsky79a}
{Priedhorsky} W., {Garmire} G.P., {Rothschild} R., et~al., 1979, ApJ
  \href{http://dx.doi.org/10.1086/157396}{233, 350}

\bibitem[\protect\astroncite{{Rapisarda} et~al.}{2017}]{Rapisarda2017a}
{Rapisarda} S., {Ingram} A., {van der Klis} M.,  2017, MNRAS
  \href{http://dx.doi.org/10.1093/mnras/stx2110}{472, 3821}

\bibitem[\protect\astroncite{{Remillard} et~al.}{2022}]{Remillard2022a}
{Remillard} R.A., {Loewenstein} M., {Steiner} J.F., et~al., 2022, AJ
  \href{http://dx.doi.org/10.3847/1538-3881/ac4ae6}{163, 130}

\bibitem[\protect\astroncite{{Revnivtsev} et~al.}{2000}]{Revnivtsev2000a}
{Revnivtsev} M., {Gilfanov} M., {Churazov} E.,  2000, A\&A
  \href{http://dx.doi.org/10.48550/arXiv.astro-ph/0007092}{363, 1013}

\bibitem[\protect\astroncite{{Shaposhnikov} et~al.}{2012}]{Shaposhnikov2012a}
{Shaposhnikov} N., {Jahoda} K., {Markwardt} C., et~al., 2012, ApJ
  \href{http://dx.doi.org/10.1088/0004-637X/757/2/159}{757, 159}

\bibitem[\protect\astroncite{{Steiner} et~al.}{2009}]{Steiner2009a}
{Steiner} J.F., {Narayan} R., {McClintock} J.E., {Ebisawa} K.,  2009, PASP
  \href{http://dx.doi.org/10.1086/648535}{121, 1279}

\bibitem[\protect\astroncite{{Stevens} et~al.}{2018}]{Stevens18a}
{Stevens} A.L., {Uttley} P., {Altamirano} D., et~al., 2018, ApJL
  \href{http://dx.doi.org/10.3847/2041-8213/aae1a4}{865, L15}

\bibitem[\protect\astroncite{{Stiele} \& {Kong}}{2018}]{StieleKong18}
{Stiele} H., {Kong} A.K.H.,  2018, ApJ
  \href{http://dx.doi.org/10.3847/1538-4357/aae7d3}{868, 71}

\bibitem[\protect\astroncite{{Sugimoto} et~al.}{2016}]{Sugimoto16a}
{Sugimoto} J., {Mihara} T., {Kitamoto} S., et~al., 2016, PASJ
  \href{http://dx.doi.org/10.1093/pasj/psw004}{68, S17}

\bibitem[\protect\astroncite{{Sunyaev} \&
  {Tr\"umper}}{1979}]{SunyaevTruemper1979a}
{Sunyaev} R.A., {Tr\"umper} J.,  1979, Nat
  \href{http://dx.doi.org/10.1038/279506a0}{279, 506}

\bibitem[\protect\astroncite{{Tananbaum} et~al.}{1972}]{Tananbaum1972a}
{Tananbaum} H., {Gursky} H., {Kellogg} E., et~al., 1972, ApJL
  \href{http://dx.doi.org/10.1086/181042}{177, L5}

\bibitem[\protect\astroncite{{Tomsick} et~al.}{2014}]{Tomsick14a}
{Tomsick} J.A., {Nowak} M.A., {Parker} M., et~al., 2014, ApJ
  \href{http://dx.doi.org/10.1088/0004-637X/780/1/78}{780, 78}

\bibitem[\protect\astroncite{{Uttley} et~al.}{2014}]{Uttley14a}
{Uttley} P., {Cackett} E.M., {Fabian} A.C., et~al., 2014, A\&AR
  \href{http://dx.doi.org/10.1007/s00159-014-0072-0}{22, 72}

\bibitem[\protect\astroncite{{Uttley} et~al.}{2011}]{Uttley11a}
{Uttley} P., {Wilkinson} T., {Cassatella} P., et~al., 2011, MNRAS
  \href{http://dx.doi.org/10.1111/j.1745-3933.2011.01056.x}{414, L60}

\bibitem[\protect\astroncite{{Vaughan} \& {Nowak}}{1997}]{VaughanNowak97a}
{Vaughan} B.A., {Nowak} M.A.,  1997, ApJL
  \href{http://dx.doi.org/10.1086/310430}{474, L43}

\bibitem[\protect\astroncite{{Vaughan} et~al.}{2003}]{Vaughan03a}
{Vaughan} S., {Edelson} R., {Warwick} R.S., {Uttley} P.,  2003, MNRAS
  \href{http://dx.doi.org/10.1046/j.1365-2966.2003.07042.x}{345, 1271}

\bibitem[\protect\astroncite{{Veledina}}{2018}]{Veledina2018a}
{Veledina} A.,  2018, MNRAS \href{http://dx.doi.org/10.1093/mnras/sty2556}{481,
  4236}

\bibitem[\protect\astroncite{{Walborn}}{1973}]{Walborn73a}
{Walborn} N.R.,  1973, ApJL \href{http://dx.doi.org/10.1086/181131}{179, L123}

\bibitem[\protect\astroncite{{Wang} et~al.}{2022}]{Wang22a}
{Wang} J., {Kara} E., {Lucchini} M., et~al., 2022, ApJ
  \href{http://dx.doi.org/10.3847/1538-4357/ac6262}{930, 18}

\bibitem[\protect\astroncite{{Wang} et~al.}{2020}]{Wang20a}
{Wang} J., {Kara} E., {Steiner} J.F., et~al., 2020, ApJ
  \href{http://dx.doi.org/10.3847/1538-4357/ab9ec3}{899, 44}

\bibitem[\protect\astroncite{{Wang} et~al.}{2021}]{Wang21a}
{Wang} J., {Mastroserio} G., {Kara} E., et~al., 2021, ApJL
  \href{http://dx.doi.org/10.3847/2041-8213/abec79}{910, L3}

\bibitem[\protect\astroncite{{Wilkinson} \&
  {Uttley}}{2009}]{WilkinsonUttley09a}
{Wilkinson} T., {Uttley} P.,  2009, MNRAS
  \href{http://dx.doi.org/10.1111/j.1365-2966.2009.15008.x}{397, 666}

\bibitem[\protect\astroncite{{Wilms} et~al.}{2000}]{Wilms00a}
{Wilms} J., {Allen} A., {McCray} R.,  2000, ApJ 542, 914

\bibitem[\protect\astroncite{{Wilms} et~al.}{2006}]{Wilms06a}
{Wilms} J., {Nowak} M.A., {Pottschmidt} K., et~al., 2006, A\&A
  \href{http://dx.doi.org/10.1051/0004-6361:20053938}{447, 245}

\bibitem[\protect\astroncite{{Wilms} et~al.}{2007}]{Wilms07a}
{Wilms} J., {Pottschmidt} K., {Pooley} G.G., et~al., 2007, ApJL
  \href{http://dx.doi.org/10.1086/520508}{663, L97}

\bibitem[\protect\astroncite{{Zdziarski} et~al.}{2021}]{Zdziarski21a}
{Zdziarski} A.A., {Dzie{\l}ak} M.A., {De Marco} B., et~al., 2021, ApJL
  \href{http://dx.doi.org/10.3847/2041-8213/abe7ef}{909, L9}

\bibitem[\protect\astroncite{{Zdziarski} et~al.}{2016}]{Zdziarski2016a}
{Zdziarski} A.A., {Segreto} A., {Pooley} G.G.,  2016, MNRAS
  \href{http://dx.doi.org/10.1093/mnras/stv2647}{456, 775}

\bibitem[\protect\astroncite{{Zhang} et~al.}{2020}]{Zhang20a}
{Zhang} L., {Altamirano} D., {C{\'u}neo} V.A., et~al., 2020, MNRAS
  \href{http://dx.doi.org/10.1093/mnras/staa2842}{499, 851}

\bibitem[\protect\astroncite{{Zhou} et~al.}{2022}]{Zhou22a}
{Zhou} M., {Grinberg} V., {Bu} Q.C., et~al., 2022, A\&A
  \href{http://dx.doi.org/10.1051/0004-6361/202244240}{666, A172}

\end{thebibliography}

\appendix

\section{Energy-dependent RMS Variability Amplitude in the Hard State}
\label{subsec:rms_spectrum}

\begin{figure}
  \resizebox{\hsize}{!}{\includegraphics{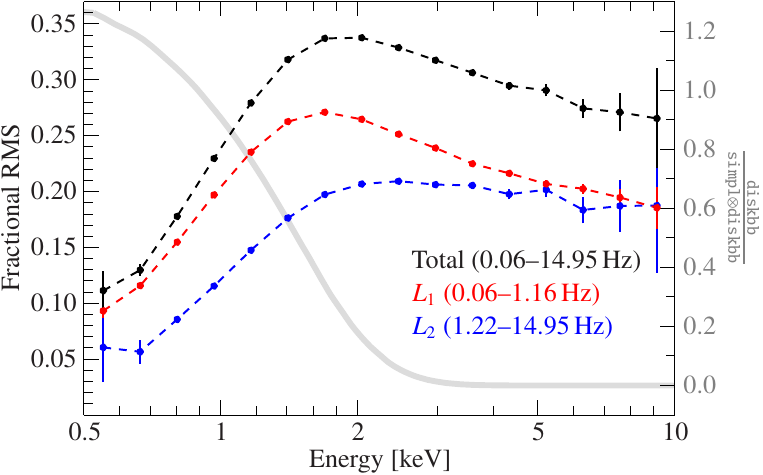}}
  \caption{RMS spectrum of observation 2636010101 of \cyg at $\Gamma\sim 1.8$. The black dashed line shows the total RMS averaged over the 0.06--15\,Hz range, which peaks at around 2\,keV. The low- (red) and high-frequency (blue dashed line) variability shows a bimodal
    behavior. The high-frequency variability from $L_2$ remains
    constant above ${\sim}2$\,keV, while the RMS of $L_1$ drops
    off. Below roughly 1.5\,keV, the variability drops at all
    frequencies. The remaining variability is dominated by $L_1$. The
    gray curve with the scale on the right axis measures the contribution from the
    accretion disk in the hard state. $L_1$ peaks at ${\sim}1.7$\,keV where
    the disk contributes ${\sim}40$\% of the flux.}
  \label{fig:2636010101_RMS}
\end{figure}

In order to better quantify the peak energy of the two main
variability components of the hard state, we use the RMS variability
amplitude spectrum. Figure~\ref{fig:2636010101_RMS} shows the
fractional RMS of lightcurves extracted in small energy bands,
integrated over three frequency ranges that sample the total PSD
(0.06--15\,Hz), the low-frequency range of $L_1$ (0.06--1.16\,Hz), and
the high-frequency range of $L_2$ (1.22--15\,Hz). Overall, most
variability comes from ${\geq}1.5$\,keV and the variability drops
significantly at low energies (Fig.~\ref{fig:energy_power_map}). We
can constrain the energy where $L_1$ peaks to be at $\sim$1.7\,keV
and derive a disk fraction of roughly 40\%. At higher energies, the
RMS of $L_1$ declines, while the high-frequency variability covering
$L_2$ stays constant above 2\,keV. This behavior is consistent with
\textsl{AstroSAT} data of \cyg, shown in \citet[][their
Fig.~4]{Misra17a}, and \hxmt data, shown in \citet[][their
Fig.~7]{Feng22a}.

\section{Influence of the Stellar Wind on the Time Lag and Coherence in the Hard State}
\label{app:sec:stellar_wind}

\begin{figure}
  \resizebox{\hsize}{!}{\includegraphics{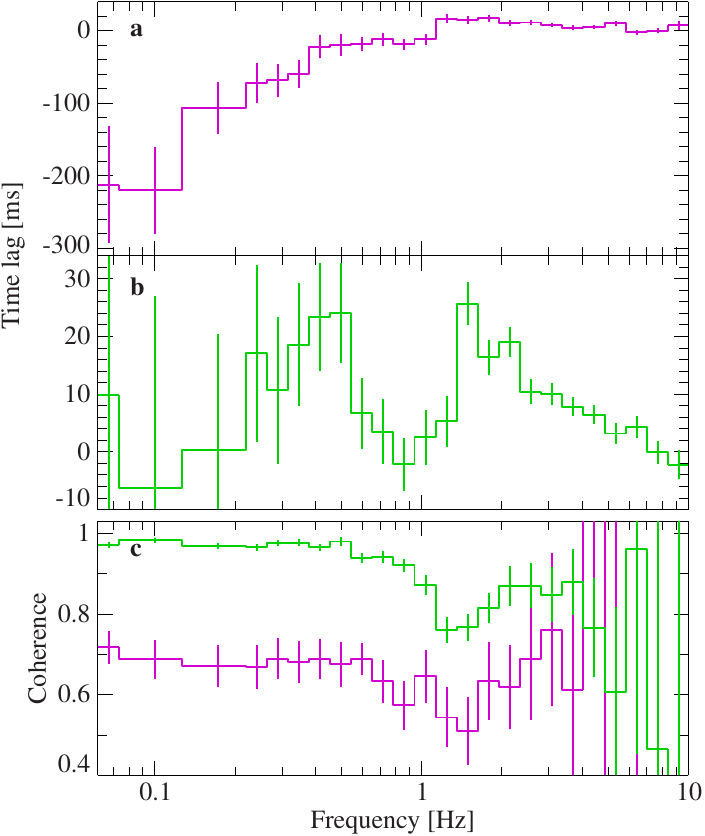}}
  \caption{Lag-frequency spectra and coherence of \cyg between 0.5--1\,keV and 2--4\,keV for observation 0100320106. Colors were chosen to match Fig.~6 of \citet{Lai22a}. \textbf{a)} Data during dipping events (magenta) show soft lags of up to ${\sim}200$\,ms. \textbf{b)} Data outside of dips (green) show a strong change in time lag around 1--2\,Hz. \textbf{c)} The coherence of the non-dipping data shows a dip at 1--2\,Hz, while data taken during dipping events show an overall reduced coherence.}
  \label{fig:wind_absorption}
\end{figure}

\citet{Lai22a} proposed that low-frequency soft lags occur during
dipping events when clumps of material pass the line-of-sight, and are
closely related to absorption variations. As a case study we analyze
observation 0100320106 during periods with and without dipping, which
were selected using cuts through the color-color diagram. We 
confirm that dipping events produce very strong soft lags up to
200\,ms at 0.1\,Hz (Fig.~\ref{fig:wind_absorption}). In fact, this is
a factor ${>}5$ longer than the ${\sim}40$\,ms found by
\citet{Lai22a}. We also confirm that the process producing these soft
lags is incoherent.

The timing feature is present in the non-dipping case, suggesting
that the feature is not related to the stellar wind. This result is
also supported by the fact that we see the timing feature in hard
state data taken at all orbital phases, in particular also at inferior
conjunction, when the influence of the stellar wind is weakest (for
reference, see Figs.~\ref{fig:app:1100320112}
and~\ref{fig:app:4690020105} taken at $\phi_\mathrm{orb}\sim 0.45$,
Fig.~\ref{fig:app:0100320107} at $\phi_\mathrm{orb}=0.23$, and
Figs.~\ref{fig:app:0100320104} and~\ref{fig:app:1100320113} at
$\phi_\mathrm{orb}=0.6$--0.8).

\section{Supplementary Data}
\label{appendix:summary_plots}

\begin{figure*}
\centering
\includegraphics[width=1\textwidth]{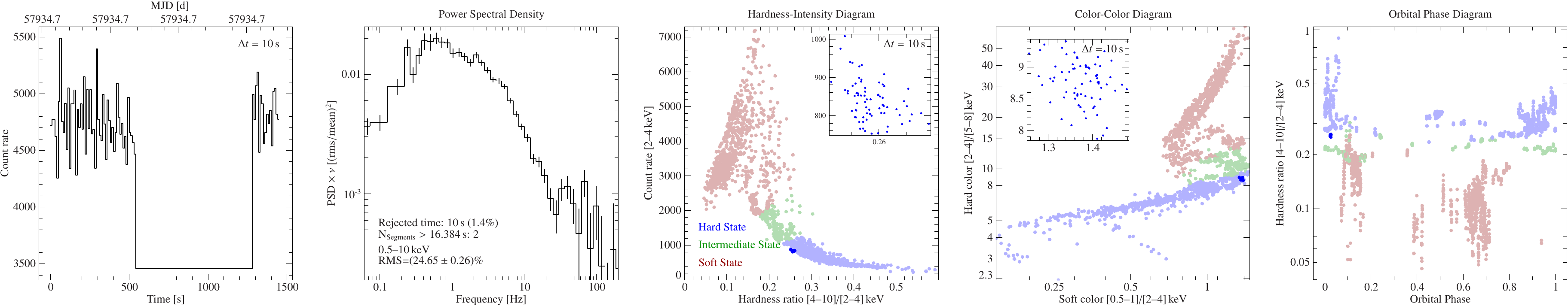}
\includegraphics[width=1\textwidth]{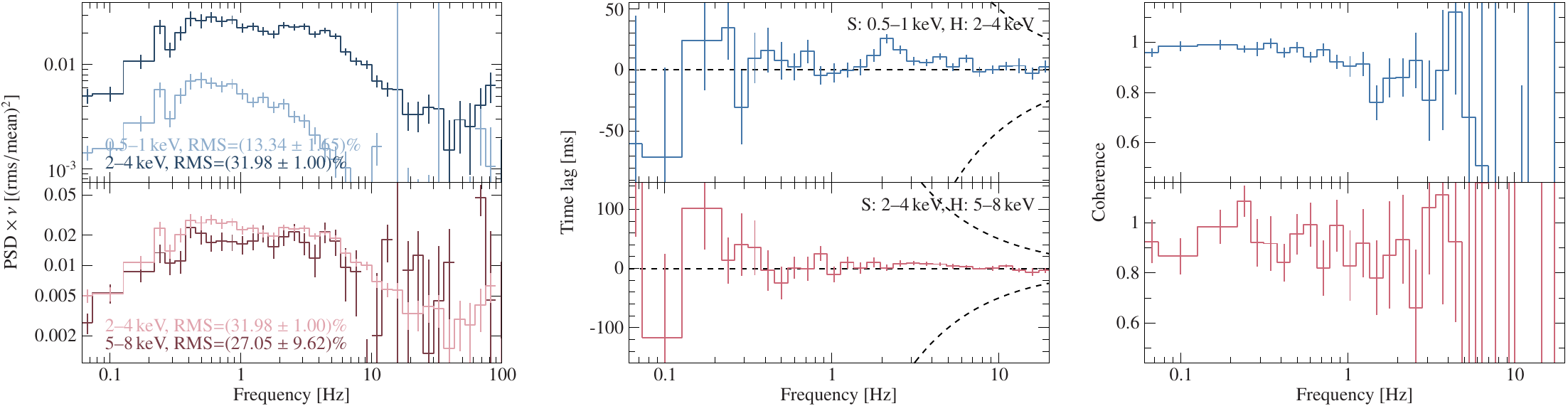}
\caption{\nicer observation 0100320101 of \cyg. $\Gamma\sim 1.8$.}
\label{fig:app:0100320101}
\end{figure*}

\begin{figure*}
\centering
\includegraphics[width=1\textwidth]{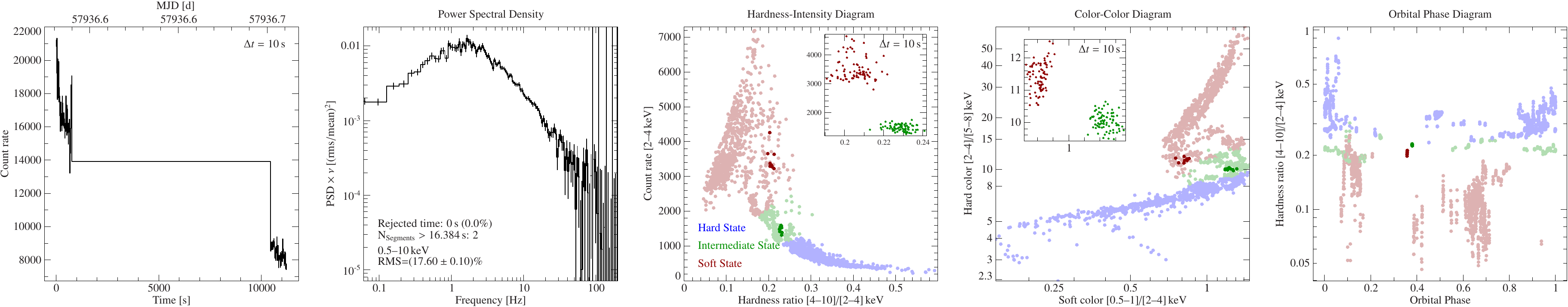}
\includegraphics[width=1\textwidth]{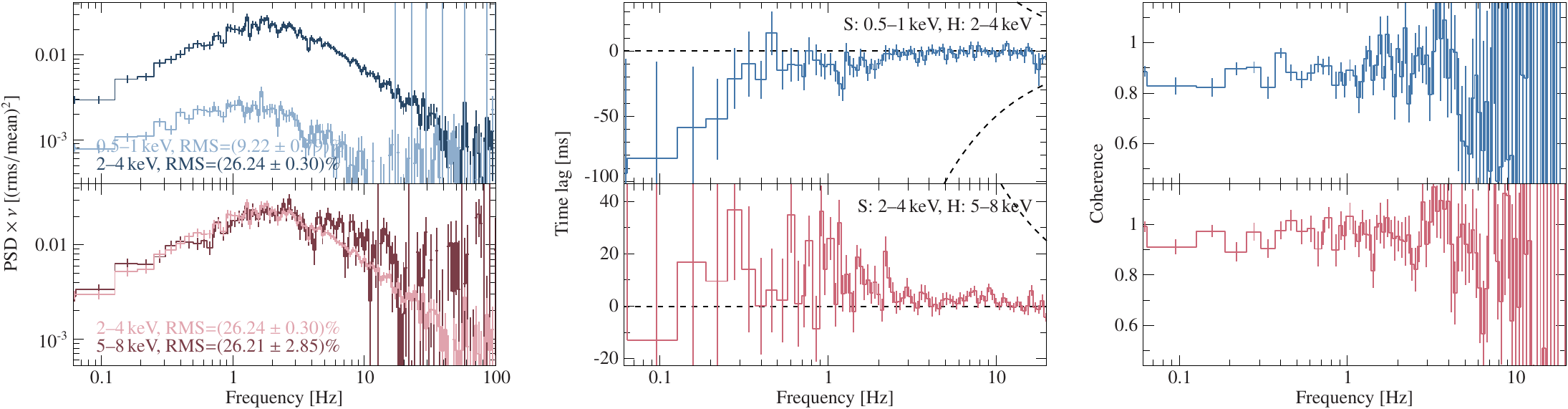}
\caption{\nicer observation 0100320102 of \cyg. $\Gamma\sim 2.2$.}
\label{fig:app:0100320102}
\end{figure*}

\begin{figure*}
\centering
\includegraphics[width=1\textwidth]{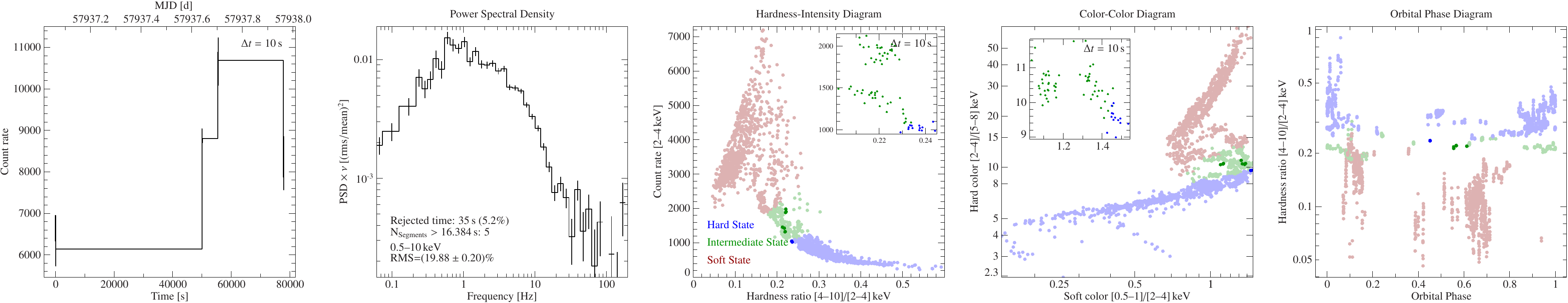}
\includegraphics[width=1\textwidth]{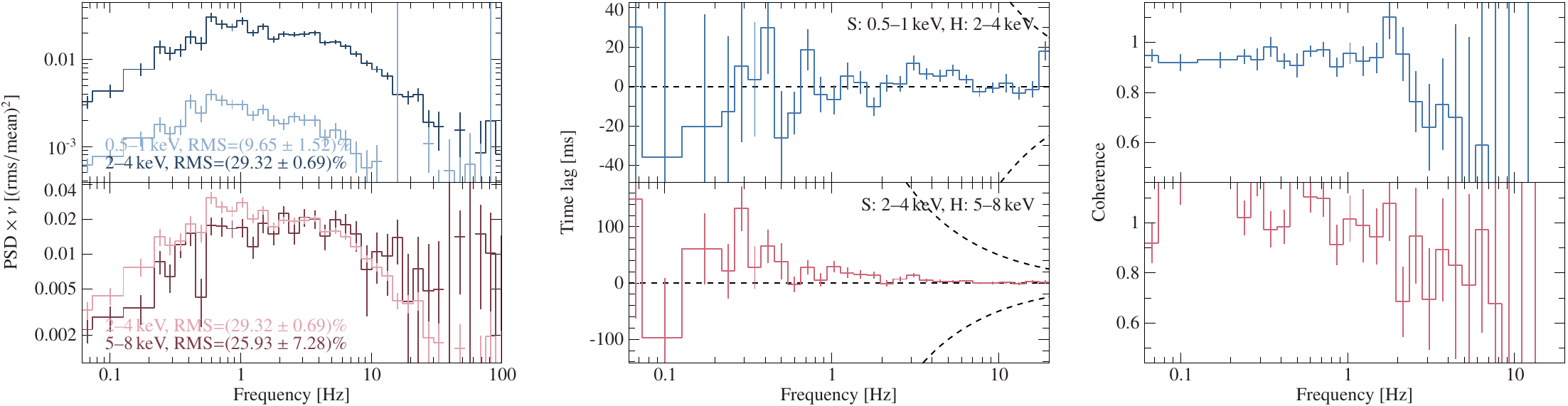}
\caption{\nicer observation 0100320103 of \cyg. $\Gamma\sim 2.1$.}
\label{fig:app:0100320103}
\end{figure*}

\begin{figure*}
\centering
\includegraphics[width=1\textwidth]{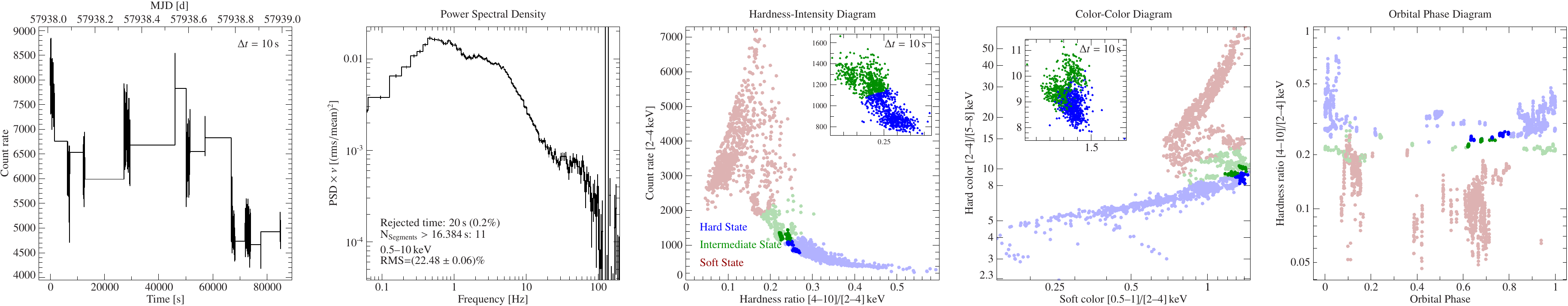}
\includegraphics[width=1\textwidth]{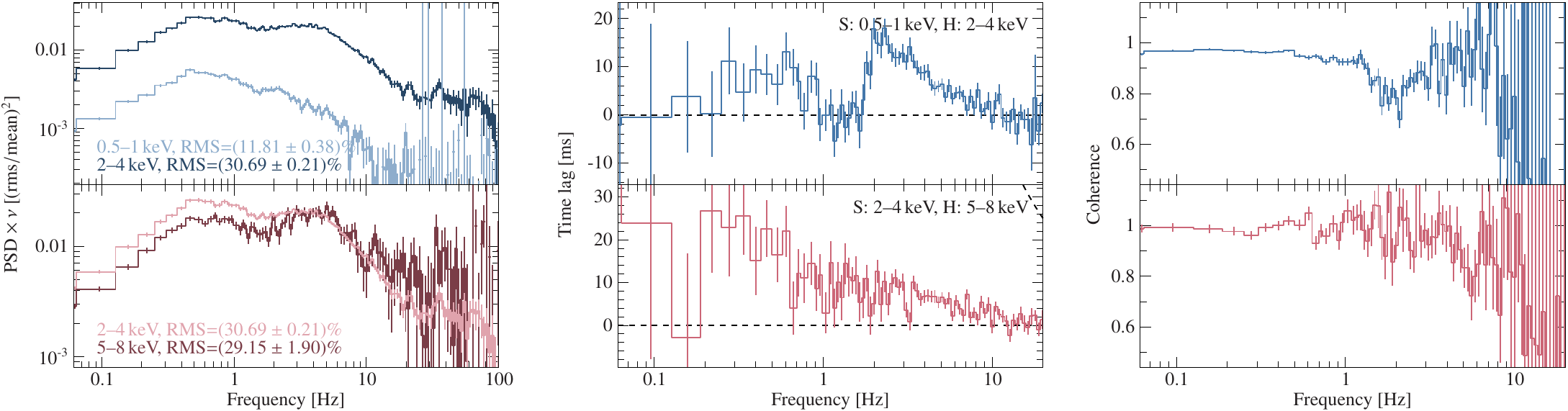}
\caption{\nicer observation 0100320104 of \cyg. $\Gamma\sim 1.9$.}
\label{fig:app:0100320104}
\end{figure*}

\begin{figure*}
\centering
\includegraphics[width=1\textwidth]{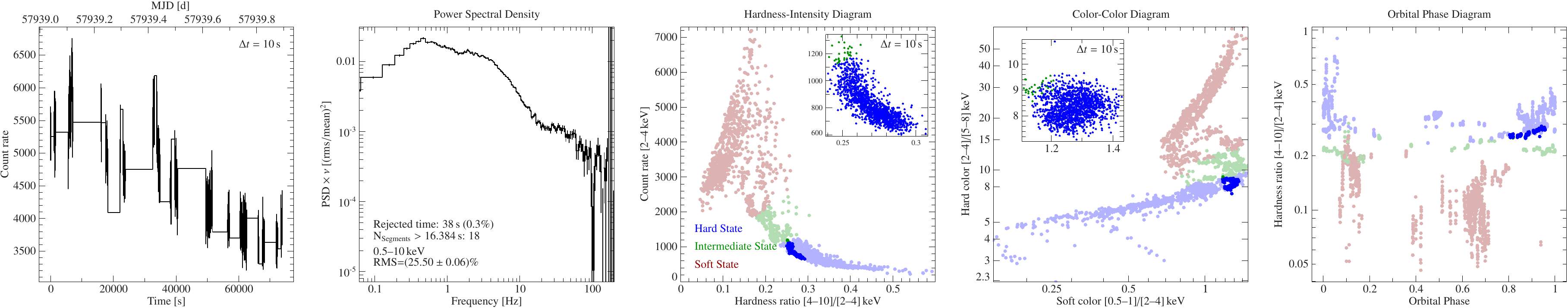}
\includegraphics[width=1\textwidth]{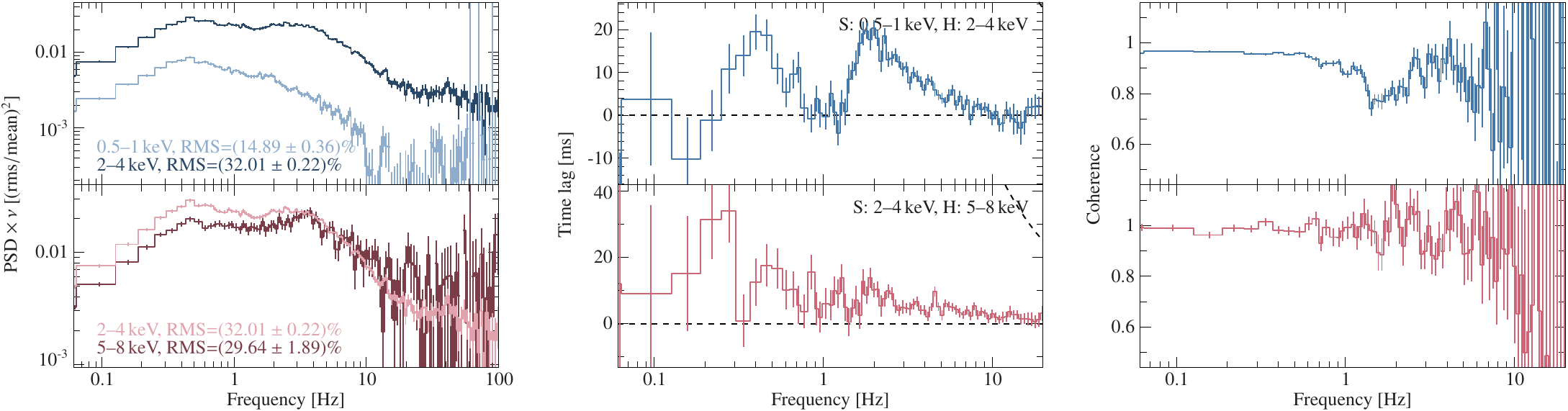}
\caption{\nicer observation 0100320105 of \cyg. $\Gamma\sim 1.9$.}
\label{fig:app:0100320105}
\end{figure*}

\begin{figure*}
\centering
\includegraphics[width=1\textwidth]{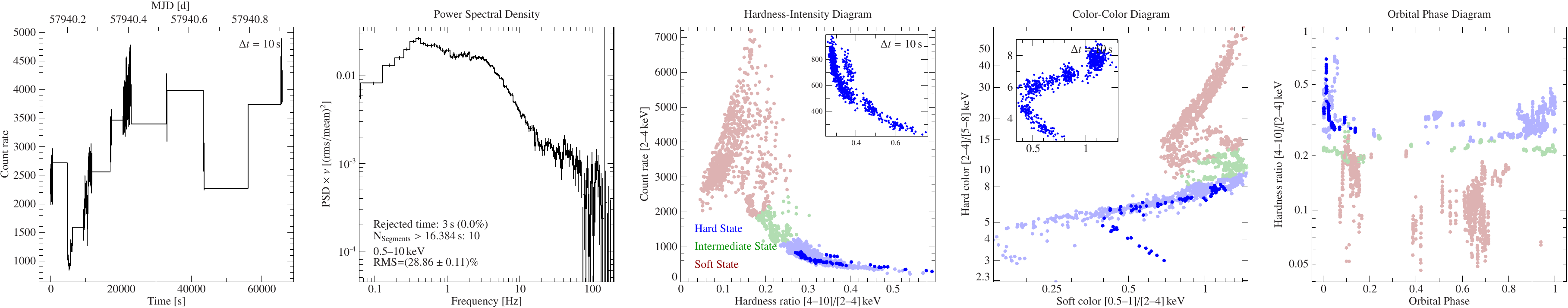}
\includegraphics[width=1\textwidth]{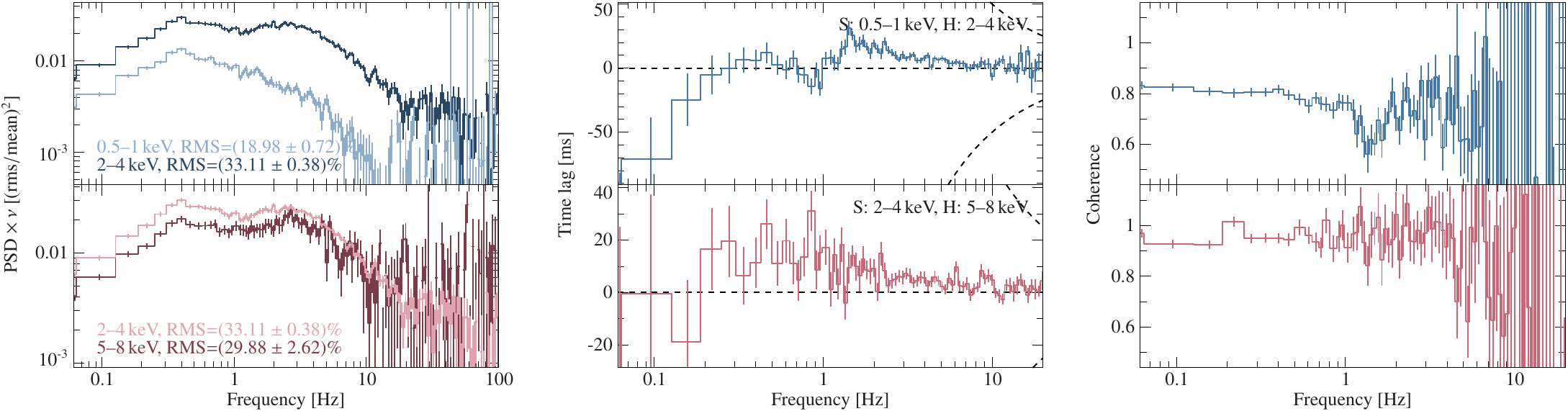}
\caption{\nicer observation 0100320106 of \cyg. $\Gamma\sim 1.8$.}
\label{fig:app:0100320106}
\end{figure*}

\begin{figure*}
\centering
\includegraphics[width=1\textwidth]{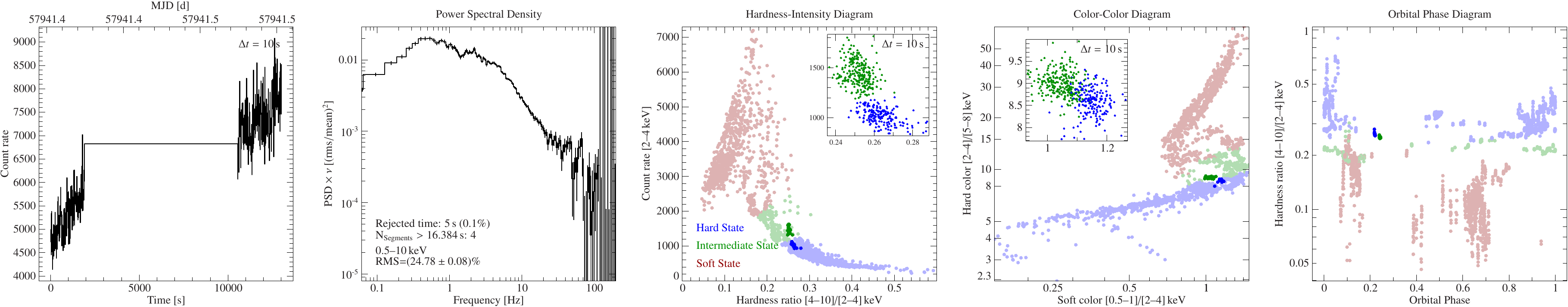}
\includegraphics[width=1\textwidth]{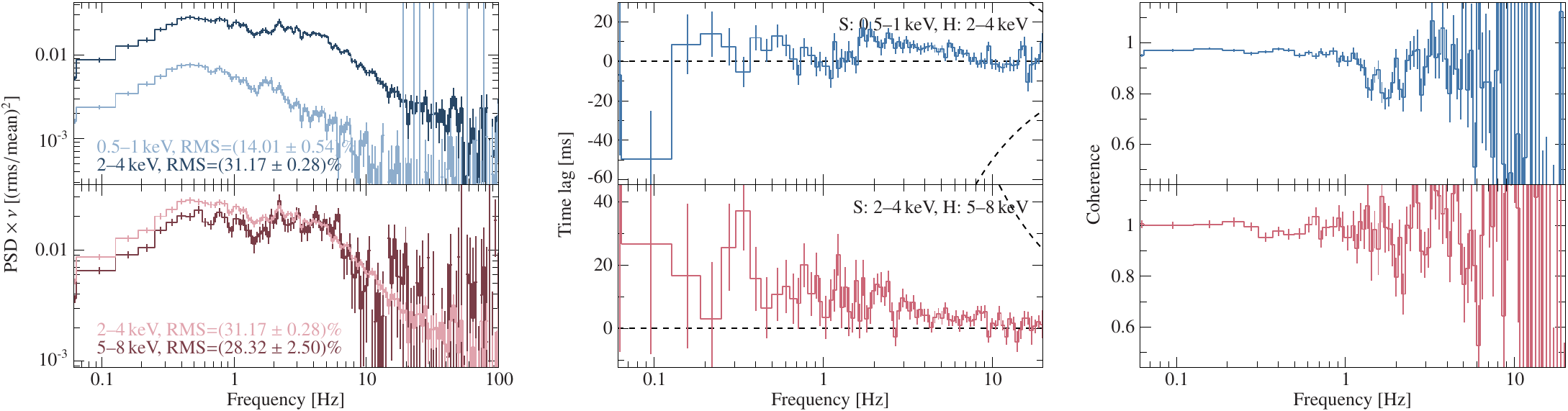}
\caption{\nicer observation 0100320107 of \cyg. $\Gamma\sim 2.0$.}
\label{fig:app:0100320107}
\end{figure*}

\begin{figure*}
\centering
\includegraphics[width=1\textwidth]{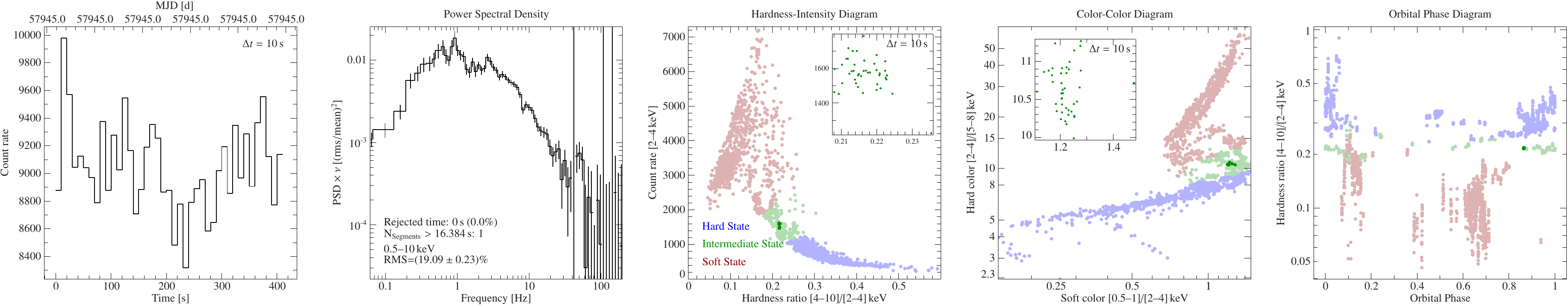}
\includegraphics[width=1\textwidth]{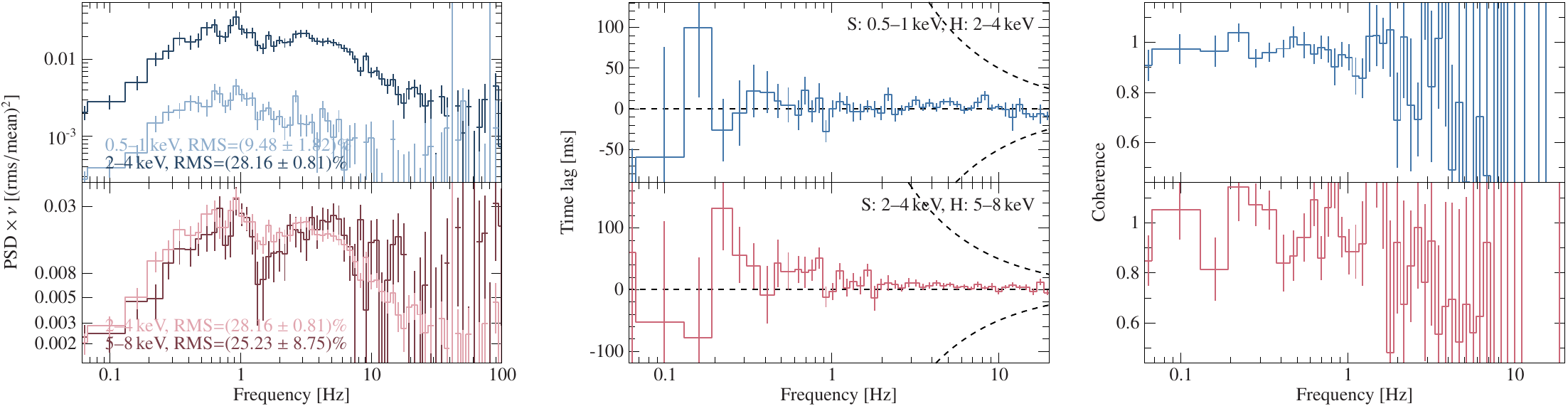}
\caption{\nicer observation 0100320108 of \cyg. $\Gamma\sim 2.1$.}
\label{fig:app:0100320108}
\end{figure*}

\begin{figure*}
\centering
\includegraphics[width=1\textwidth]{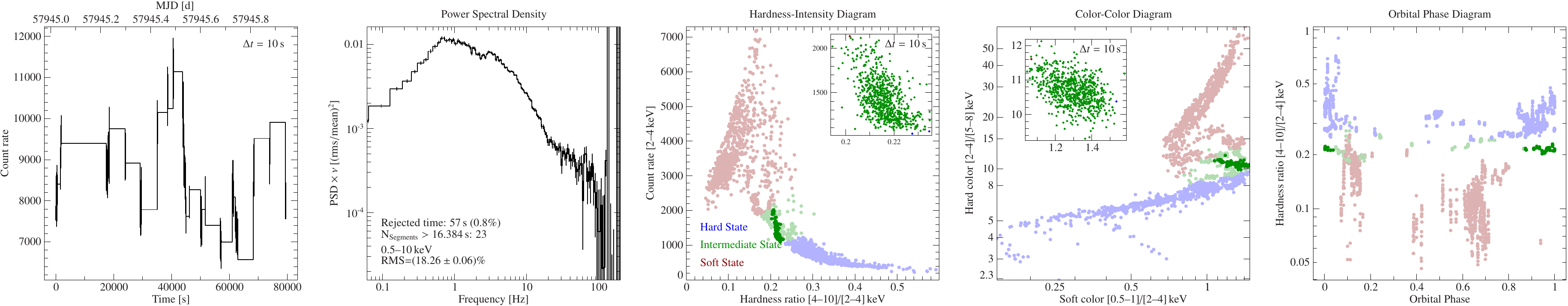}
\includegraphics[width=1\textwidth]{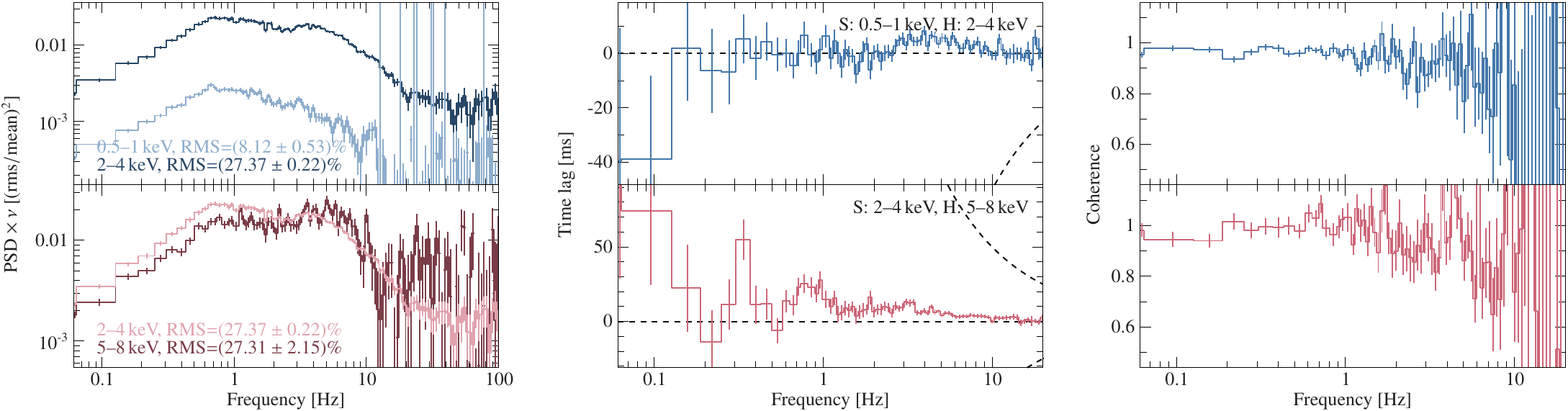}
\caption{\nicer observation 0100320109 of \cyg. $\Gamma\sim 2.1$.}
\label{fig:app:0100320109}
\end{figure*}

\begin{figure*}
\centering
\includegraphics[width=1\textwidth]{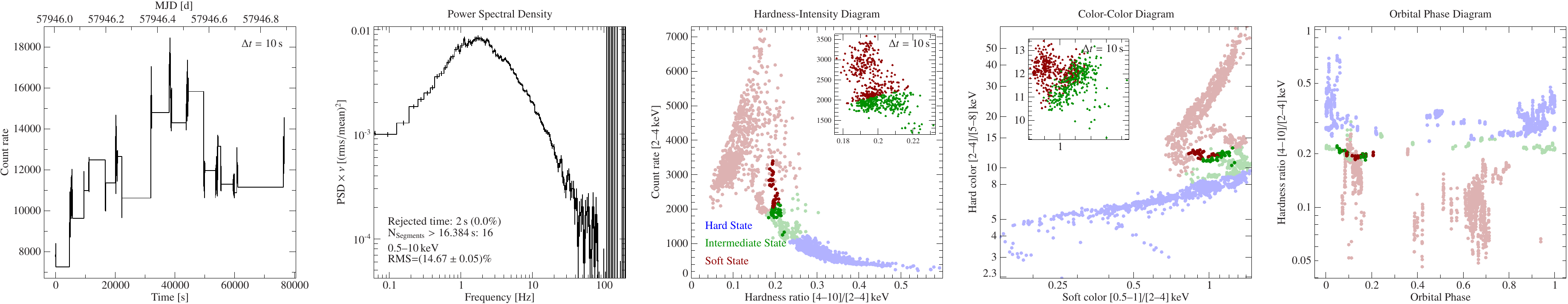}
\includegraphics[width=1\textwidth]{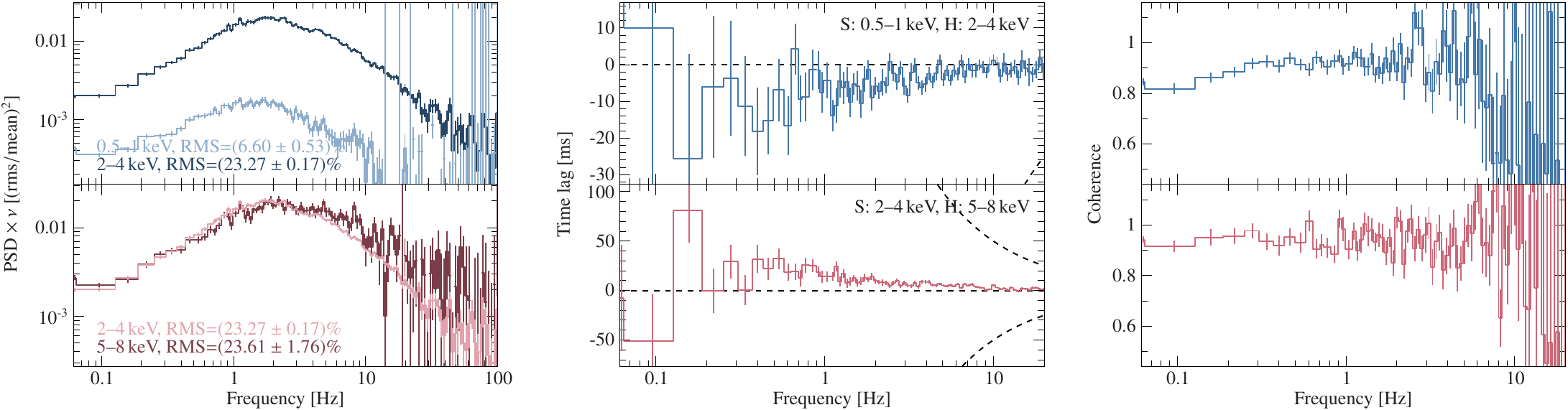}
\caption{\nicer observation 0100320110 of \cyg. $\Gamma\sim 2.3$.}
\label{fig:app:0100320110}
\end{figure*}

\begin{figure*}
\centering
\includegraphics[width=1\textwidth]{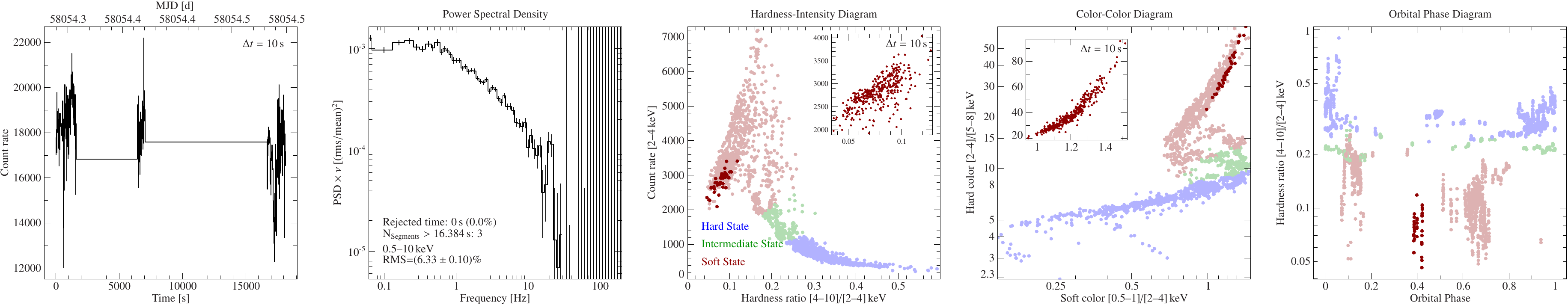}
\includegraphics[width=1\textwidth]{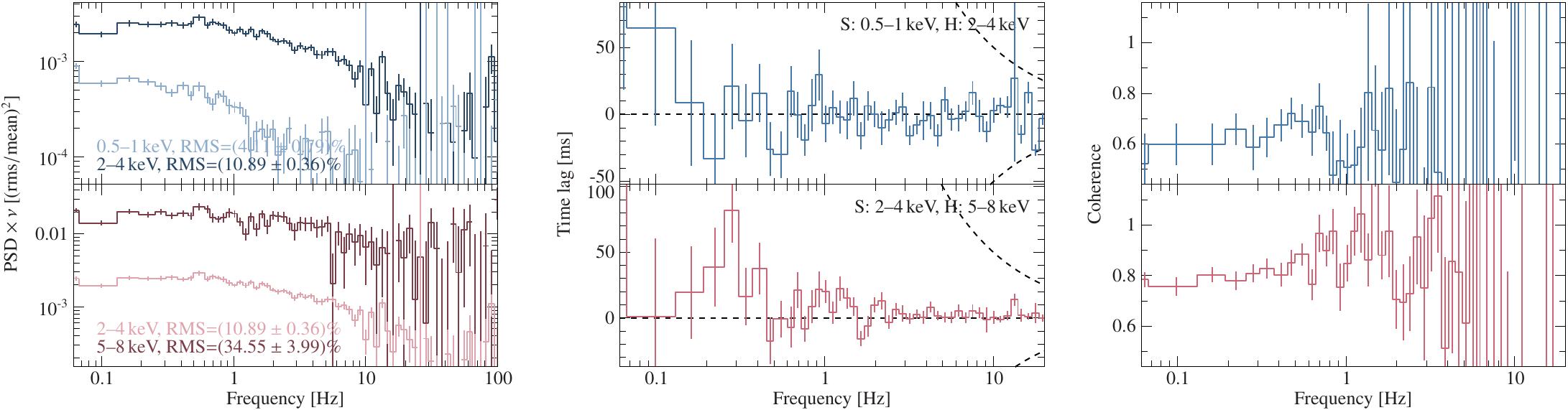}
\caption{\nicer observation 1100320101 of \cyg. $\Gamma\sim 3.3$.}
\label{fig:app:1100320101}
\end{figure*}

\begin{figure*}
\centering
\includegraphics[width=1\textwidth]{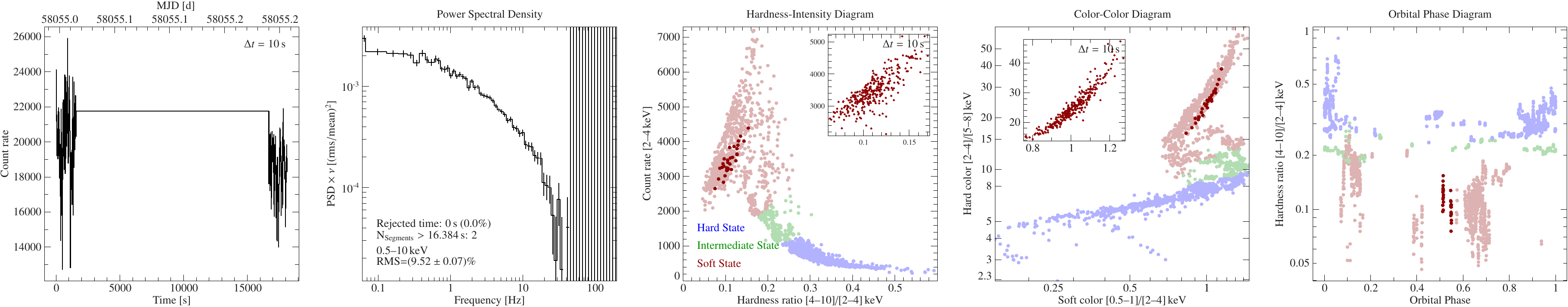}
\includegraphics[width=1\textwidth]{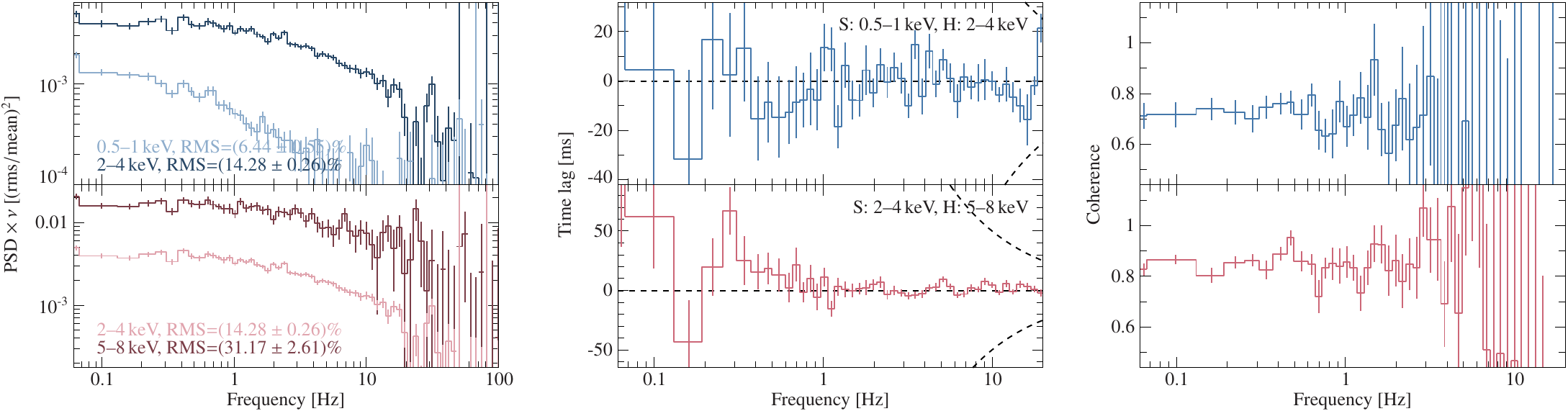}
\caption{\nicer observation 1100320102 of \cyg. $\Gamma\sim 3.0$.}
\label{fig:app:1100320102}
\end{figure*}

\begin{figure*}
\centering
\includegraphics[width=1\textwidth]{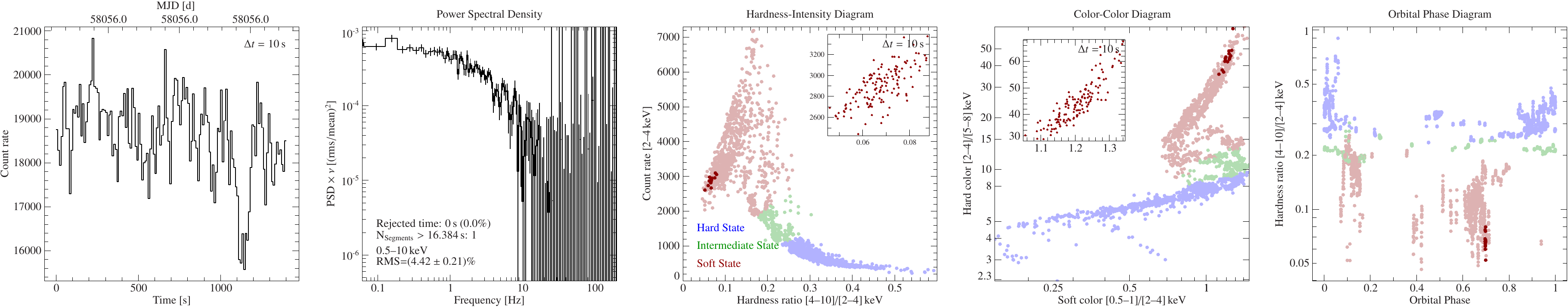}
\includegraphics[width=1\textwidth]{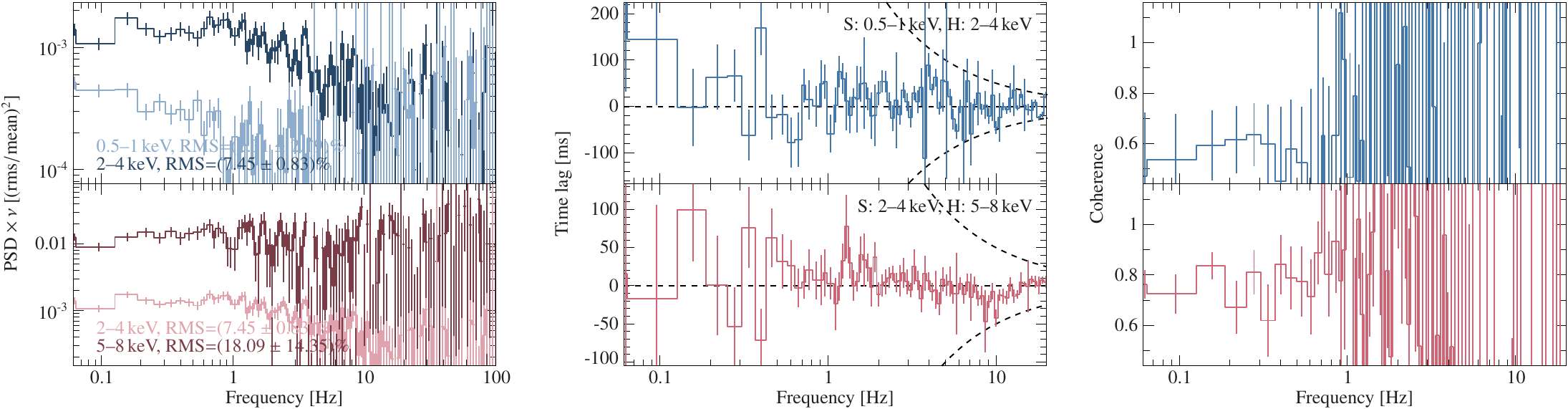}
\caption{\nicer observation 1100320103 of \cyg. $\Gamma\sim 3.4$.}
\label{fig:app:1100320103}
\end{figure*}

\begin{figure*}
\centering
\includegraphics[width=1\textwidth]{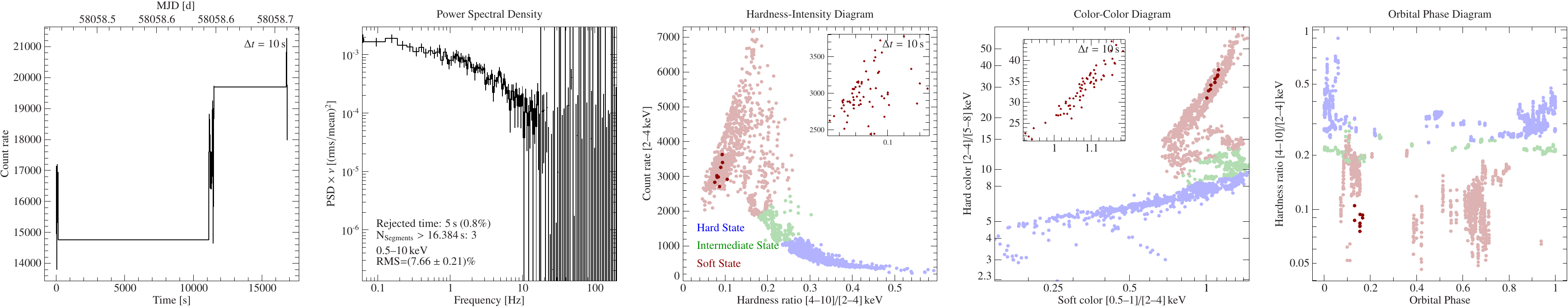}
\includegraphics[width=1\textwidth]{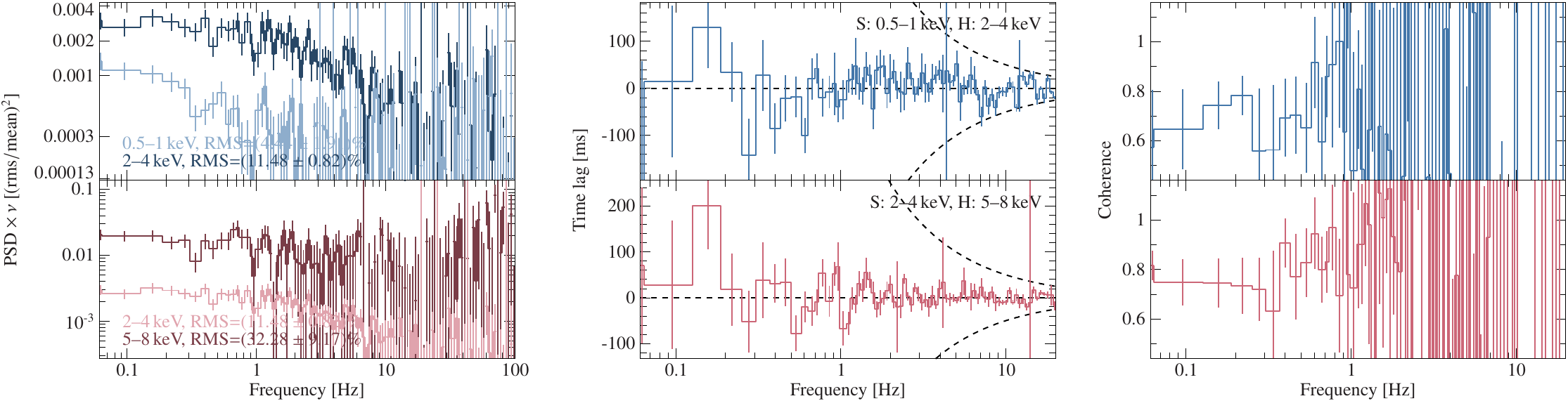}
\caption{\nicer observation 1100320104 of \cyg. $\Gamma\sim 3.0$.}
\label{fig:app:1100320104}
\end{figure*}

\begin{figure*}
\centering
\includegraphics[width=1\textwidth]{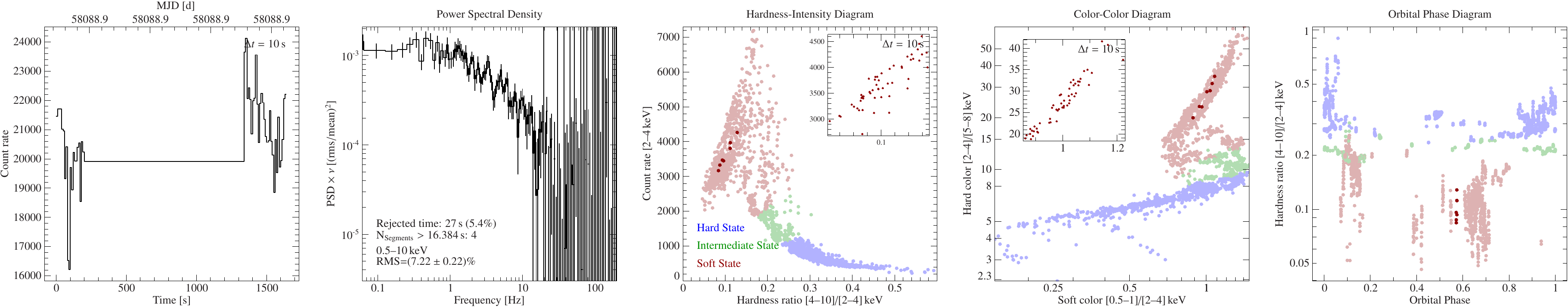}
\includegraphics[width=1\textwidth]{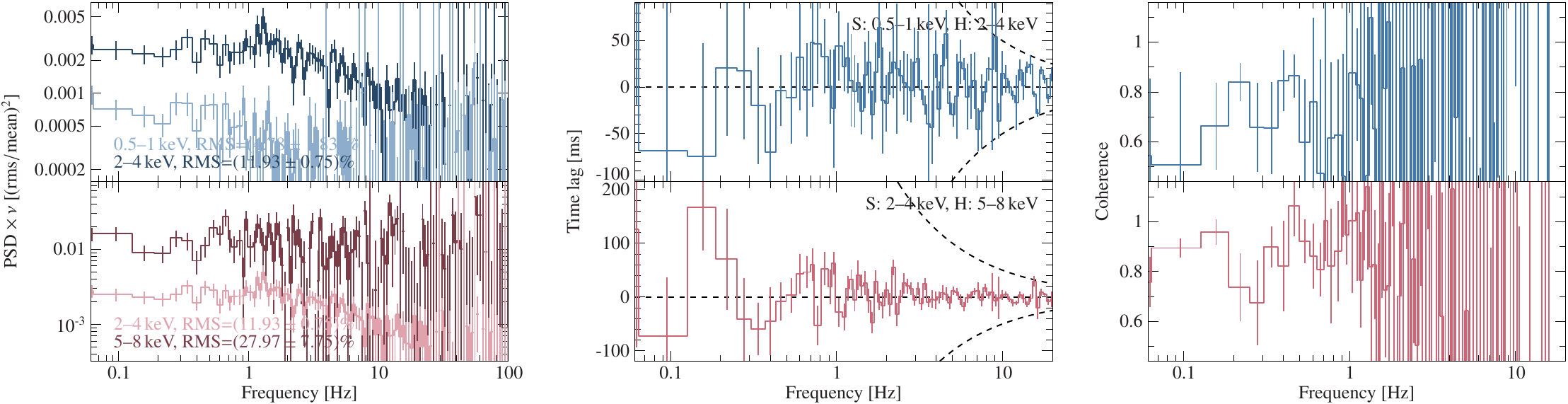}
\caption{\nicer observation 1100320106 of \cyg. $\Gamma\sim 3.0$.}
\label{fig:app:1100320106}
\end{figure*}

\begin{figure*}
\centering
\includegraphics[width=1\textwidth]{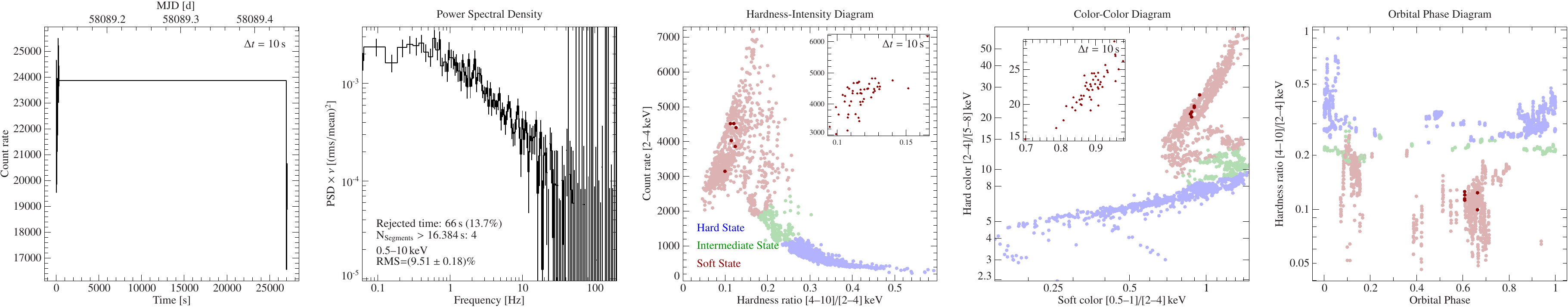}
\includegraphics[width=1\textwidth]{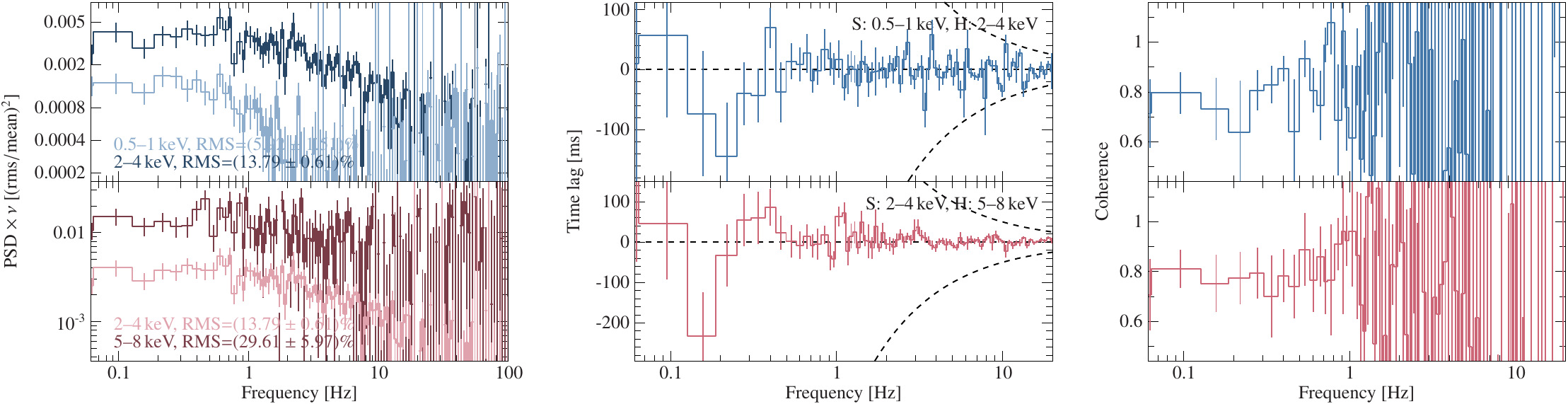}
\caption{\nicer observation 1100320107 of \cyg. $\Gamma\sim 2.9$.}
\label{fig:app:1100320107}
\end{figure*}

\begin{figure*}
\centering
\includegraphics[width=1\textwidth]{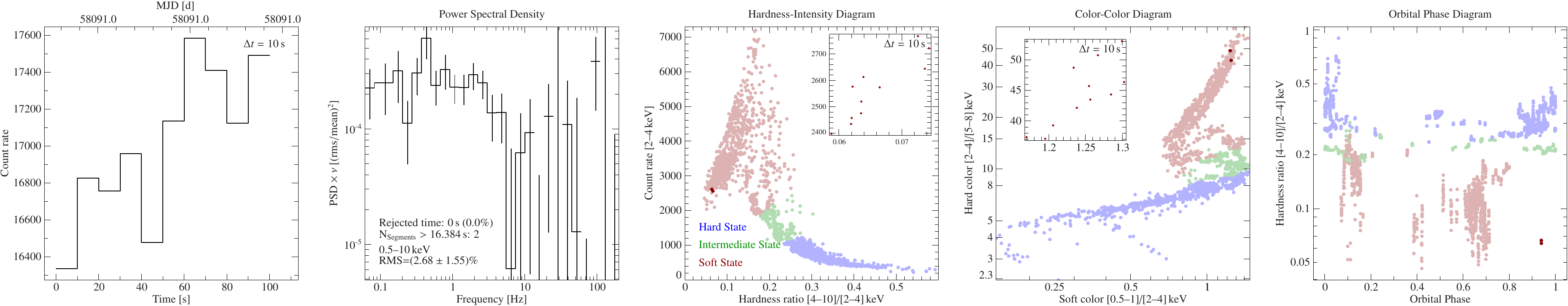}
\includegraphics[width=1\textwidth]{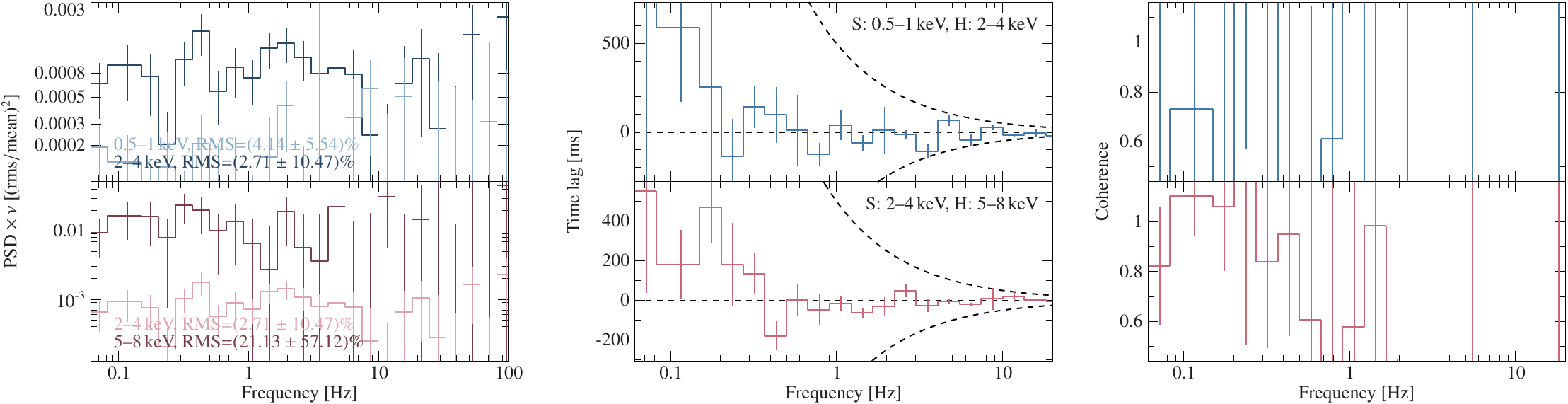}
\caption{\nicer observation 1100320108 of \cyg. $\Gamma\sim 3.5$.}
\label{fig:app:1100320108}
\end{figure*}

\begin{figure*}
\centering
\includegraphics[width=1\textwidth]{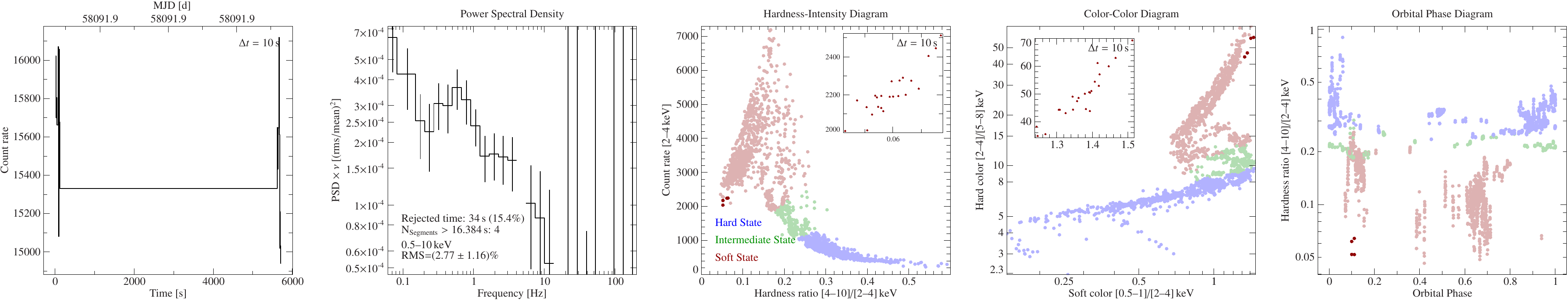}
\includegraphics[width=1\textwidth]{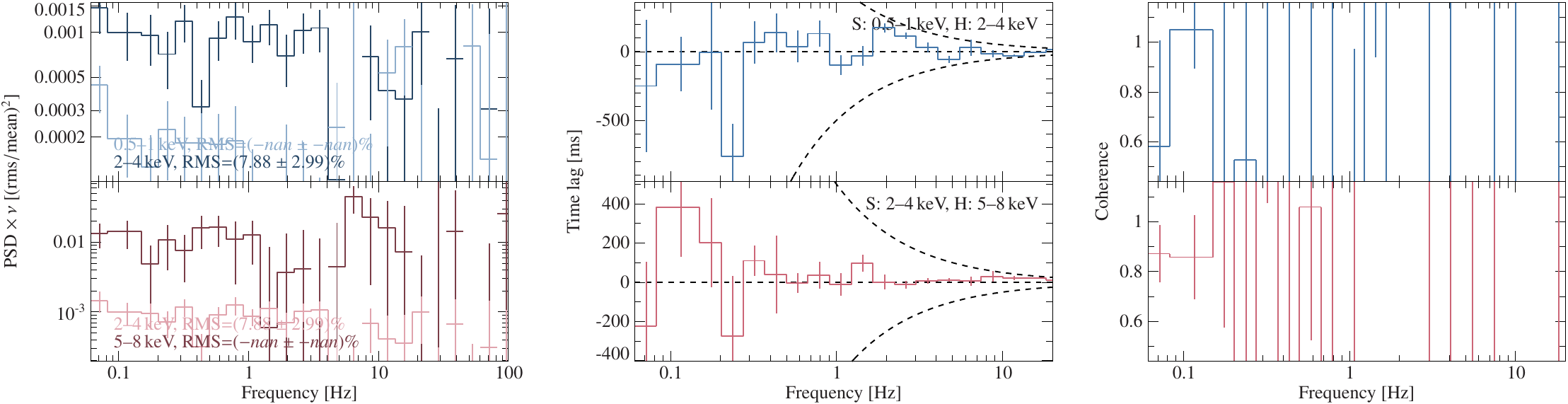}
\caption{\nicer observation 1100320109 of \cyg. $\Gamma\sim 3.3$.}
\label{fig:app:1100320109}
\end{figure*}

\begin{figure*}
\centering
\includegraphics[width=1\textwidth]{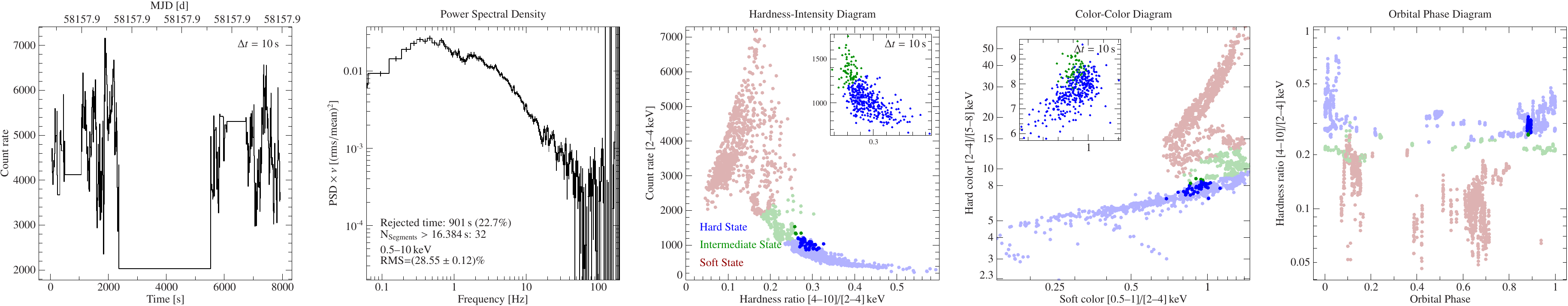}
\includegraphics[width=1\textwidth]{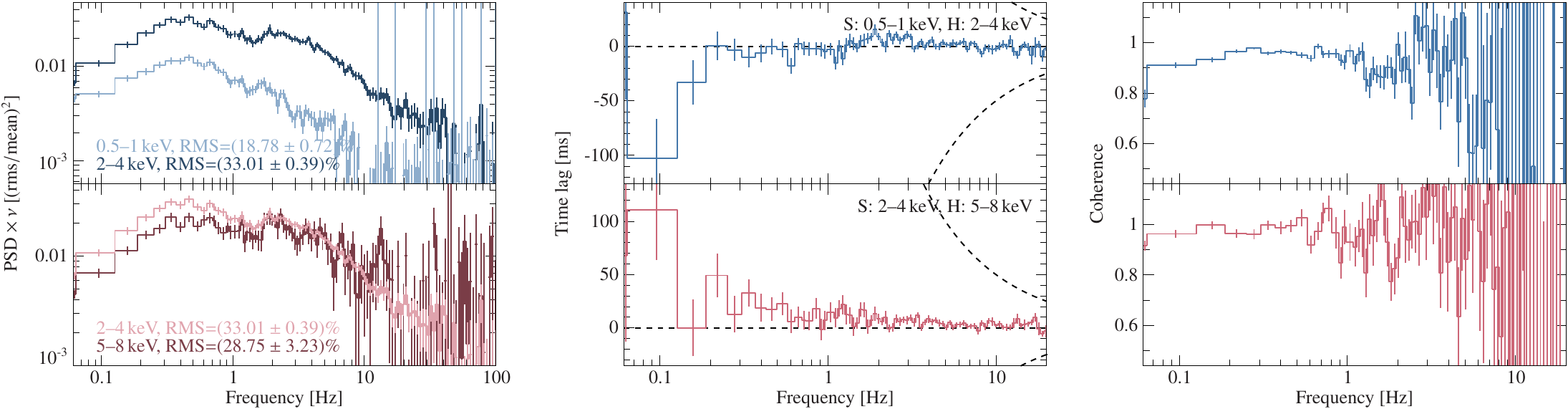}
\caption{\nicer observation 1100320110 of \cyg. $\Gamma\sim 1.6$.}
\label{fig:app:1100320110}
\end{figure*}

\begin{figure*}
\centering
\includegraphics[width=1\textwidth]{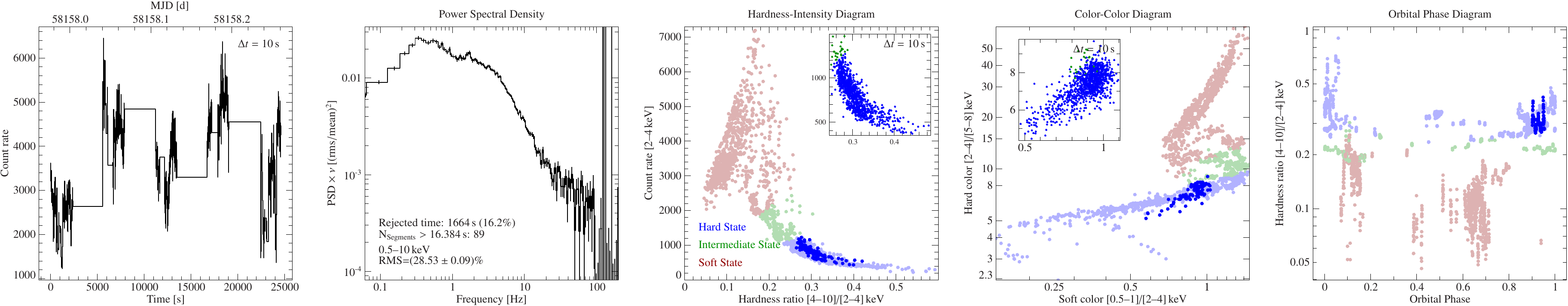}
\includegraphics[width=1\textwidth]{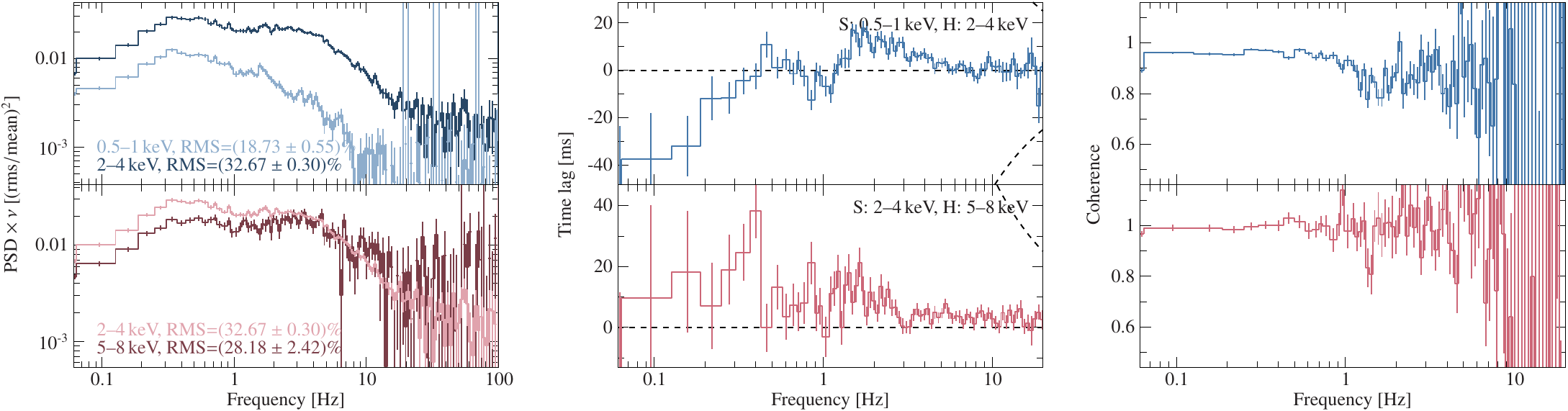}
\caption{\nicer observation 1100320111 of \cyg. $\Gamma\sim 1.8$.}
\label{fig:app:1100320111}
\end{figure*}

\begin{figure*}
\centering
\includegraphics[width=1\textwidth]{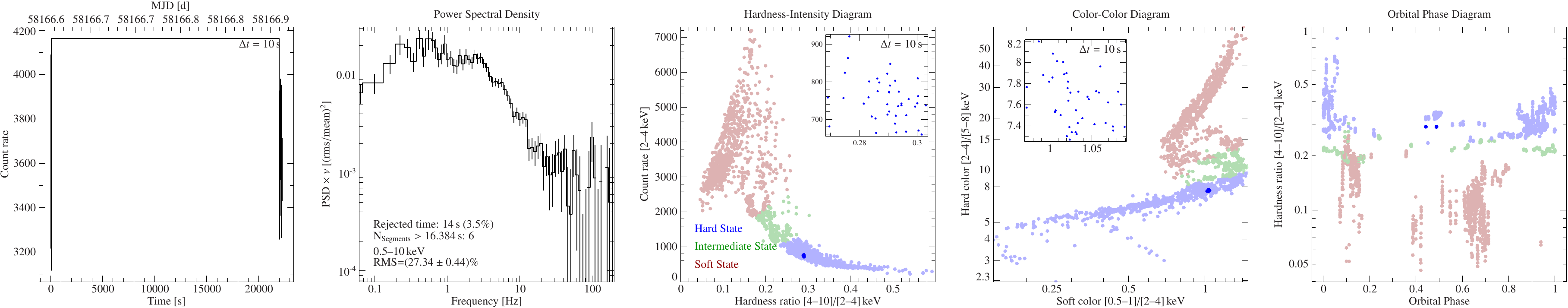}
\includegraphics[width=1\textwidth]{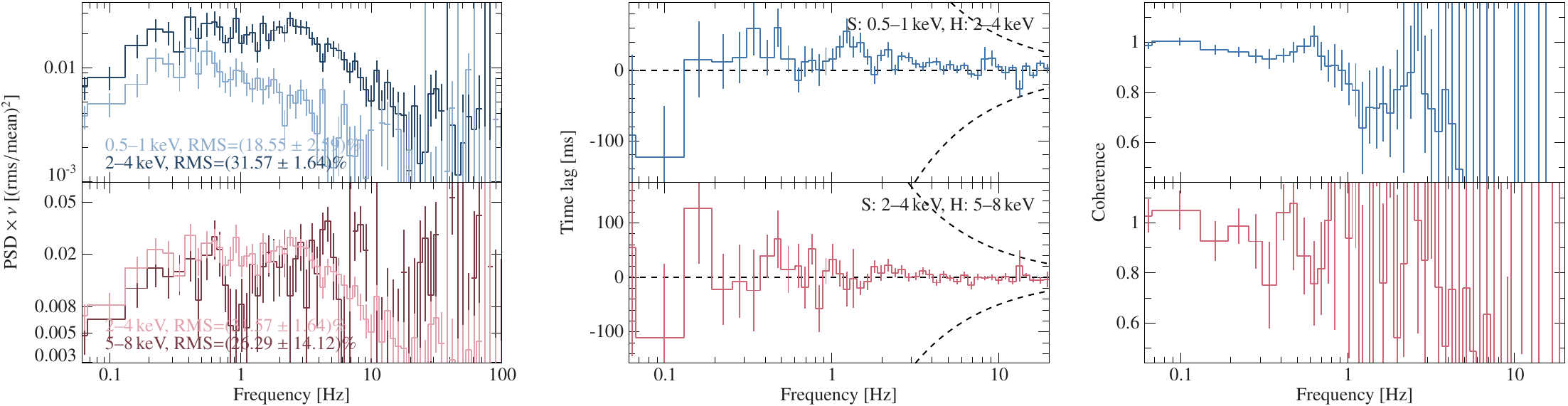}
\caption{\nicer observation 1100320112 of \cyg. $\Gamma\sim 1.8$.}
\label{fig:app:1100320112}
\end{figure*}

\begin{figure*}
\centering
\includegraphics[width=1\textwidth]{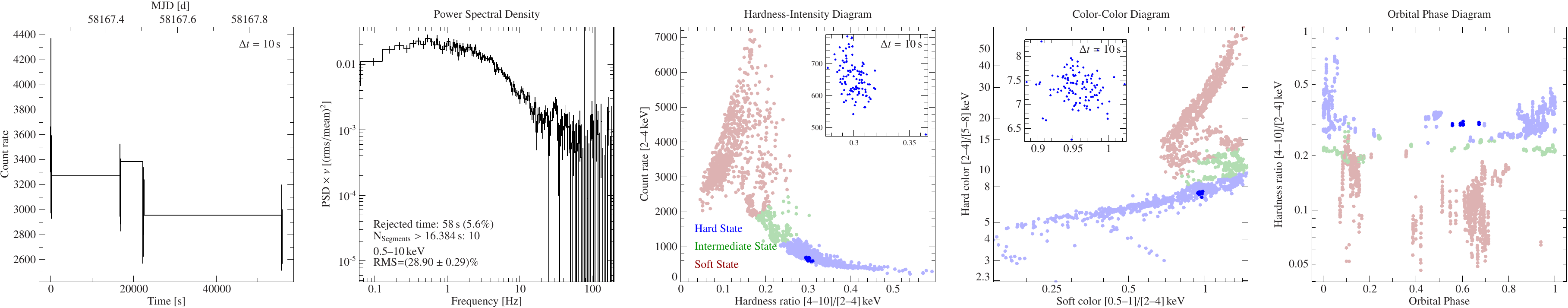}
\includegraphics[width=1\textwidth]{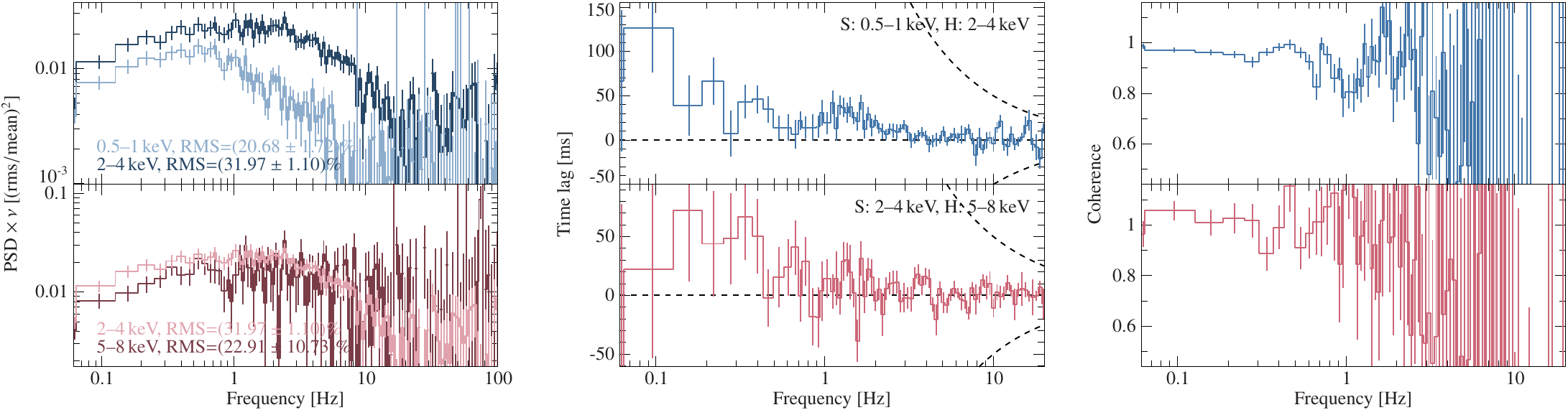}
\caption{\nicer observation 1100320113 of \cyg. $\Gamma\sim 1.7$.}
\label{fig:app:1100320113}
\end{figure*}

\begin{figure*}
\centering
\includegraphics[width=1\textwidth]{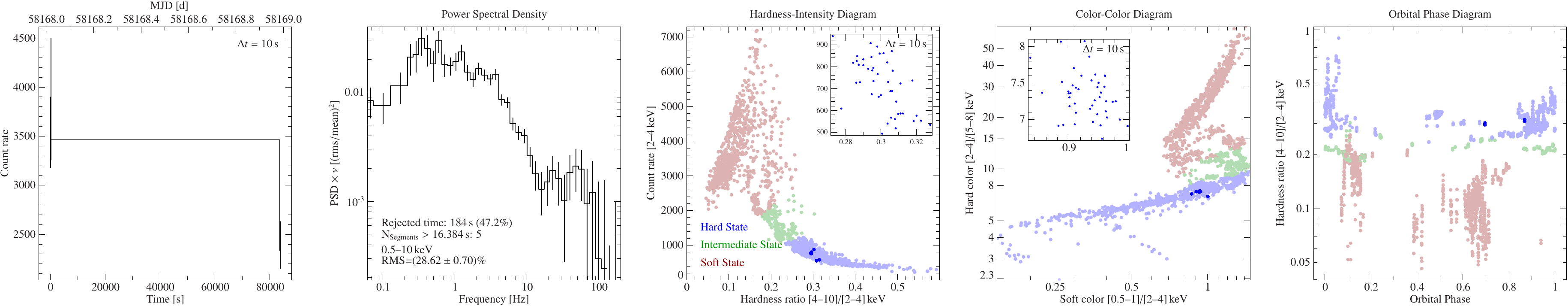}
\includegraphics[width=1\textwidth]{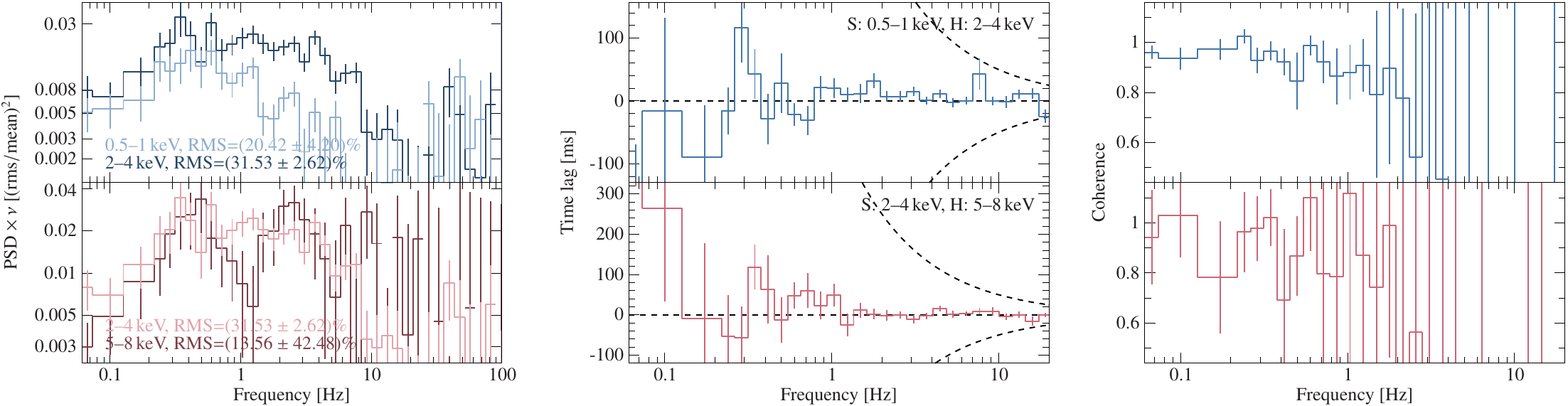}
\caption{\nicer observation 1100320114 of \cyg. $\Gamma\sim 1.8$.}
\label{fig:app:1100320114}
\end{figure*}

\begin{figure*}
\centering
\includegraphics[width=1\textwidth]{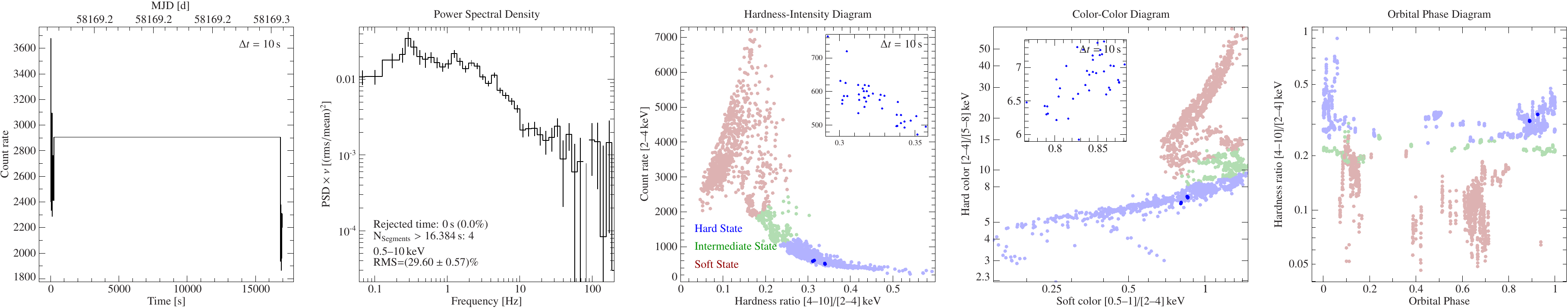}
\includegraphics[width=1\textwidth]{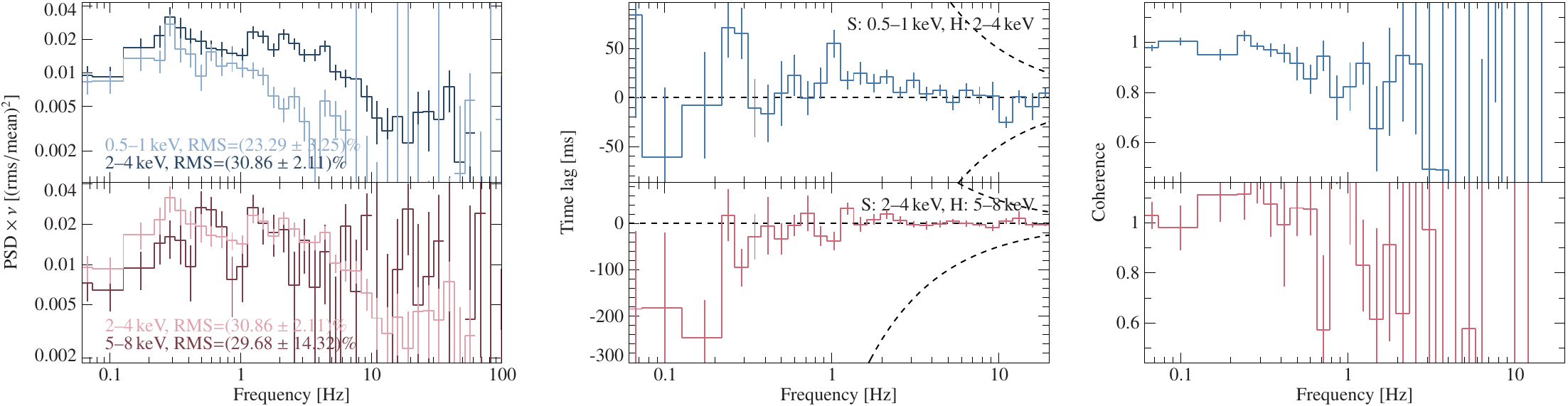}
\caption{\nicer observation 1100320115 of \cyg. $\Gamma\sim 1.7$.}
\label{fig:app:1100320115}
\end{figure*}

\begin{figure*}
\centering
\includegraphics[width=1\textwidth]{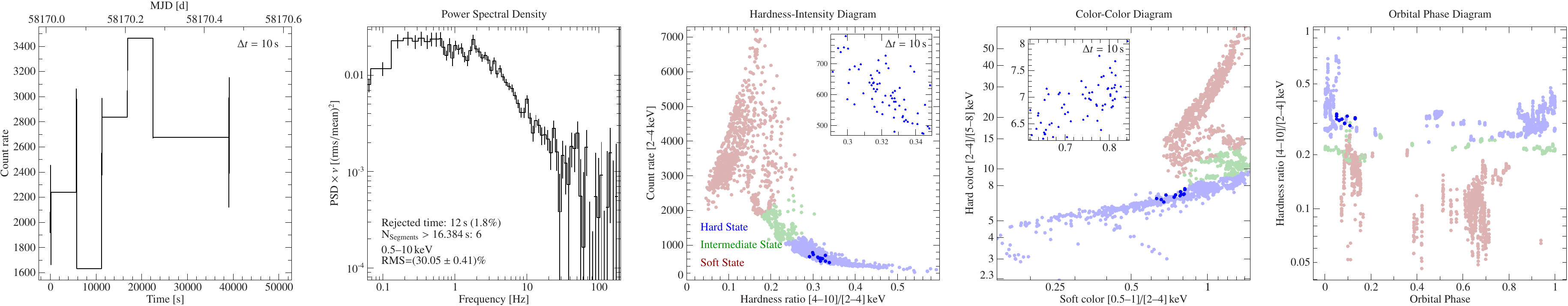}
\includegraphics[width=1\textwidth]{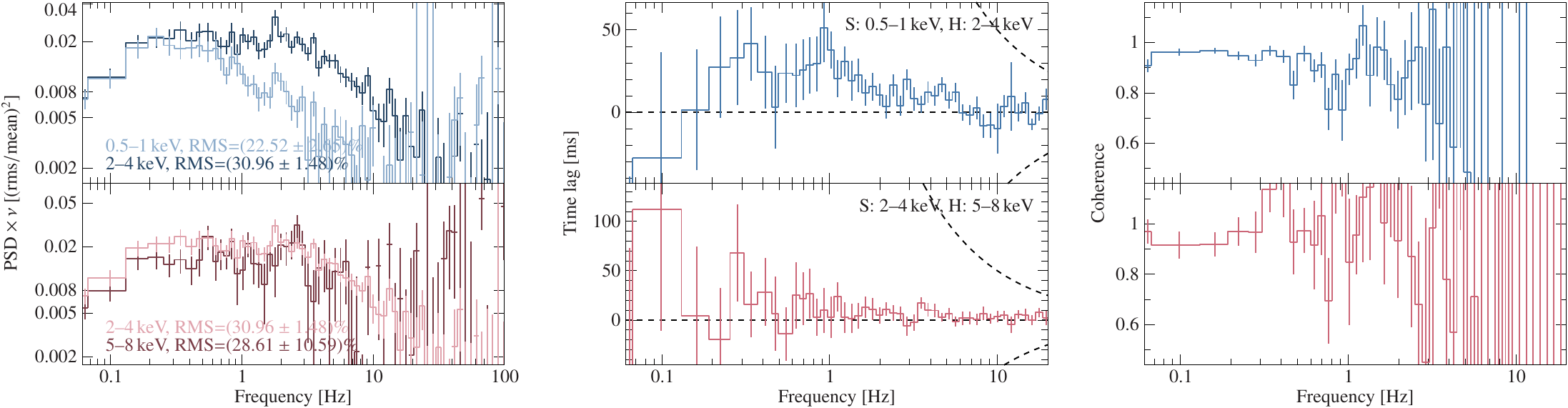}
\caption{\nicer observation 1100320116 of \cyg. $\Gamma\sim 1.7$.}
\label{fig:app:1100320116}
\end{figure*}

\begin{figure*}
\centering
\includegraphics[width=1\textwidth]{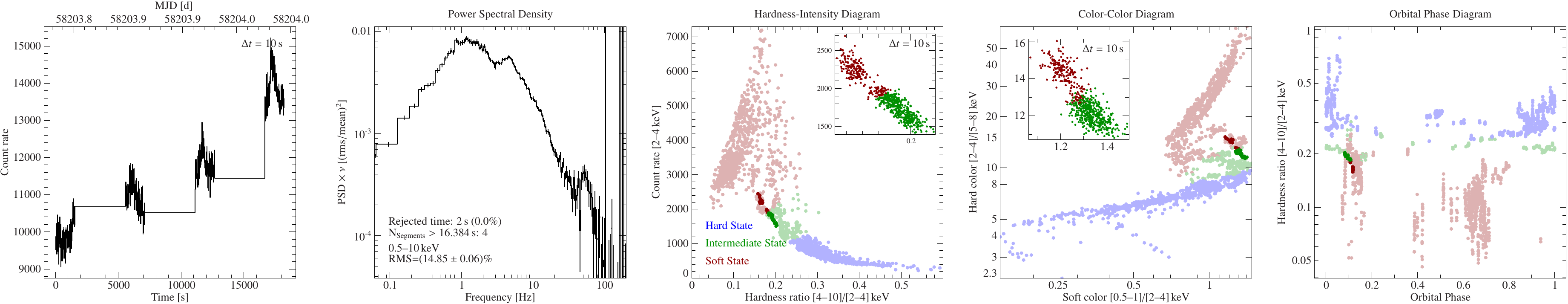}
\includegraphics[width=1\textwidth]{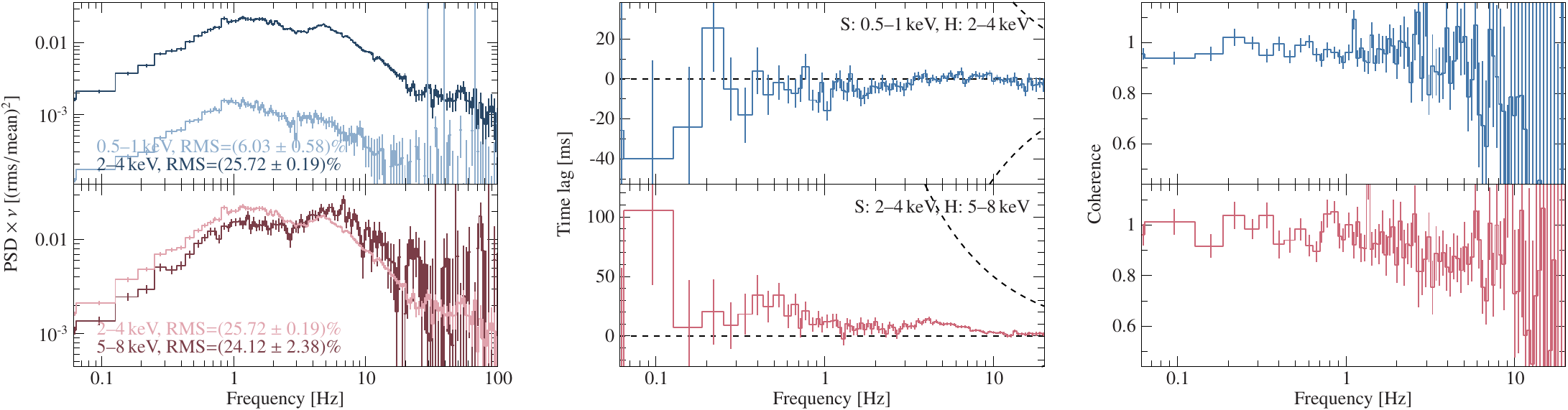}
\caption{\nicer observation 1100320117 of \cyg. $\Gamma\sim 2.2$.}
\label{fig:app:1100320117}
\end{figure*}

\begin{figure*}
\centering
\includegraphics[width=1\textwidth]{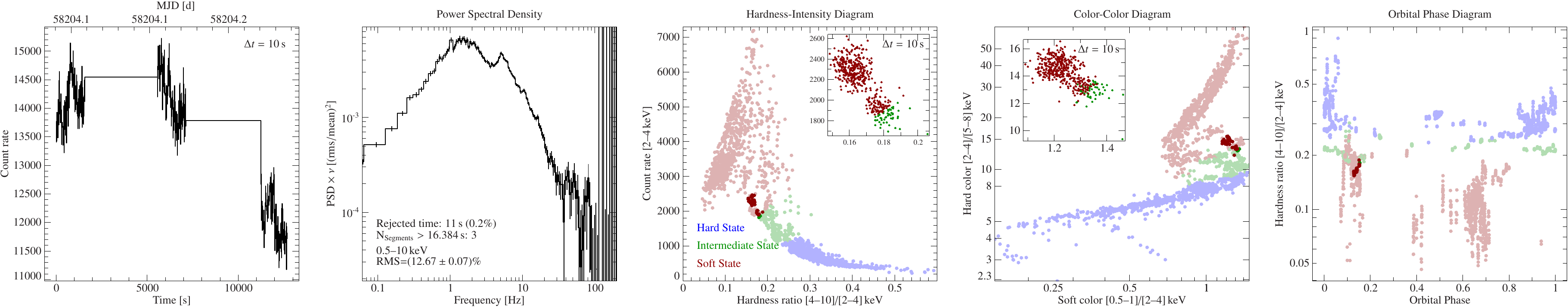}
\includegraphics[width=1\textwidth]{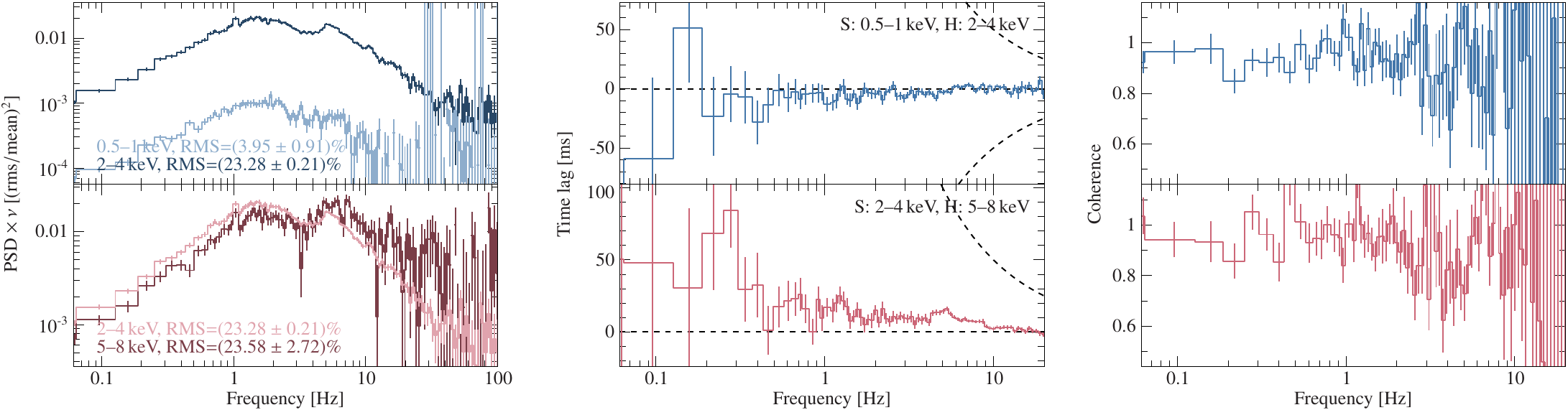}
\caption{\nicer observation 1100320118 of \cyg. $\Gamma\sim 2.3$.}
\label{fig:app:1100320118}
\end{figure*}

\begin{figure*}
\centering
\includegraphics[width=1\textwidth]{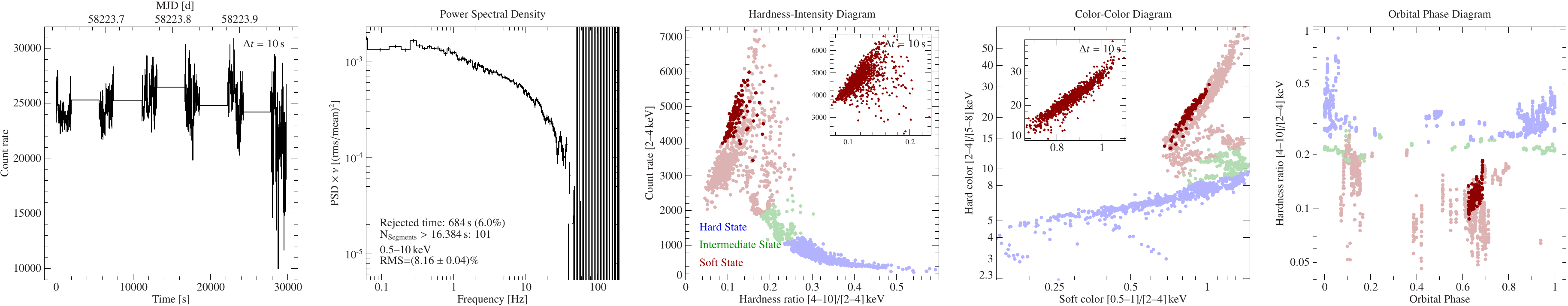}
\includegraphics[width=1\textwidth]{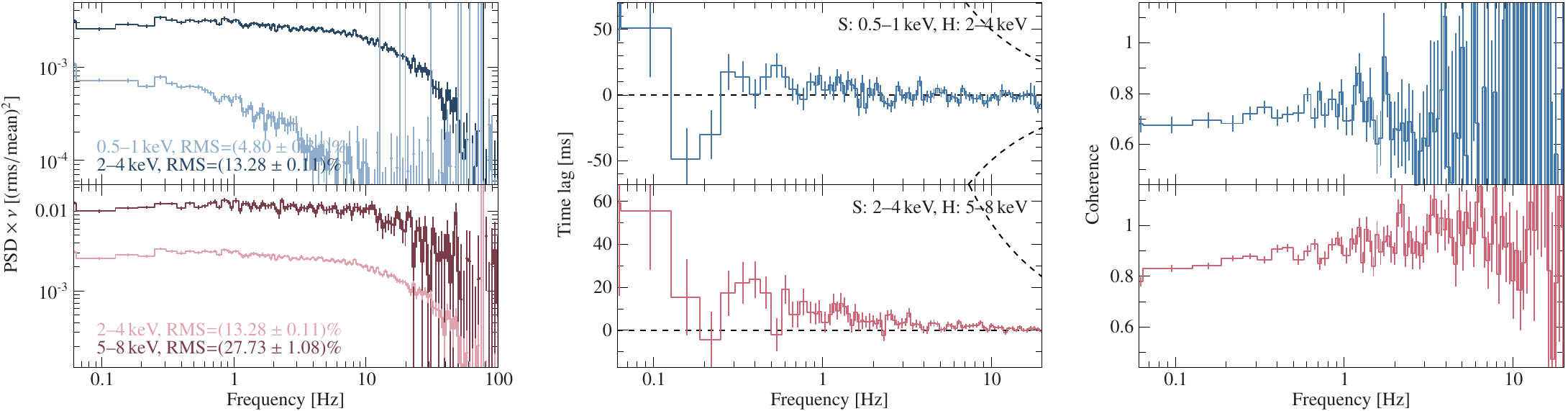}
\caption{\nicer observation 1100320119 of \cyg. $\Gamma\sim 2.7$.}
\label{fig:app:1100320119}
\end{figure*}

\begin{figure*}
\centering
\includegraphics[width=1\textwidth]{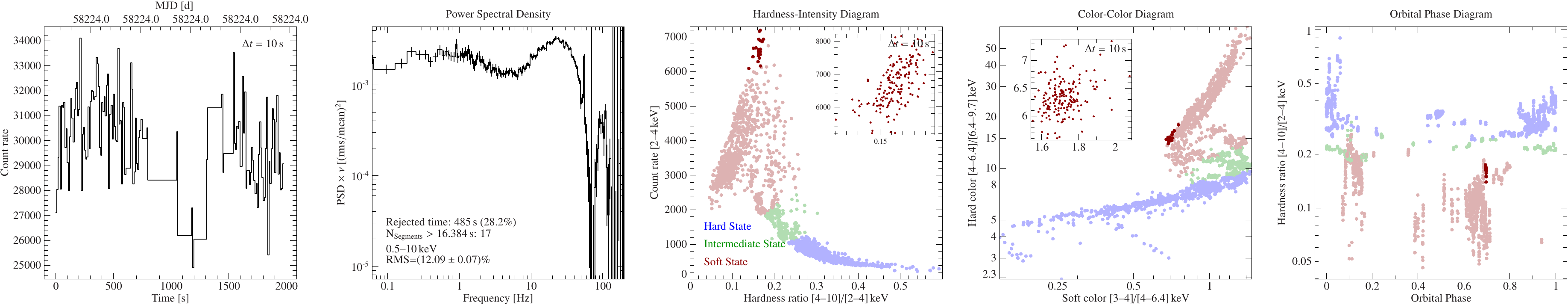}
\includegraphics[width=1\textwidth]{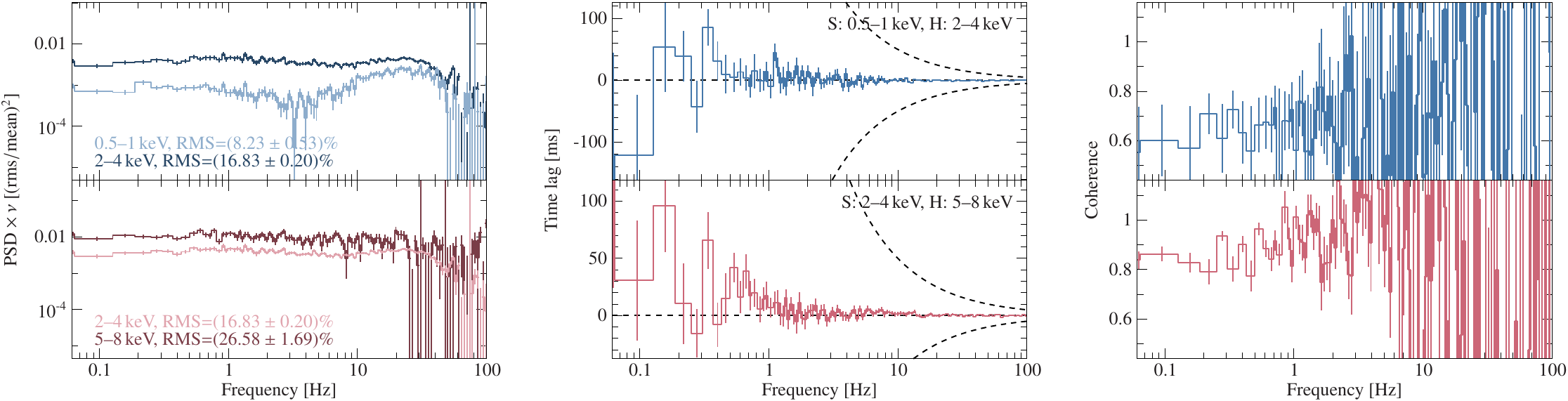}
\caption{\nicer observation 1100320120 of \cyg. $\Gamma\sim 2.7$.}
\label{fig:app:1100320120}
\end{figure*}

\begin{figure*}
\centering
\includegraphics[width=1\textwidth]{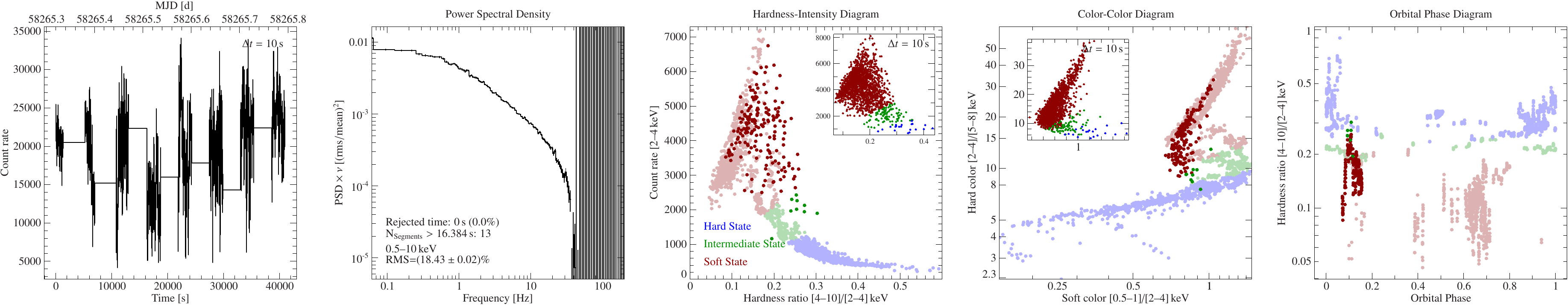}
\includegraphics[width=1\textwidth]{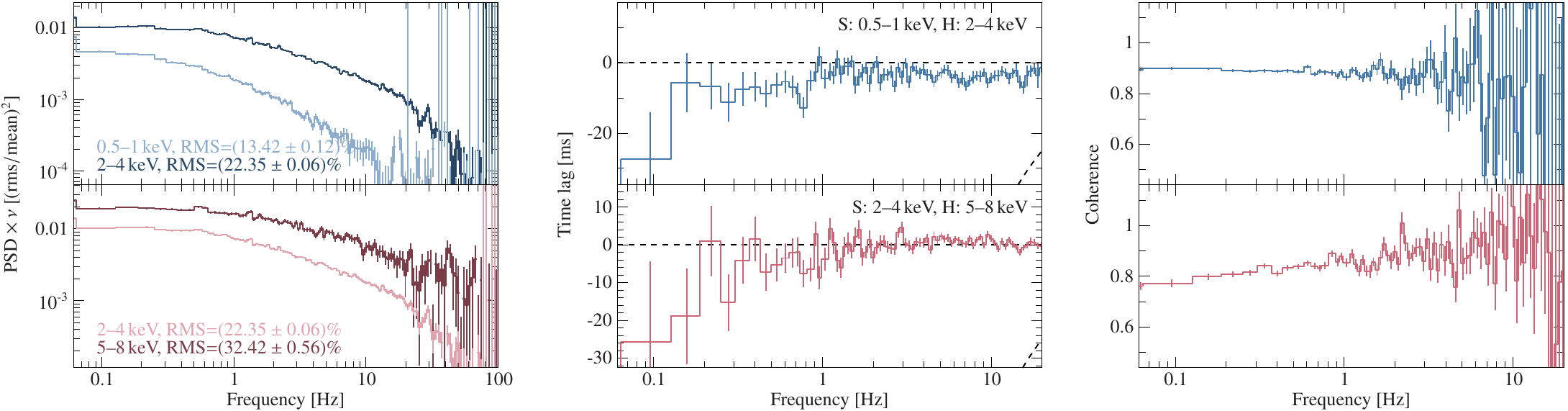}
\caption{\nicer observation 1100320121 of \cyg. $\Gamma\sim 2.7$.}
\label{fig:app:1100320121}
\end{figure*}

\begin{figure*}
\centering
\includegraphics[width=1\textwidth]{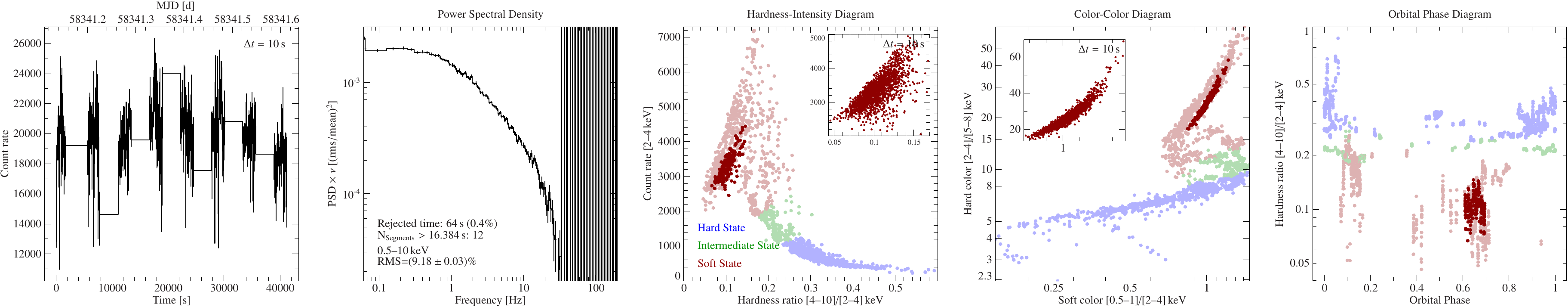}
\includegraphics[width=1\textwidth]{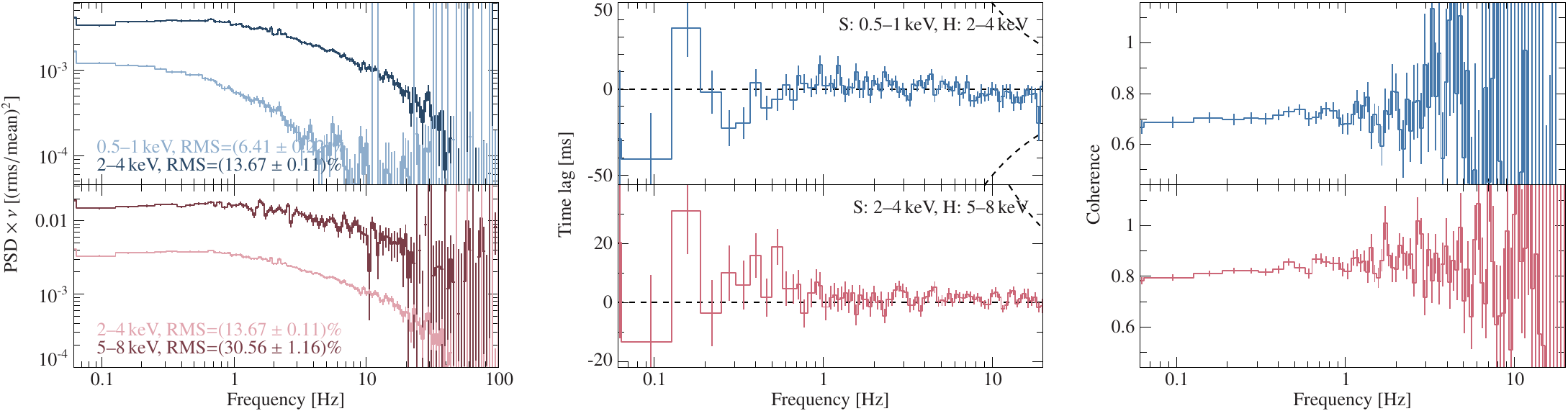}
\caption{\nicer observation 1100320122 of \cyg. $\Gamma\sim 3.1$.}
\label{fig:app:1100320122}
\end{figure*}

\begin{figure*}
\centering
\includegraphics[width=1\textwidth]{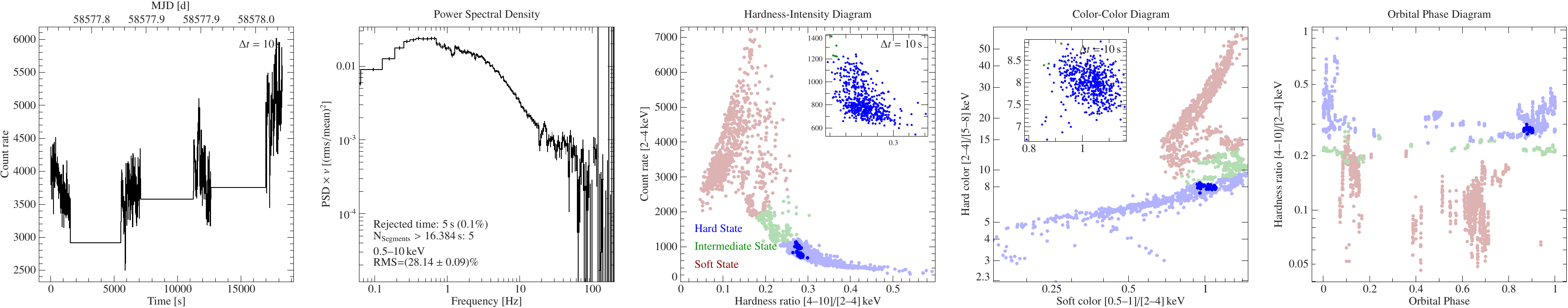}
\includegraphics[width=1\textwidth]{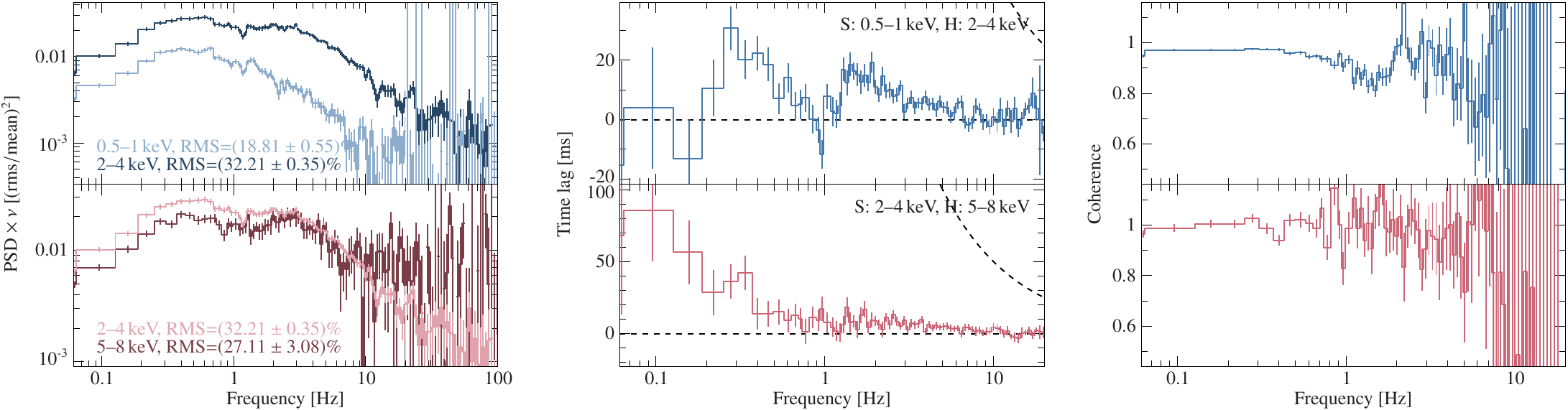}
\caption{\nicer observation 2100320101 of \cyg. $\Gamma\sim 1.8$.}
\label{fig:app:2100320101}
\end{figure*}

\begin{figure*}
\centering
\includegraphics[width=1\textwidth]{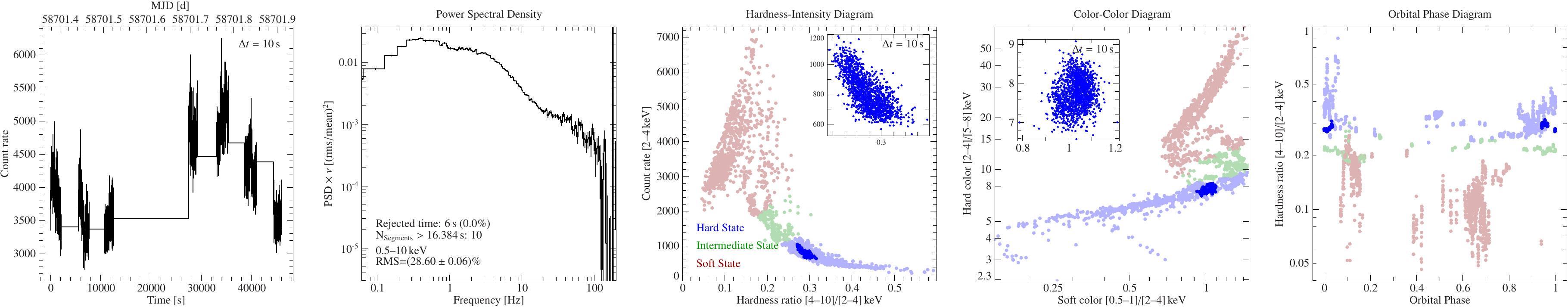}
\includegraphics[width=1\textwidth]{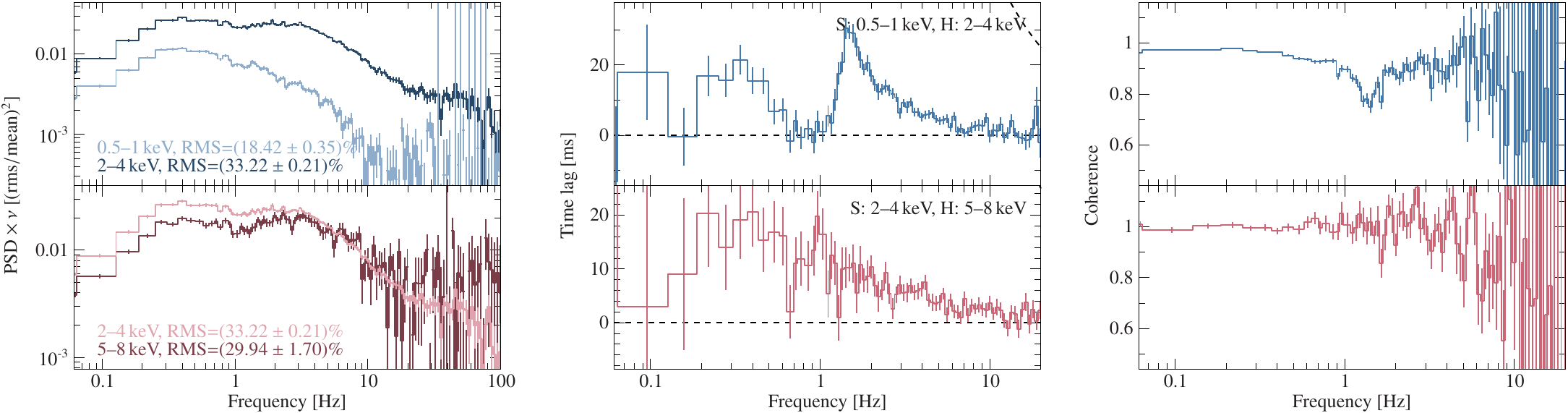}
\caption{\nicer observation 2636010101 of \cyg. $\Gamma\sim 1.8$.}
\label{fig:app:2636010101}
\end{figure*}

\begin{figure*}
\centering
\includegraphics[width=1\textwidth]{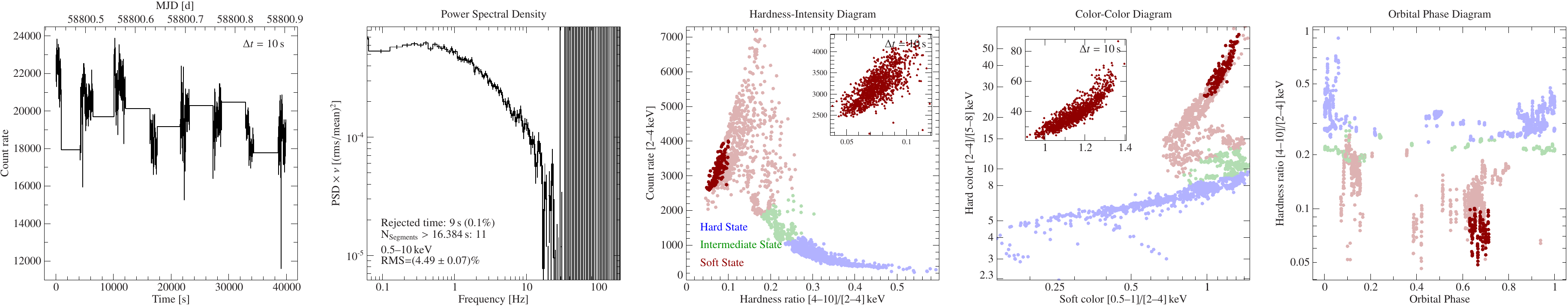}
\includegraphics[width=1\textwidth]{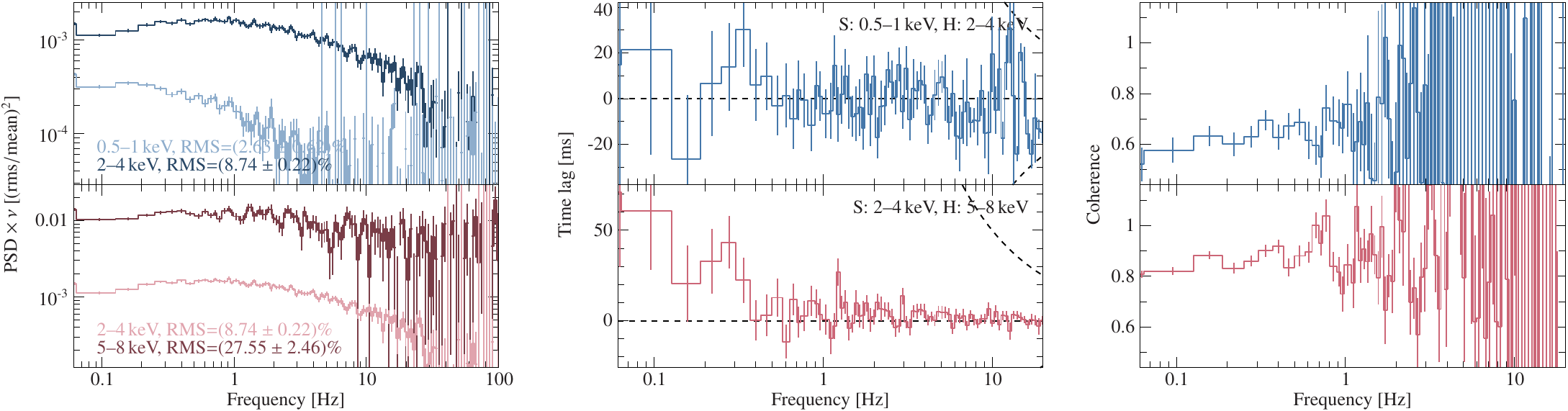}
\caption{\nicer observation 2636010102 of \cyg. $\Gamma\sim 3.1$.}
\label{fig:app:2636010102}
\end{figure*}

\begin{figure*}
\centering
\includegraphics[width=1\textwidth]{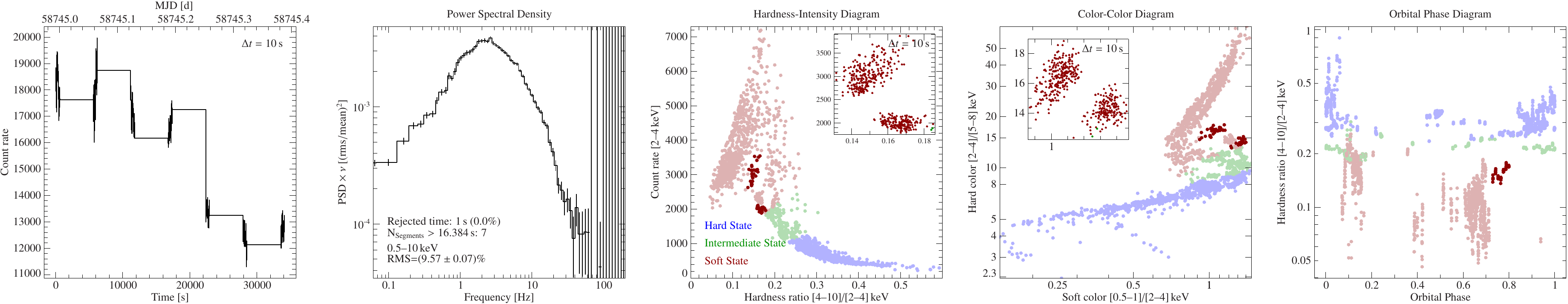}
\includegraphics[width=1\textwidth]{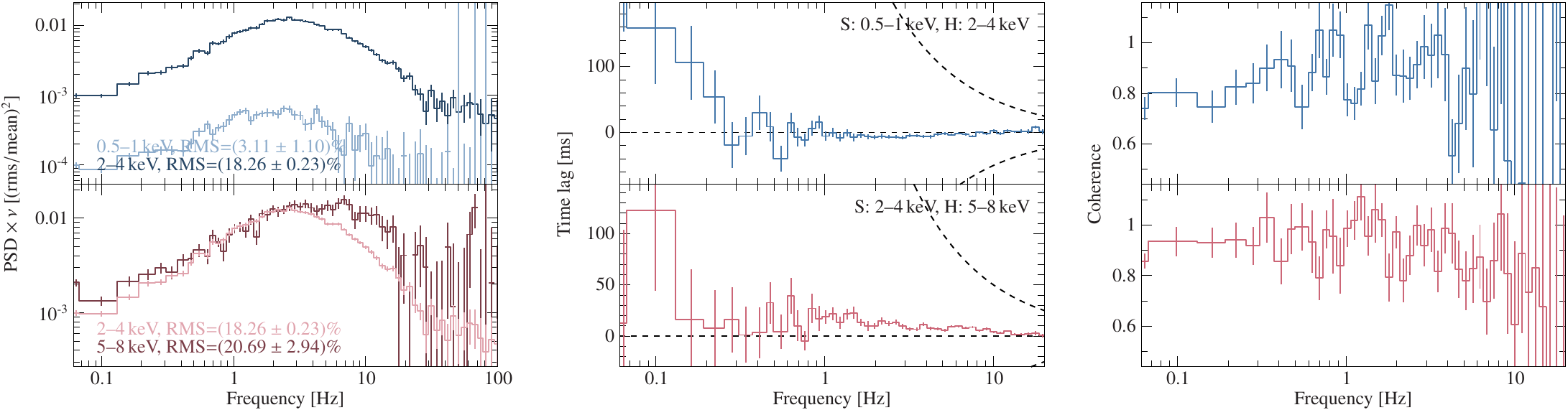}
\caption{\nicer observation 2636010201 of \cyg. $\Gamma\sim 2.4$.}
\label{fig:app:2636010201}
\end{figure*}

\begin{figure*}
\centering
\includegraphics[width=1\textwidth]{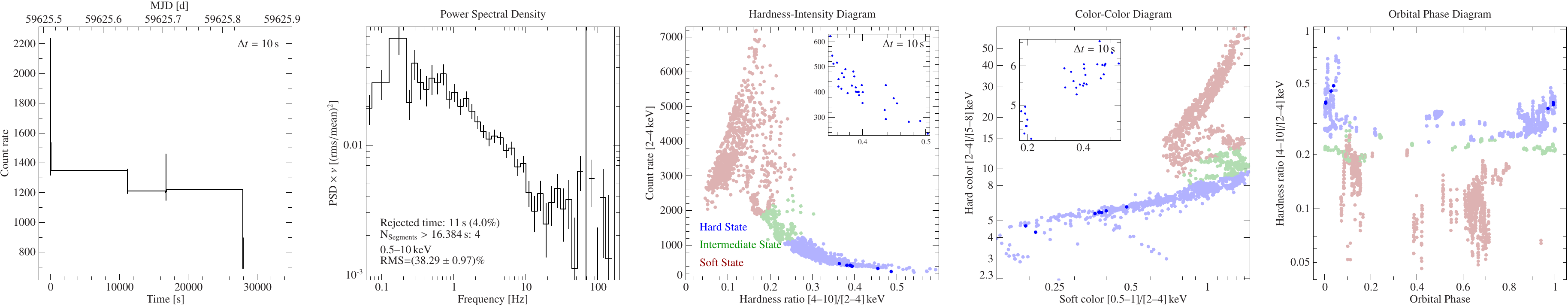}
\includegraphics[width=1\textwidth]{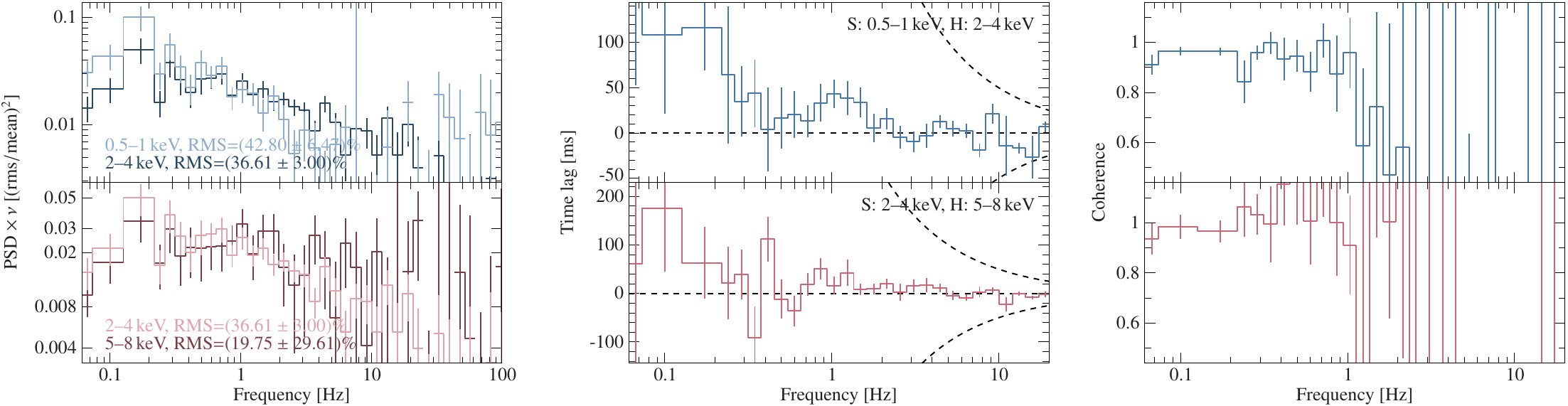}
\caption{\nicer observation 4690010103 of \cyg. $\Gamma\sim 1.7$.}
\label{fig:app:4690010103}
\end{figure*}

\begin{figure*}
\centering
\includegraphics[width=1\textwidth]{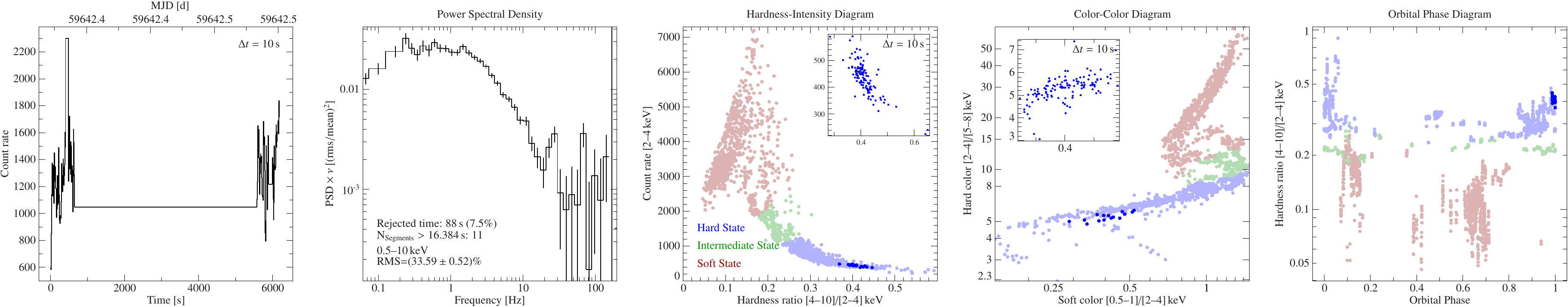}
\includegraphics[width=1\textwidth]{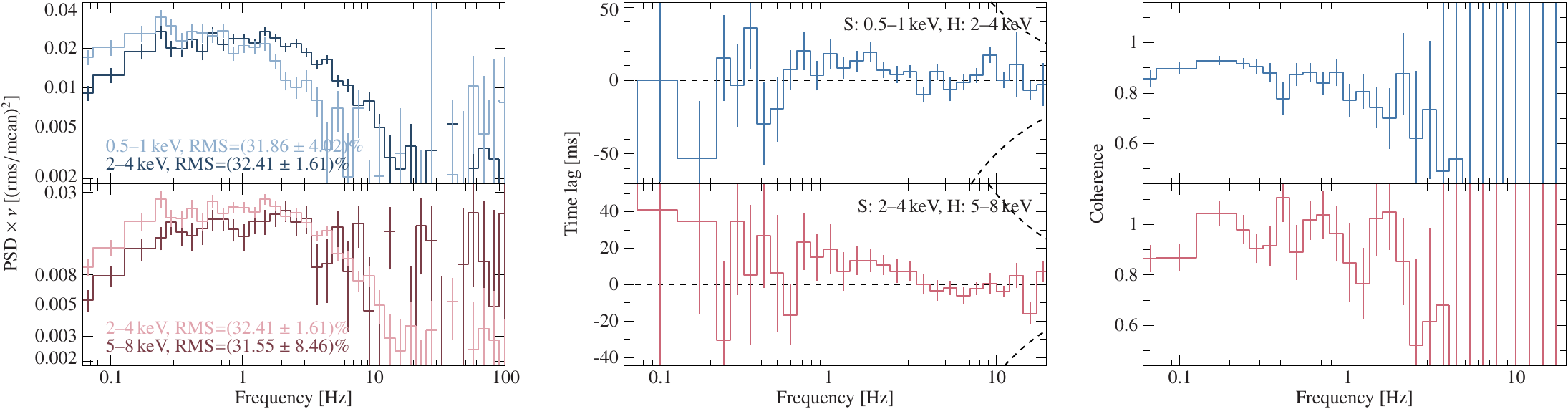}
\caption{\nicer observation 4690010104 of \cyg. $\Gamma\sim 1.5$.}
\label{fig:app:4690010104}
\end{figure*}

\begin{figure*}
\centering
\includegraphics[width=1\textwidth]{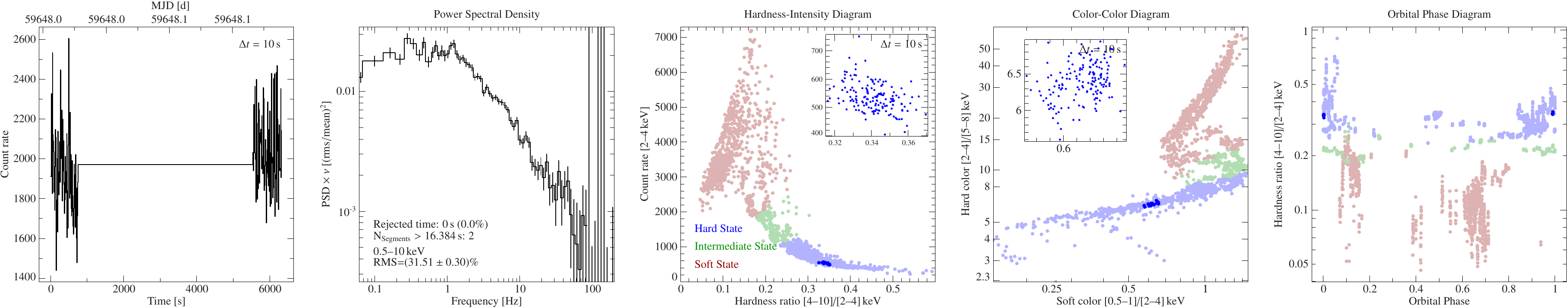}
\includegraphics[width=1\textwidth]{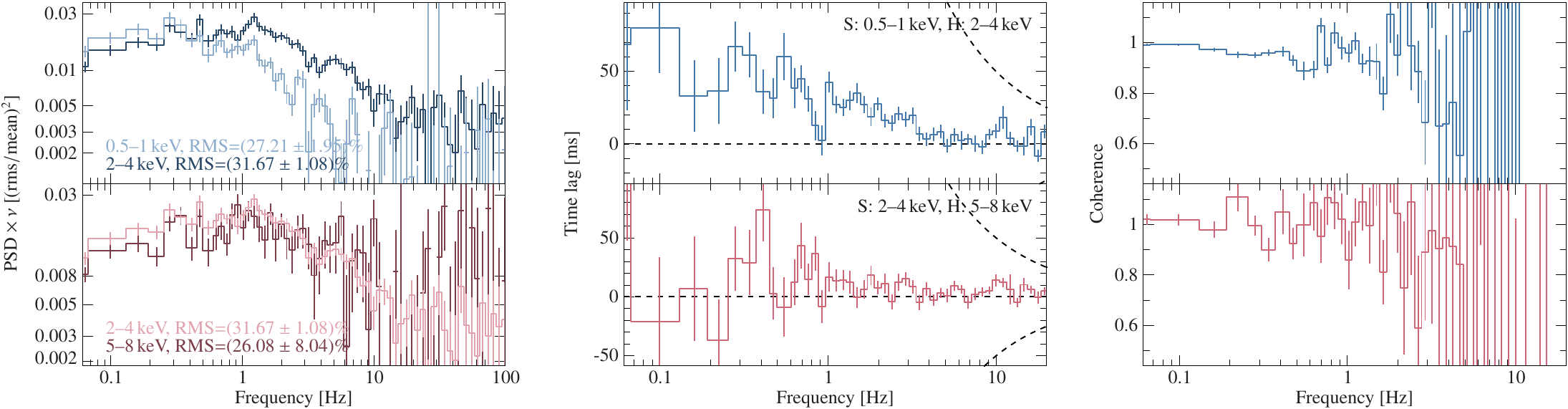}
\caption{\nicer observation 4690010105 of \cyg. $\Gamma\sim 1.6$.}
\label{fig:app:4690010105}
\end{figure*}

\begin{figure*}
\centering
\includegraphics[width=1\textwidth]{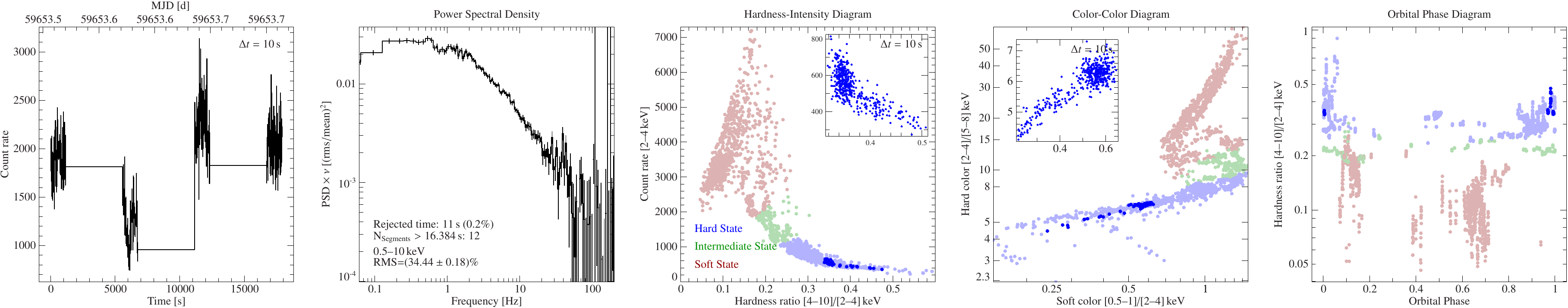}
\includegraphics[width=1\textwidth]{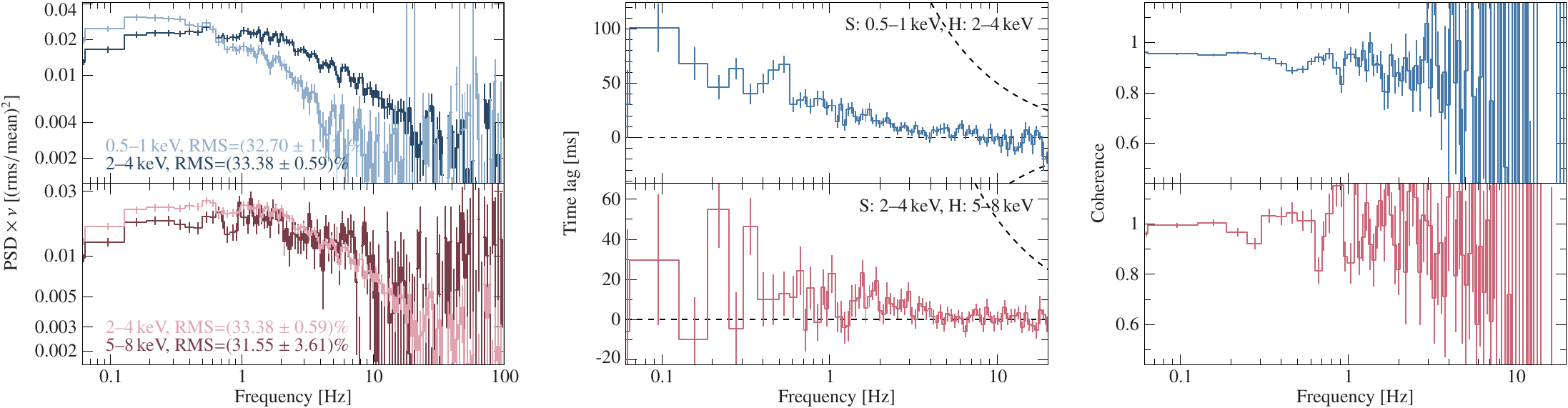}
\caption{\nicer observation 4690010106 of \cyg. $\Gamma\sim 1.6$.}
\label{fig:app:4690010106}
\end{figure*}

\begin{figure*}
\centering
\includegraphics[width=1\textwidth]{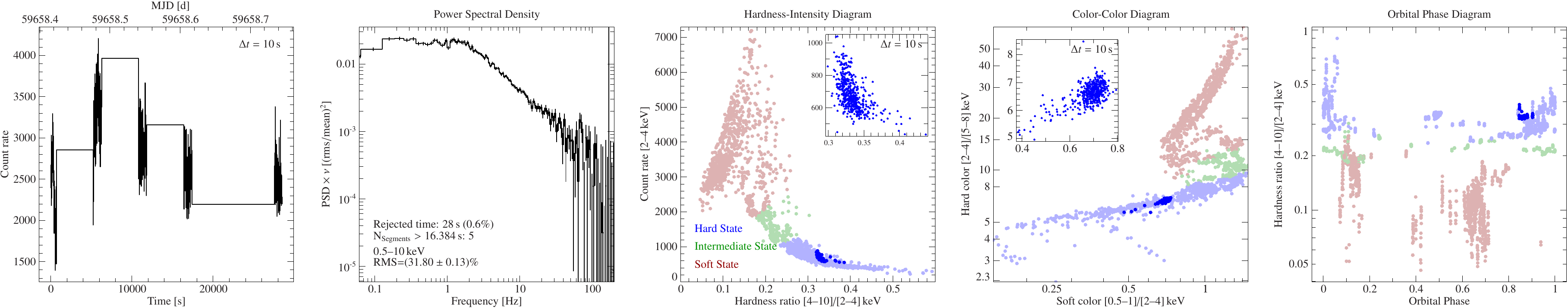}
\includegraphics[width=1\textwidth]{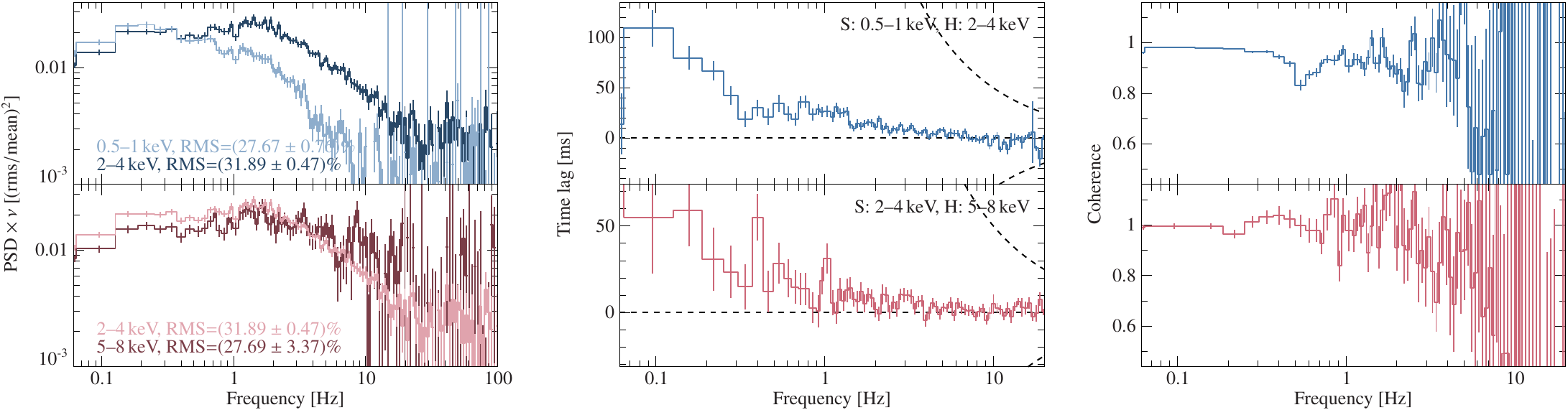}
\caption{\nicer observation 4690010107 of \cyg. $\Gamma\sim 1.7$.}
\label{fig:app:4690010107}
\end{figure*}

\begin{figure*}
\centering
\includegraphics[width=1\textwidth]{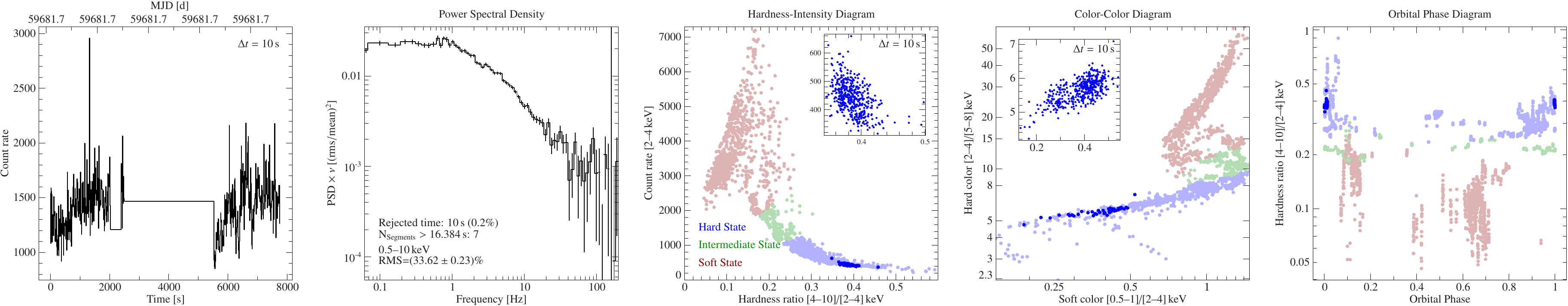}
\includegraphics[width=1\textwidth]{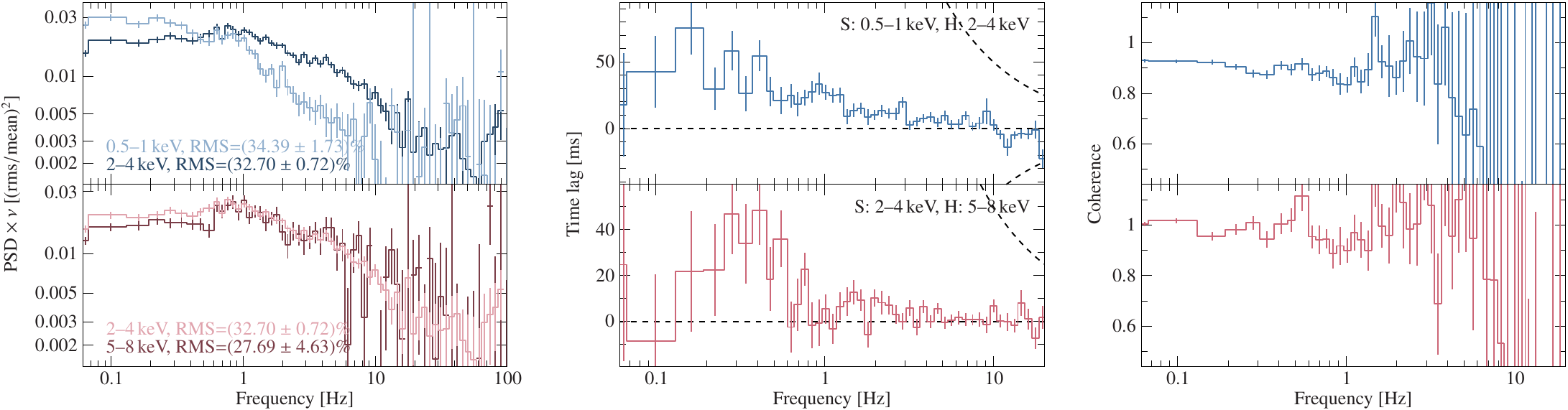}
\caption{\nicer observation 4690010109 of \cyg. $\Gamma\sim 1.6$.}
\label{fig:app:4690010109}
\end{figure*}

\begin{figure*}
\centering
\includegraphics[width=1\textwidth]{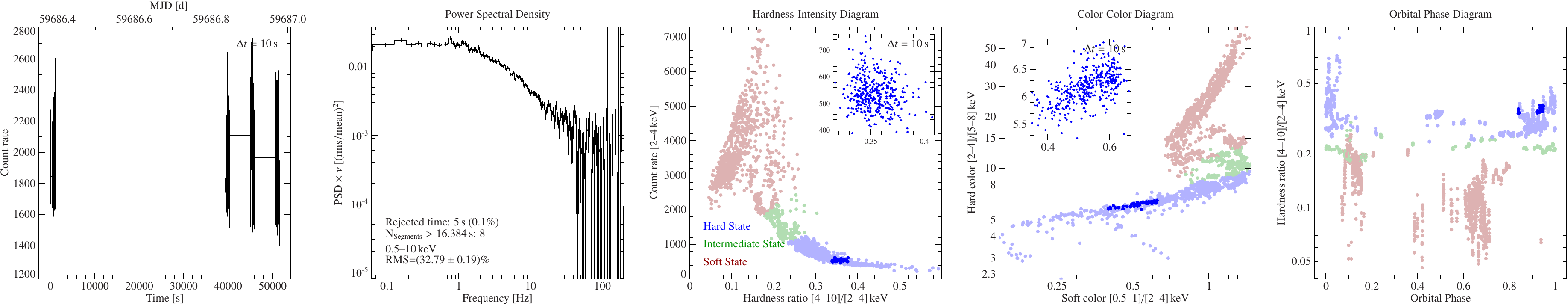}
\includegraphics[width=1\textwidth]{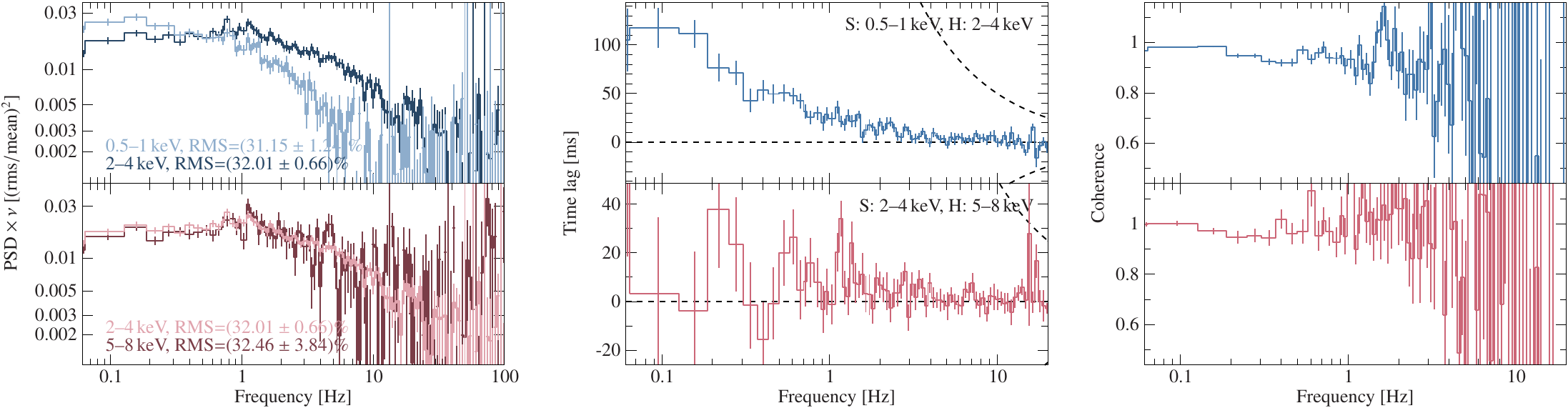}
\caption{\nicer observation 4690010110 of \cyg. $\Gamma\sim 1.7$.}
\label{fig:app:4690010110}
\end{figure*}

\begin{figure*}
\centering
\includegraphics[width=1\textwidth]{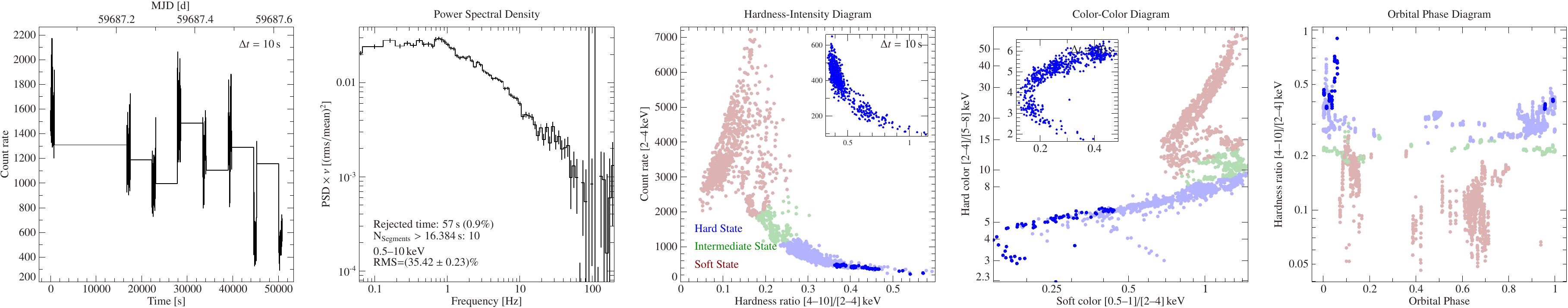}
\includegraphics[width=1\textwidth]{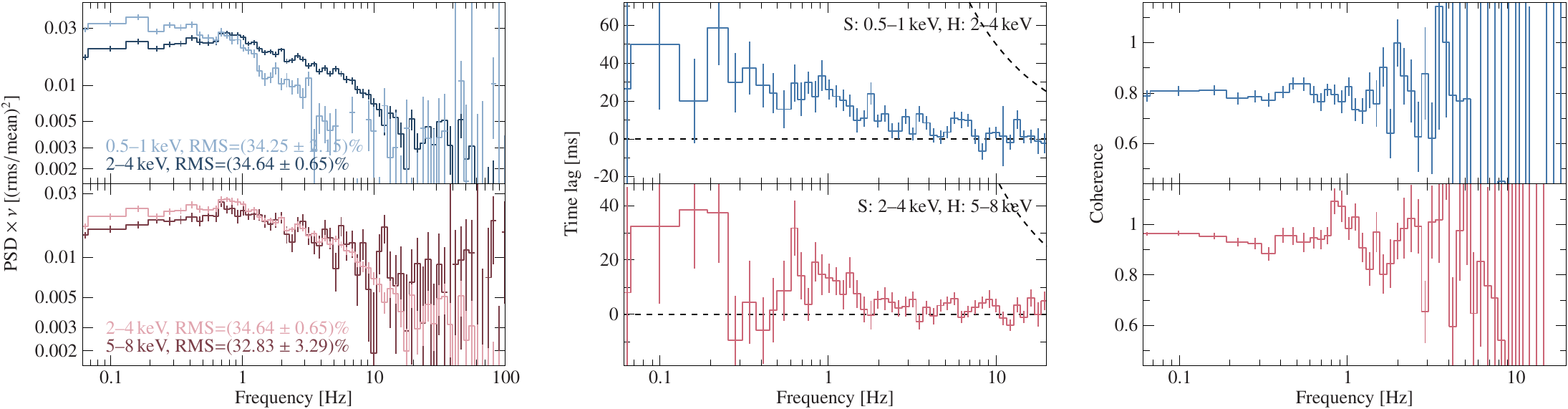}
\caption{\nicer observation 4690010111 of \cyg. $\Gamma\sim 1.6$.}
\label{fig:app:4690010111}
\end{figure*}

\begin{figure*}
\centering
\includegraphics[width=1\textwidth]{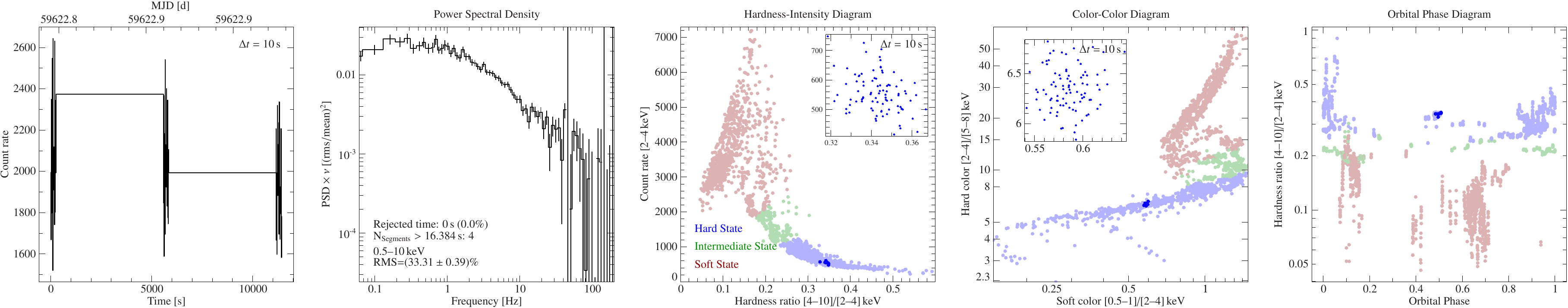}
\includegraphics[width=1\textwidth]{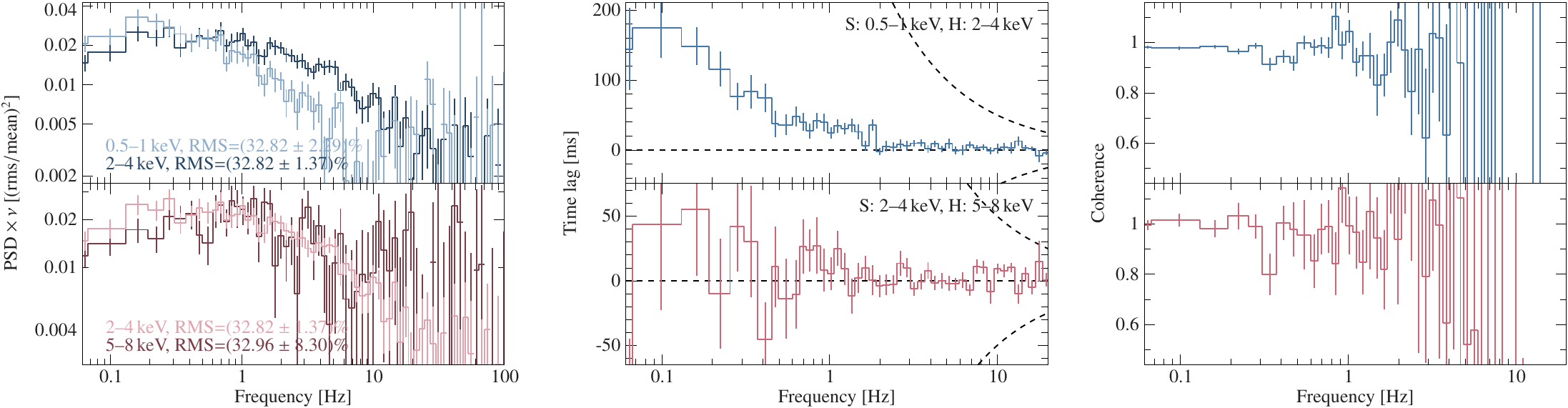}
\caption{\nicer observation 4690020101 of \cyg. $\Gamma\sim 1.6$.}
\label{fig:app:4690020101}
\end{figure*}

\begin{figure*}
\centering
\includegraphics[width=1\textwidth]{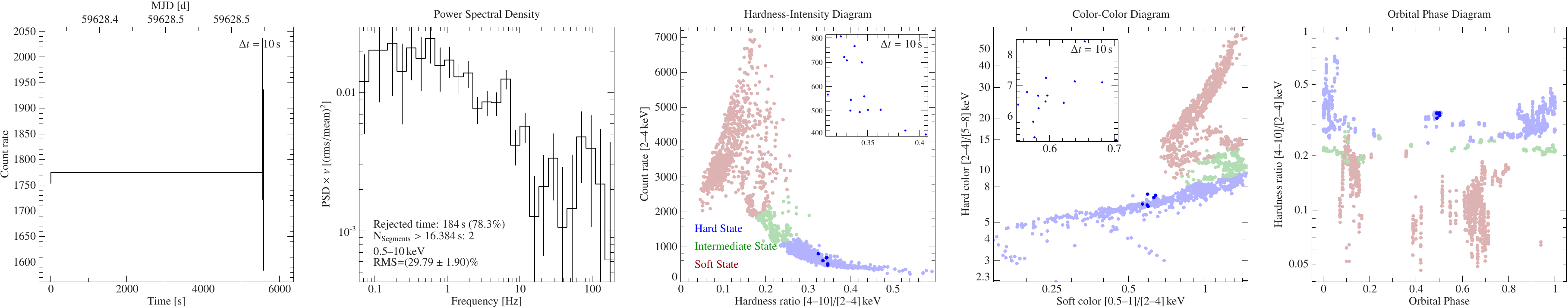}
\includegraphics[width=1\textwidth]{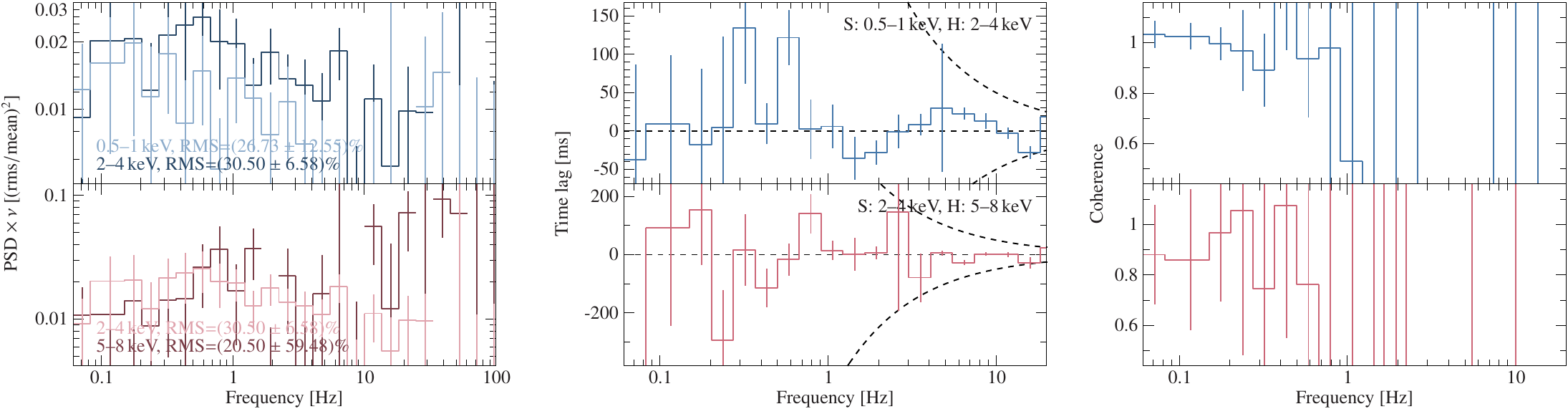}
\caption{\nicer observation 4690020102 of \cyg. $\Gamma\sim 1.6$.}
\label{fig:app:4690020102}
\end{figure*}

\begin{figure*}
\centering
\includegraphics[width=1\textwidth]{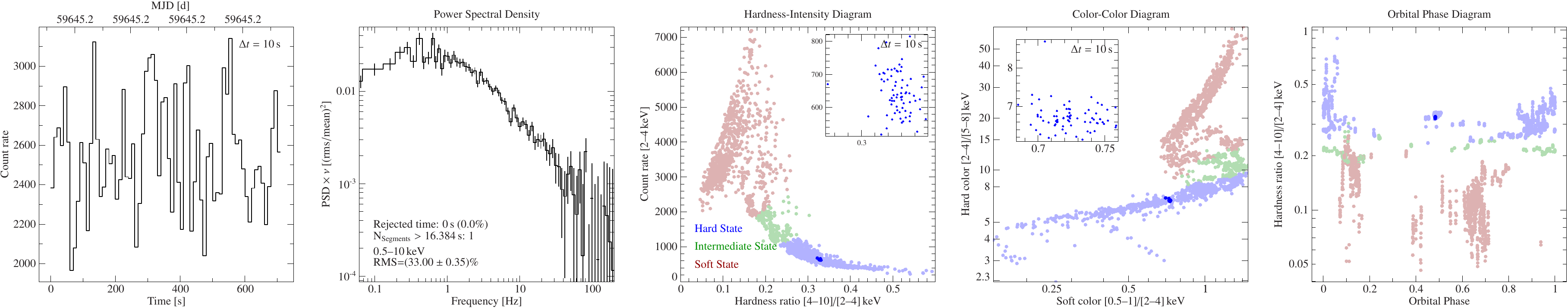}
\includegraphics[width=1\textwidth]{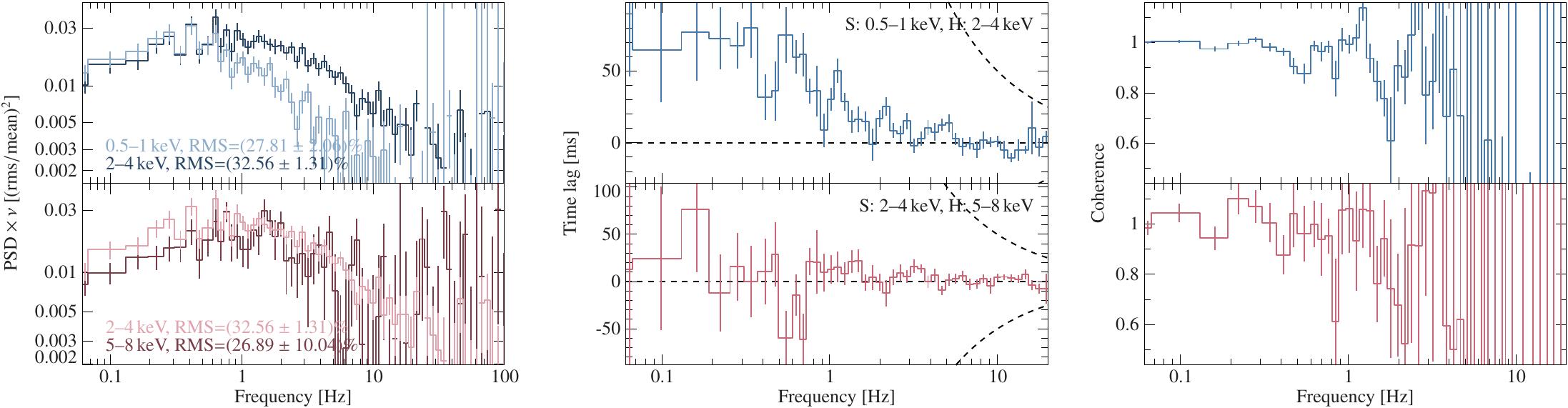}
\caption{\nicer observation 4690020103 of \cyg. $\Gamma\sim 1.7$.}
\label{fig:app:4690020103}
\end{figure*}

\begin{figure*}
\centering
\includegraphics[width=1\textwidth]{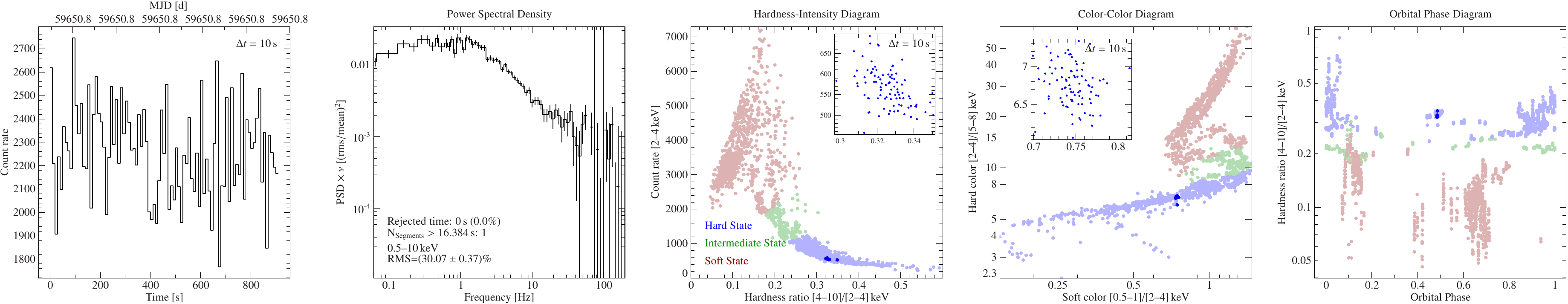}
\includegraphics[width=1\textwidth]{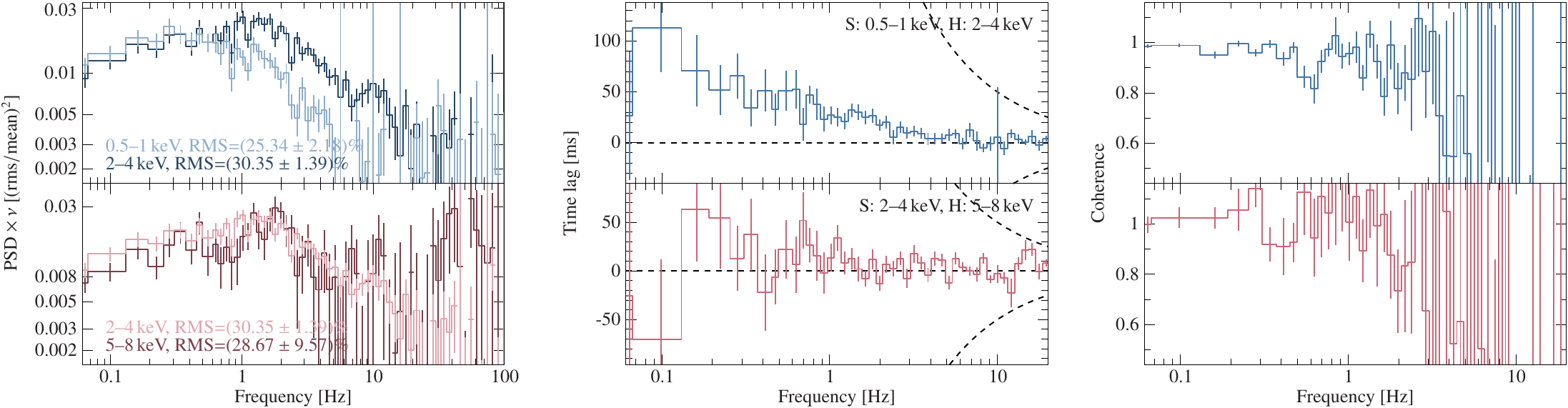}
\caption{\nicer observation 4690020104 of \cyg. $\Gamma\sim 1.7$.}
\label{fig:app:4690020104}
\end{figure*}

\begin{figure*}
\centering
\includegraphics[width=1\textwidth]{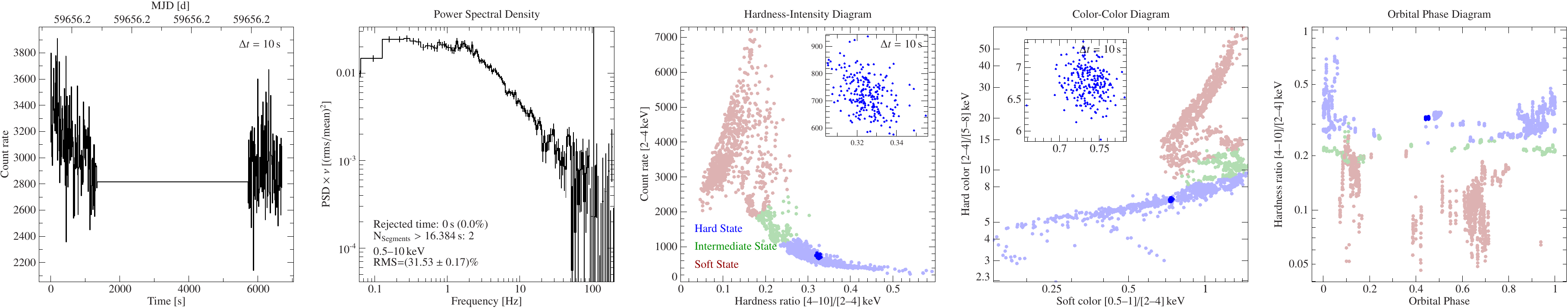}
\includegraphics[width=1\textwidth]{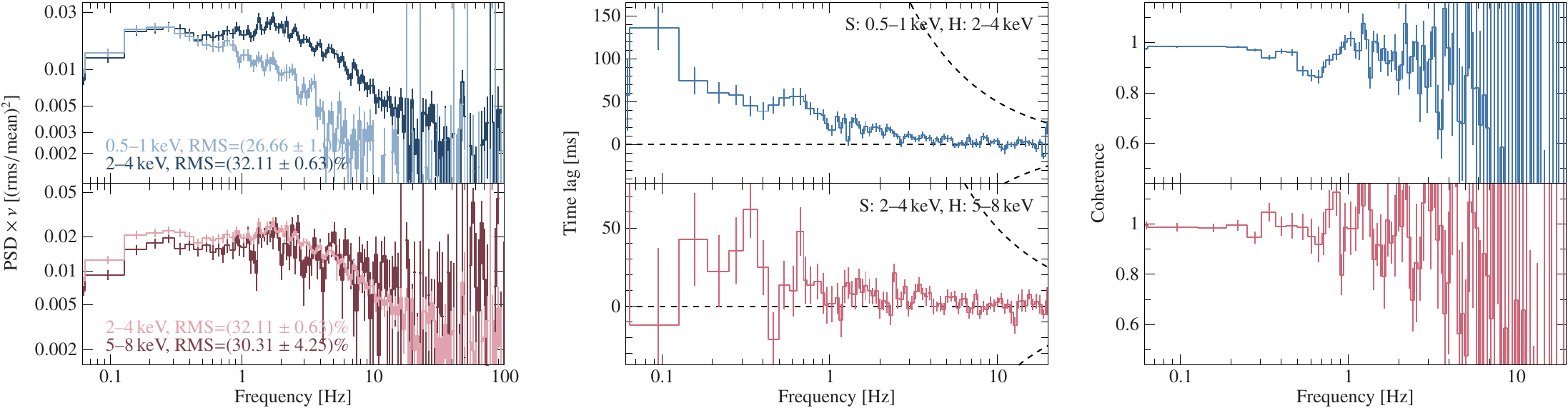}
\caption{\nicer observation 4690020105 of \cyg. $\Gamma\sim 1.7$.}
\label{fig:app:4690020105}
\end{figure*}

\begin{figure*}
\centering
\includegraphics[width=1\textwidth]{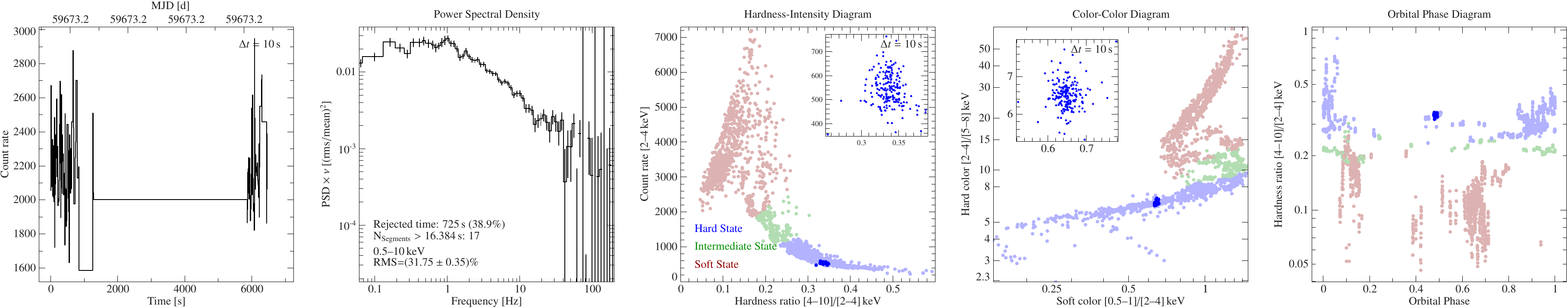}
\includegraphics[width=1\textwidth]{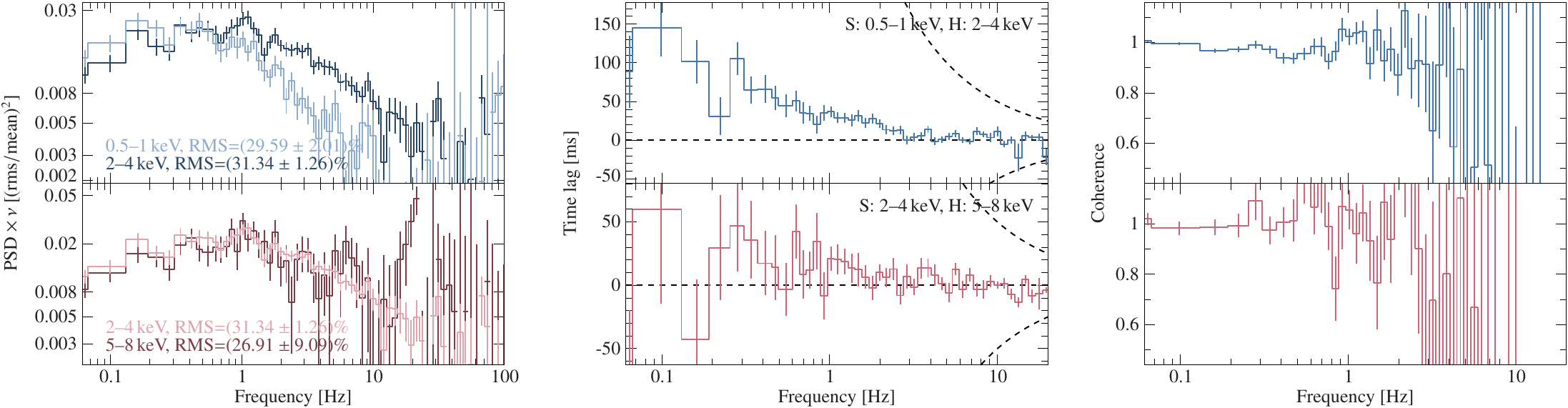}
\caption{\nicer observation 4690020107 of \cyg. $\Gamma\sim 1.6$.}
\label{fig:app:4690020107}
\end{figure*}

\begin{figure*}
\centering
\includegraphics[width=1\textwidth]{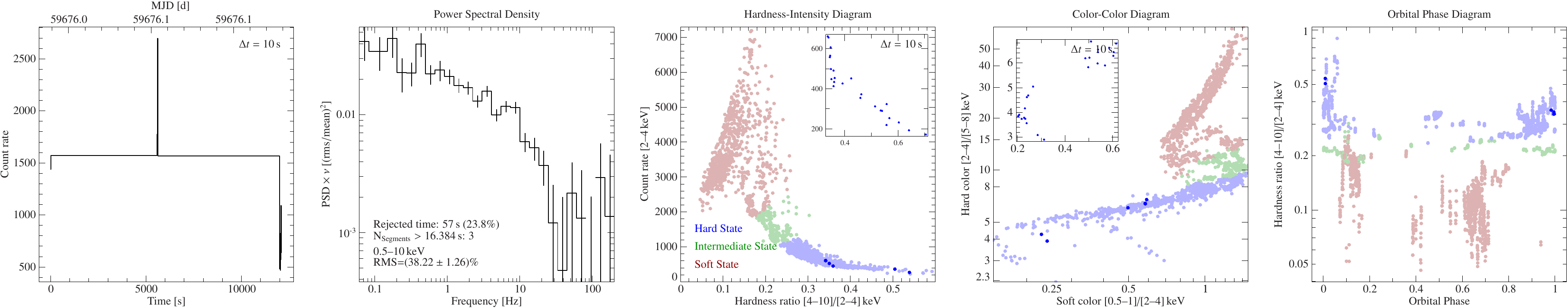}
\includegraphics[width=1\textwidth]{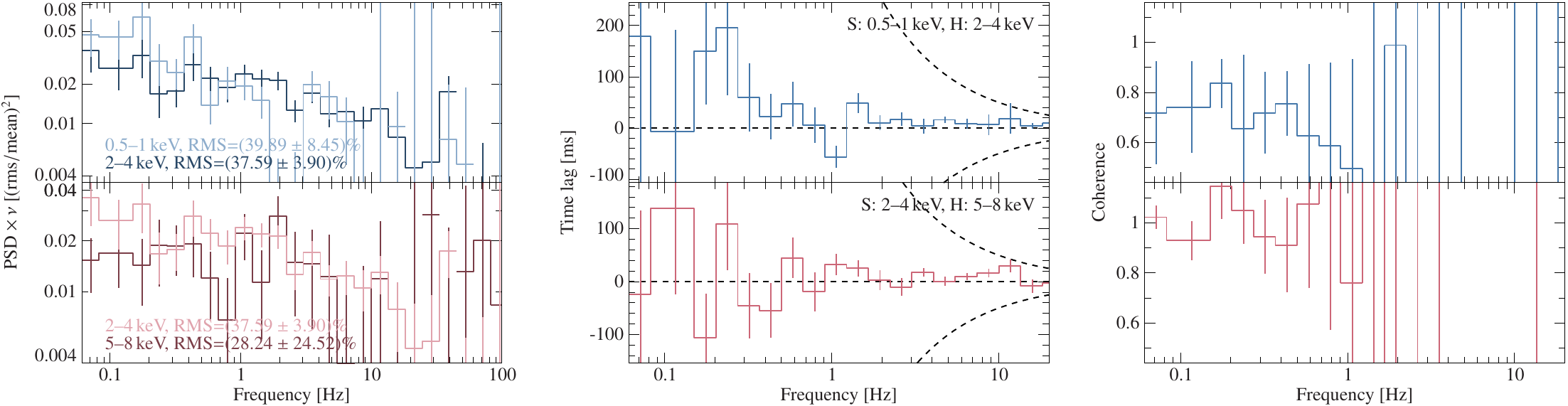}
\caption{\nicer observation 4690020108 of \cyg. $\Gamma\sim 1.5$.}
\label{fig:app:4690020108}
\end{figure*}

\begin{figure*}
\centering
\includegraphics[width=1\textwidth]{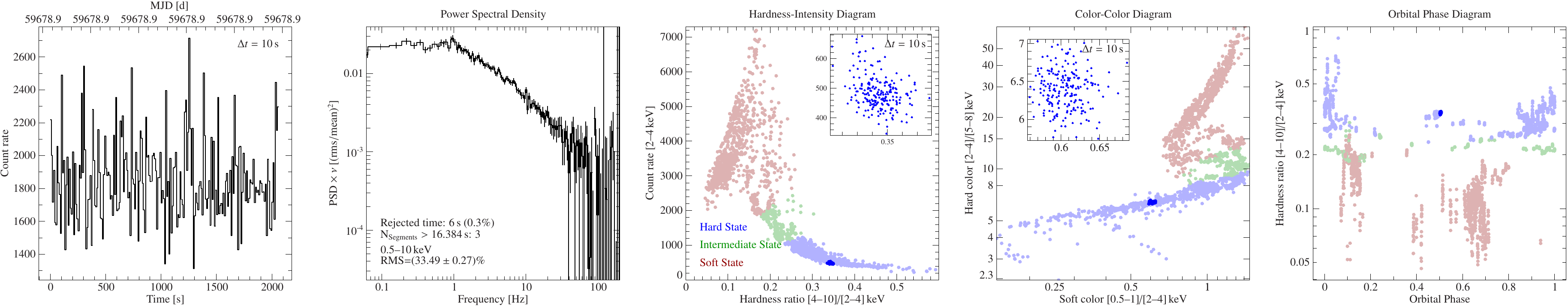}
\includegraphics[width=1\textwidth]{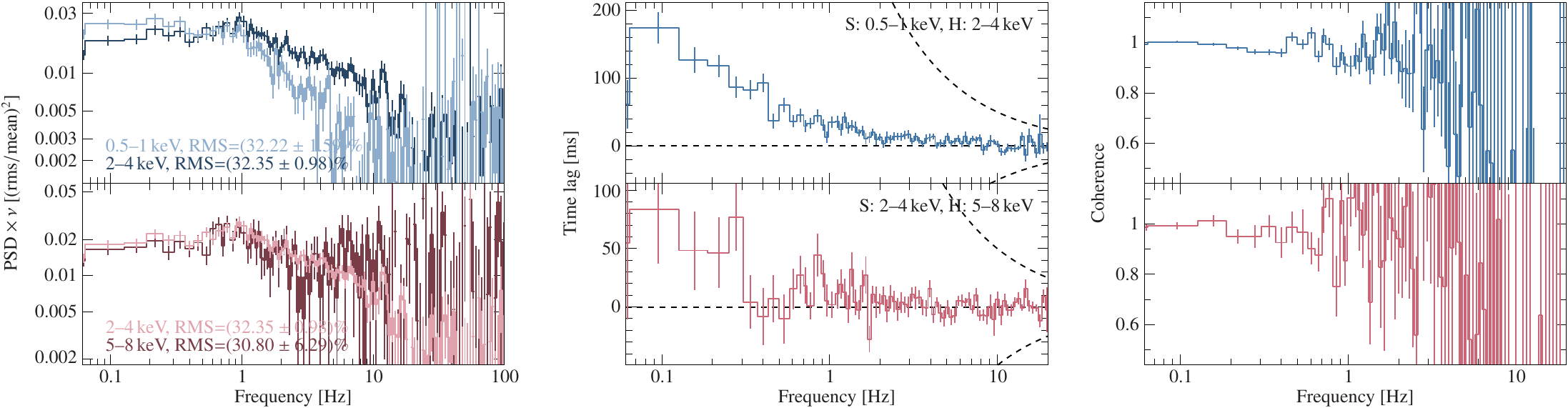}
\caption{\nicer observation 4690020109 of \cyg. $\Gamma\sim 1.6$.}
\label{fig:app:4690020109}
\end{figure*}

\begin{figure*}
\centering
\includegraphics[width=1\textwidth]{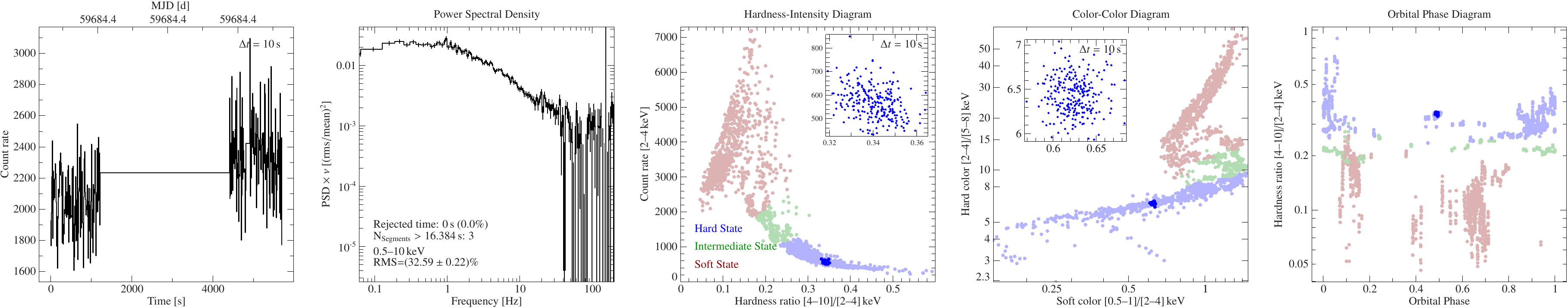}
\includegraphics[width=1\textwidth]{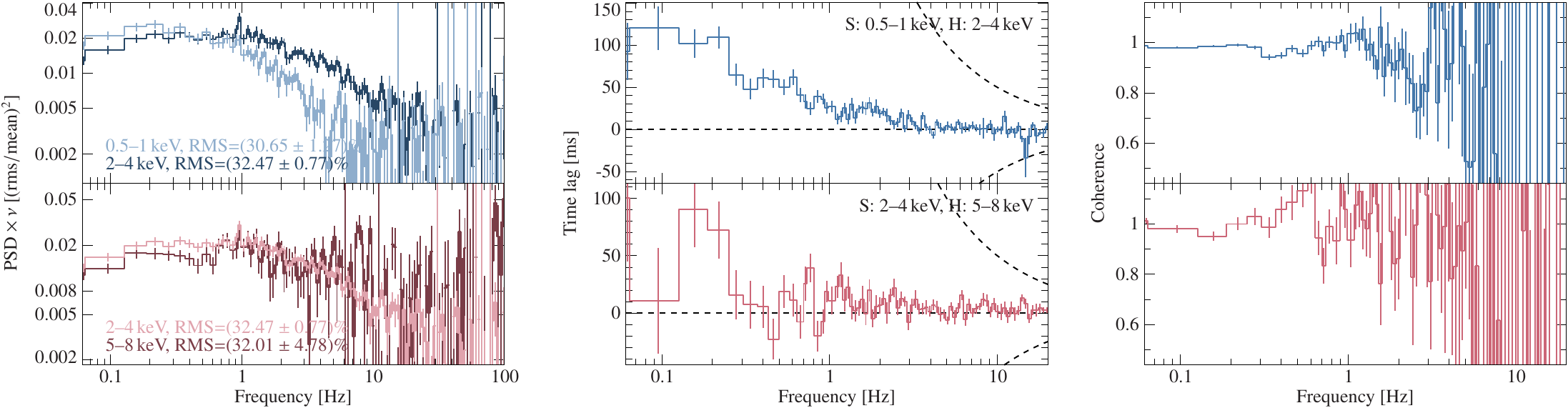}
\caption{\nicer observation 4690020110 of \cyg. $\Gamma\sim 1.7$.}
\label{fig:app:4690020110}
\end{figure*}

In this section, we provide supplementary information on the 
analysis. 
Table~\ref{tab:observation_log} contains the
observation log of the data analyzed in this paper. 
Table~\ref{tab:2636010101_multifit_4lor_plag} provides the parameters of the fit model used in
Sect.~\ref{subsec:lags}. Figures~\ref{fig:app:0100320101}--\ref{fig:app:4690020110}
show the timing products of each observation in the \nicer archive
of \cyg up to April 2022 (cycle~4). Each collection of figures is arranged as
follows.  The top row shows overview figures of the lightcurve, the
PSD in 0.5--10\,keV, and the location of the observation in the
hardness-intensity, orbital phase, and color-color diagrams. The data
in the top rows are colored as in the main text: red is the soft,
green the intermediate, and blue the hard state. The bottom panel
shows timing properties for low and high energies. The left column
shows PSDs computed in three energy bands. The center column shows the
lag-frequency spectrum for two correlated bands: Blue uses 0.5--1\,keV
and 2--4\,keV, and red uses 2--4\,keV and 5--8\,keV. Dashed lines
denote the phase-wrapping limit. The right column shows the coherence
between these energy bands.

\begin{table*}
  \caption{Overview of the \cyg observations in the \nicer
    archive up to April 2022 (cycle~4). Photon indices are derived from diskbb+power law fits, described in Sect.~\ref{subsec:cygx1_in_q-diagram}.}
  \label{tab:observation_log}
  \centering
  \begin{tabular}{lllllll}
  \hline\hline
Obs. ID & Start date & Exposure [s] & $\Gamma$ & $\phi_\mathrm{orb}$ & timing feature & wind-affected\\ 
\hline 
0100320101 & 2017-06-30T15:55:11 & 719 & 1.8 & 0.02--0.03 & -- & -- \\ 
0100320102 & 2017-07-02T12:41:37 & 1580 & 2.2 & 0.36--0.38 & -- & -- \\ 
0100320103 & 2017-07-03T01:27:00 & 668 & 2.1 & 0.45--0.61 & -- & -- \\ 
0100320104 & 2017-07-04T00:00:28 & 10682 & 1.9 & 0.62--0.80 & \checkmark & -- \\ 
0100320105 & 2017-07-05T00:49:32 & 14153 & 1.9 & 0.80--0.96 & \checkmark & -- \\ 
0100320106 & 2017-07-06T01:21:25 & 7945 & 1.8 & 0.00--0.14 & \checkmark & \checkmark \\ 
0100320107 & 2017-07-07T08:23:28 & 4369 & 2.0 & 0.22--0.24 & \checkmark & -- \\ 
0100320108 & 2017-07-10T23:19:38 & 415 & 2.1 & 0.86--0.87 & -- & -- \\ 
0100320109 & 2017-07-11T00:19:49 & 7304 & 2.1 & 0.87--0.04 & -- & -- \\ 
0100320110 & 2017-07-12T00:03:49 & 6584 & 2.3 & 0.05--0.21 & -- & -- \\ 
1100320101 & 2017-10-28T06:53:58 & 3709 & 3.3 & 0.39--0.42 & -- & -- \\ 
1100320102 & 2017-10-28T23:53:17 & 3054 & 3.0 & 0.51--0.55 & -- & -- \\ 
1100320103 & 2017-10-30T00:35:37 & 1406 & 3.4 & 0.70--0.70 & -- & -- \\ 
1100320104 & 2017-11-01T11:05:00 & 626 & 3.0 & 0.13--0.17 & -- & -- \\ 
1100320106 & 2017-12-01T22:13:28 & 501 & 3.0 & 0.57--0.57 & -- & -- \\ 
1100320107 & 2017-12-02T03:07:00 & 483 & 2.9 & 0.61--0.66 & -- & -- \\ 
1100320108 & 2017-12-03T23:37:15 & 106 & 3.5 & 0.94--0.94 & -- & -- \\ 
1100320109 & 2017-12-04T21:15:20 & 221 & 3.3 & 0.10--0.11 & -- & -- \\ 
1100320110 & 2018-02-08T20:16:33 & 3970 & 1.6 & 0.88--0.90 & -- & -- \\ 
1100320111 & 2018-02-08T23:21:40 & 10251 & 1.8 & 0.90--0.95 & \checkmark & \checkmark \\ 
1100320112 & 2018-02-17T14:29:47 & 400 & 1.8 & 0.44--0.49 & -- & -- \\ 
1100320113 & 2018-02-18T05:51:40 & 1044 & 1.7 & 0.56--0.67 & -- & -- \\ 
1100320114 & 2018-02-19T00:21:00 & 390 & 1.8 & 0.69--0.87 & -- & -- \\ 
1100320115 & 2018-02-20T02:38:00 & 386 & 1.7 & 0.89--0.93 & -- & -- \\ 
1100320116 & 2018-02-21T00:10:00 & 678 & 1.7 & 0.05--0.13 & -- & -- \\ 
1100320117 & 2018-03-26T18:41:40 & 6203 & 2.2 & 0.08--0.12 & -- & -- \\ 
1100320118 & 2018-03-27T00:52:40 & 4612 & 2.3 & 0.13--0.15 & -- & -- \\ 
1100320119 & 2018-04-15T14:55:08 & 11373 & 2.7 & 0.63--0.69 & -- & -- \\ 
1100320121 & 2018-05-27T07:38:00 & 17450 & 2.7 & 0.07--0.16 & -- & (\checkmark) \\ 
1100320122 & 2018-08-11T03:12:34 & 17954 & 3.1 & 0.61--0.69 & -- & -- \\ 
2100320101 & 2019-04-04T18:14:20 & 5801 & 1.8 & 0.87--0.90 & \checkmark & -- \\ 
2636010101 & 2019-08-06T09:01:20 & 14207 & 1.8 & 0.94--0.04 & \checkmark & -- \\ 
2636010102 & 2019-11-13T10:32:00 & 12355 & 3.1 & 0.63--0.71 & -- & -- \\ 
2636010201 & 2019-09-18T23:42:59 & 4272 & 2.4 & 0.73--0.80 & -- & -- \\ 
4690010103 & 2022-02-15T12:10:45 & 274 & 1.7 & 0.97--0.04 & -- & \checkmark \\ 
4690010104 & 2022-03-04T09:46:04 & 1176 & 1.5 & 0.99--1.00 & -- & \checkmark \\ 
4690010105 & 2022-03-10T00:29:20 & 1564 & 1.6 & 0.99--0.00 & -- & -- \\ 
4690010106 & 2022-03-15T12:07:00 & 4759 & 1.6 & 0.97--0.01 & shelf & \checkmark \\ 
4690010107 & 2022-03-20T09:48:00 & 4795 & 1.7 & 0.85--0.90 & shelf & (\checkmark) \\ 
4690010109 & 2022-04-12T15:50:21 & 4386 & 1.6 & 1.00--0.01 & -- & \checkmark \\ 
4690010110 & 2022-04-17T09:04:37 & 3798 & 1.7 & 0.84--0.95 & -- & \checkmark \\ 
4690010111 & 2022-04-18T00:44:17 & 6180 & 1.6 & 0.96--0.06 & -- & \checkmark \\ 
4690020101 & 2022-02-12T19:01:14 & 879 & 1.6 & 0.48--0.51 & -- & -- \\ 
4690020102 & 2022-02-18T10:05:41 & 235 & 1.6 & 0.49--0.50 & -- & -- \\ 
4690020103 & 2022-03-07T04:20:20 & 715 & 1.7 & 0.48--0.48 & -- & -- \\ 
4690020104 & 2022-03-12T19:04:58 & 912 & 1.7 & 0.48--0.49 & -- & -- \\ 
4690020105 & 2022-03-18T03:35:50 & 2320 & 1.7 & 0.44--0.46 & shelf & -- \\ 
4690020107 & 2022-04-04T03:37:06 & 1865 & 1.6 & 0.48--0.49 & -- & -- \\ 
4690020108 & 2022-04-06T23:41:40 & 239 & 1.5 & 0.98--0.01 & -- & \checkmark \\ 
4690020109 & 2022-04-09T21:20:00 & 2063 & 1.6 & 0.50--0.51 & -- & -- \\ 
4690020110 & 2022-04-15T09:14:00 & 2441 & 1.7 & 0.48--0.50 & -- & -- \\ 
    \hline
    \end{tabular}
\end{table*}

\begin{table}
  \caption{Lorentzian model parameters of the simultaneous fit of observation 2636010101 of \cyg (Fig.~\ref{fig:2636010101_multifit_4lor_plag}). Each non-zero-centered Lorentzian adds 6 free parameters: Frequency, width, and two normalizations for the low- and high-energy PSD plus the lag and modulus of the cross vector. The norm of the real/imaginary part is the modulus of the cross vector times the cosine/sine of the phase lag, respectively. The frequency of the zeroth Lorentzian ($L_0$) is fixed to 0\,Hz and phase zero is anchored to the strongest component ($L_1$). $L_n$ denotes the narrow component at the frequency of the timing feature. Uncertainties denote 90\% confidence limits.}
  \label{tab:2636010101_multifit_4lor_plag}
  \centering

\begin{tabular}{ll}
 \hline
 \hline
 Parameter & Value \\
 \hline
 $\nu_{L_0}$ [Hz] & 0 (frozen) \\
 $\nu_{L_1}$ [Hz] & $0.186\pm0.010$ \\
 $\nu_{L_2}$ [Hz] & $1.76\pm0.12$ \\
 $\nu_{L_n}$ [Hz] & $1.57\pm0.04$ \\
 $\Delta_{L_0}$ [Hz] & $\left(8\pm6\right)\times10^{-3}$ \\
 $\Delta_{L_1}$ [Hz] & $0.703\pm0.018$ \\
 $\Delta_{L_2}$ [Hz] & $4.79\pm0.10$ \\
 $\Delta_{L_n}$ [Hz] & $0.62^{+0.14}_{-0.12}$ \\
 $\Delta\phi_{L_0}$ [rad] & $-0.13^{+0.09}_{-0.13}$ \\
 $\Delta\phi_{L_1}$ [rad] & 0 (frozen) \\
 $\Delta\phi_{L_2}$ [rad] & $0.197^{+0.021}_{-0.022}$ \\
 $\Delta\phi_{L_n}$ [rad] & $1.01^{+0.16}_{-0.14}$ \\
 \small $N (L_0, \mathrm{CPD})$ & $\left(6.1^{+4.0\,\mathrm{(p)}}_{-2.3}\right)\times10^{-4}$ \\
 \small $N (L_0, \text{0.5--1\,keV})$ & $\left(6.6^{+3.5\,\mathrm{(p)}}_{-2.0}\right)\times10^{-4}$ \\
 \small $N (L_0, \text{2--4\,keV})$ & $\left(6\pm4\right)\times10^{-4}\,^\mathrm{(p)}$ \\
 \small $N (L_1, \mathrm{CPD})$ & $0.0448\pm0.0009$ \\
 \small $N (L_1, \text{0.5--1\,keV})$ & $0.0307\pm0.0006$ \\
 \small $N (L_1, \text{2--4\,keV})$ & $0.0665^{+0.0016}_{-0.0017}$ \\
 \small $N (L_2, \mathrm{CPD})$ & $0.0129\pm0.0007$ \\
 \small $N (L_2, \text{0.5--1\,keV})$ & $\left(3.7\pm0.4\right)\times10^{-3}$ \\
 \small $N (L_2, \text{2--4\,keV})$ & $0.0428\pm0.0016$ \\
 \small $N (L_n, \mathrm{CPD})$ & $\left(1.59^{+0.34}_{-0.28}\right)\times10^{-3}$ \\
 \small $N (L_n, \text{0.5--1\,keV})$ & $\left(9.6^{+2.3}_{-2.0}\right)\times10^{-4}$ \\
 \small $N (L_n, \text{2--4\,keV})$ & $\left(10^{+8}_{-7}\right)\times10^{-4}$ \\
 \hline
\end{tabular}

\end{table}

\clearpage

\end{document}